\documentclass[twocolumn,superscriptaddress,nofootinbib,notitlepage,longbibliography,aps]{revtex4-1}
\pdfoutput=1

\usepackage{graphicx}
\usepackage{amsmath}
\usepackage{amssymb}
\usepackage{amsfonts}
\usepackage{physics}
\usepackage[dvipsnames]{xcolor}
\usepackage[linktoc=none]{hyperref}
\usepackage[utf8]{inputenc}
\usepackage{braket}
\usepackage{bbold}
\usepackage{graphicx}
\usepackage{hhline}

\hypersetup{colorlinks,linkcolor={blue},citecolor={teal},urlcolor={violet}}

\newcommand{\documenttitle}{Starburst Nuclei as Light Dark Matter Laboratories}

\newcommand{\INFN}{INFN - Sezione di Napoli, Complesso Univ. Monte S. Angelo, I-80126 Napoli, Italy}
\newcommand{\UNINA}{Dipartimento di Fisica ``Ettore Pancini'', Università degli studi di Napoli ``Federico II'', Complesso Univ. Monte S. Angelo, I-80126 Napoli, Italy}
\newcommand{\SSM}{Scuola Superiore Meridionale, Università degli studi di Napoli ``Federico II'', Largo San Marcellino 10, 80138 Napoli, Italy}
\newcommand{\NBIA}{Niels Bohr International Academy, Niels Bohr Institute, University of Copenhagen, Copenhagen, Denmark}

\newcommand{\INAF}{INAF-Osservatorio Astronomico di Capodimonte, Salita Moiariello 16,
I-80131 Naples, Italy}

\begin{document}

\title{\documenttitle}

\author{Antonio Ambrosone}
\email{aambrosone@na.infn.it}
\affiliation{\UNINA}
\affiliation{\INFN}
\author{Marco Chianese}
\email{chianese@na.infn.it}
\affiliation{\UNINA}
\affiliation{\INFN}
\author{Damiano F.G. Fiorillo}
\email{damiano.fiorillo@nbi.ku.dk}
\affiliation{\NBIA}
\author{Antonio Marinelli}
\email{antonio.marinelli@na.infn.it}
\affiliation{\UNINA}
\affiliation{\INFN}
\affiliation{\INAF}
\author{Gennaro Miele}
\email{miele@na.infn.it}
\affiliation{\UNINA}
\affiliation{\INFN}
\affiliation{\SSM}

\date{\today}
\begin{abstract}
Starburst galaxies are well-motivated astrophysical emitters of high-energy gamma-rays. They are well-known cosmic-ray ``reservoirs'', thanks to their large magnetic fields which confine high-energy protons for $\sim 10^5$~years. Over such long times, cosmic-ray transport can be significantly affected by scatterings with sub-GeV dark matter. Here we point out that this scattering distorts the cosmic-ray spectrum, and the distortion can be indirectly observed by measuring the gamma-rays produced by cosmic-rays via hadronic collisions. Present gamma-ray data show no sign of such a distortion, leading to stringent bounds on the cross section between protons and dark matter. These are highly complementary with current bounds and have large room for improvement with the future gamma-ray measurements in the 0.1--10~TeV range from the Cherenkov Telescope Array, which can strengthen the limits by as much as two orders of magnitude.
\end{abstract}

\maketitle
{\bf Introduction.~---}
The existence of Dark Matter (DM) is a milestone of the cosmological standard model~\cite{Planck:2018vyg}. However, its nature has not been identified yet~\cite{Bertone:2018krk,Kahlhoefer:2017dnp,PerezdelosHeros:2020qyt,Billard:2021uyg}. Astrophysical and cosmological observations reveal that galaxies, including the Milky Way (MW), posses a halo of non-relativistic DM particles~\cite{Iocco:2015xga,Bertone:2016nfn,Salucci:2018hqu,Werhahn:2021bal,Werhahn:2021gvl}. This has allowed direct-detection experiments to place powerful limits on the properties of DM particles which may elastically scatter off target nuclei~\cite{Billard:2021uyg}. However, due to poor sensitivity at low nuclear recoil energies, such searches are typically limited to DM masses higher than 1 GeV, leaving sub-GeV DM largely unexplored by direct measurements. To probe such light DM particles, novel approaches are required in addition to standard astrophysical~\cite{Arina:2018zcq,Cotner:2016dhw,Bertone:2016nfn,Green:2021jrr, Iocco:2015xga,DelPopolo:2007dna}, cosmological\cite{Ali-Haimoud:2015pwa,Gluscevic:2017ywp,Xu:2018efh,PhysRevD.98.023013,DES:2020fxi,Rogers:2021byl}, and collider~\cite{Daci:2015hca} searches. Ref.~\cite{Cappiello:2018hsu} proposed one such approach, pointing out that the spectrum of MW Cosmic-Rays (CRs) can be altered by DM-CR elastic interactions. Soon after, Refs.~\cite{Bringmann:2018cvk,Ema:2018bih} showed that this interaction produces Boosted Dark Matter (BDM) particles, which can then be probed in direct-detection experiments due to their large energies (see Refs.~\cite{Ema:2020ulo,Wang:2021jic,Granelli:2022ysi,Calabrese:2021src,Calabrese:2022rfa,Agashe:2014yua,Giudice:2017zke,Cappiello:2019qsw,Alvey:2019zaa,Dent:2019krz,Berger:2019ttc,Wang:2019jtk,Guo:2020drq,Jho:2020sku,Guo:2020oum,Dent:2020syp,Bell:2021xff,Feng:2021hyz,Das:2021lcr,Xia:2021vbz,Xia:2022tid,Super-Kamiokande:2017dch,Bondarenko:2019vrb,PROSPECT:2021awi,PandaX-II:2021kai,CDEX:2022fig,Maity:2022exk} for other BDM studies). 

Up until now, the impact of DM-CR interaction has been mainly analyzed in the context of our own Galaxy (few exceptions are Ref.~\cite{Wang:2021jic,Ferrer:2022kei,PhysRevD.84.069903,Cermeno:2022rni}). However, CRs suffer a larger effect in environments which confine CRs for long times, so that they traverse through the DM halo longer. Therefore, in this Letter we propose to use \textit{cosmic reservoirs}, namely sources which confine cosmic-rays, as a probe of DM-CR interactions. We focus on the nuclei of starburst galaxies (hereafter denoted as SBNi), which confine CRs~\cite{Peretti:2018tmo,Ambrosone:2020evo,Ambrosone:2021aaw} for $\sim 10^5$~years even at energies as large as $100$~TeV. While these CRs cannot be directly observed, they produce gamma-rays and neutrinos via hadronic collisions~\cite{Peretti:2018tmo,Peretti:2019vsj,Kornecki:2020riv,Kornecki:2021xiy,Ambrosone:2020evo,Ambrosone:2021aaw,Ambrosone:2022fip}. Therefore, DM-CR interaction can distort the CR spectrum, and in turn the gamma-ray flux observed from SBNi (see Fig.~\ref{fig:M82_timescale_fluxes}). Here we show that the gamma-ray data from two nearby starburst galaxies, M82 and NGC 253, do not exhibit such a distortion, allowing us to bound the DM-CR cross section at the level of $10^{-34}$~cm$^2$ for DM with $10$~keV masses, as shown in Fig.~\ref{fig:constraints}. The bounds can be substantially improved with a better knowledge of the gamma-ray flux at energies~0.1--10~TeV. We show that the future Cherenkov Telescope Array (CTA)~\cite{CTAConsortium:2018tzg} will be able to strengthen these bounds by as much as two orders of magnitude.

{\bf Cosmic-Ray transport in SBNi.~---}
High-energy gamma-rays in SBNi are produced by CRs, here assumed to be injected by supernova remnants. CR protons collide with interstellar gas, hadronically producing $\pi^0$ which decay to gamma-rays, while CR electrons leptonically produce gamma-rays via bremsstrahlung and inverse Compton scattering. Following Refs.~\cite{Peretti:2018tmo,Ambrosone:2020evo}, we assume steady balance between CR injection and cooling, advective, and diffusive escape from the SBN, modeled as a compact sphere with radius $R_{\rm SBN}\sim 10^2 \, \rm{pc}$. The CR momentum distribution $f_{\rm CR}(p)$ is
\begin{equation}\label{eq:leaky}
   f_{\rm CR}(p) = \left(\frac{1}{\tau_{\rm adv}} + \frac{1}{\tau_{\rm diff}} + \frac{1}{\tau_{\rm loss}^{\rm{eff}}}\right)^{-1}Q_{\rm CR}(p) \,,
\end{equation}
where $Q_{\rm CR}(p)$ is the injection rate from supernova remnants, and $\tau_i$ are the timescales for the various processes. We assume injection of primary protons and electrons with a power-law spectrum of spectral index $\Gamma+2$, as expected from diffusive shock acceleration. In principle, there might be a contamination of heavier nuclei e.g. Helium nuclei (see the Supplemental Material~VI, which includes Refs.~\cite{Gaisser:2013bla,Blasi_2012, Evoli:2008dv, Joshi:2013aua, Luque:2022buq, Kafexhiu:2014cua}) We also assume the injection rates to be directly proportional to the star formation rate of the source $\dot{M}_{*}$. Nevertheless, our results are independent of the specific acceleration mechanism, provided that the cosmic-rays follow a power law. We introduce an exponential cut-off at $10\, \rm{PeV}$ for protons and a gaussian cut-off at $10 \, \rm{TeV}$ for electrons. The advection timescale is $\tau_{\rm{adv}} = R_{\rm{SBN}}/v_{\rm{wind}}$, where $v_{\rm{wind}}$ is the wind velocity. Even though, for these sources, diffusion is expected to be irrelevant below $1$~PeV~\cite{Yoast-Hull:2013wwa,Lacki:2010vs,Lacki:2013ry}, we introduce it according to Ref.~\cite{Peretti:2018tmo}. Finally, the energy-loss timescale is
\begin{equation}
    \tau_{\rm loss}^{\rm{eff}} = \frac{1}{\Gamma - 1}\left[\sum_i\left(- \frac{1}{E}\frac{{\rm d}E}{{\rm d}t}\right)_i\right]^{-1}\,,
    \label{eq:loss}
\end{equation}
where the sum comprises radiative and collision processes (for further details see the Supplemental Material~I,  which includes Refs.~\cite{Peretti:2021yhc,10.1093/mnras/182.3.443}). In Eq.~\eqref{eq:loss}, we consider for protons ionization, Coulomb interactions, and proton-proton collisions, while for electrons ionization, synchrotron, bremsstrahlung, and inverse Compton scatterings off low-energy photons. From the CR distribution in  Eq.~\eqref{eq:leaky}, we obtain the gamma-ray spectrum, accounting both for pion production from proton-proton collisions and its subsequent decay, and for primary and secondary bremsstrahlung and Inverse Compton scattering (for further details see the Supplemental Material~II, which includes Refs.~\cite{Kelner:2006tc,Franceschini:2017iwq}).

{\bf DM-proton scatterings inside SBNi.~---}
If nucleons are coupled to DM (hereafter called $\chi$), CRs confined in the SBN are trapped for such a long time that they can collide with DM. Elastic DM-CR scatterings cause an additional energy-loss in Eq.~\eqref{eq:loss}, competing with the others for sufficiently large DM-proton cross sections:
\begin{equation}
    \tau^\mathrm{el}_{\chi p} = \left[-\frac{1}{E}   \left(\frac{{\rm d}E}{{\rm d} t}\right)_{\chi p}\right]^{-1}\,,
\end{equation}
with
\begin{equation}\label{eq:dmenergyloss}
    \left(\frac{{\rm d}E}{{\rm d} t}\right)_{\chi p} = \frac{\rho_\chi}{{m_\chi}}\,\int_{0}^{T^{\rm max}_\chi} {\rm d}T_\chi\, T_\chi \frac{{\rm d}\sigma_\mathrm{el}}{{\rm d}T_\chi} \,,
\end{equation}
where $m_\chi$ is the DM mass, $\rho_\chi$ is the spherically-symmetric DM density within the SBN, and ${\rm d}\sigma_\mathrm{el}/{\rm d}T_\chi$ is the differential elastic DM-proton cross section as a function of the final DM kinetic energy $T_\chi$. The maximal allowed value $T^{\rm max}_\chi$ for $T_\chi$ in a collision with a proton with kinetic energy $T = E-m_p$ is
\begin{equation}\label{eq:energygain}
    T^{\rm max}_\chi =   \frac{2T^2 + 4 m_p T}{m_\chi}\left[\left(1+\frac{m_p}{m_\chi}\right)^2 + \frac{2 T}{m_\chi}\right]^{-1} \,.
\end{equation}
The differential cross section depends on the DM-proton interaction. For definiteness, we consider Dirac fermion DM particles interacting with protons via a scalar mediator with a mass much larger than the transfer momentum $q^2 = 2m_\chi T_\chi$. Differently from Refs.~\cite{Cappiello:2018hsu,Wang:2021jic}, that assume a constant cross section with a flat spectrum in recoil energy, for $T_\chi \leq T^{\rm max}_\chi$ we have~\cite{Ema:2020ulo}
\begin{equation}\label{eq:differentialcrosssection}
    \frac{{\rm d}\sigma_\mathrm{el}}{{\rm d}T_\chi} = \frac{\sigma_{\chi p}}{ T^{\rm max}_\chi} \frac{F_{p}^{2} (q^2)}{16 \,\mu_{\chi p}^2\,s} \,  (q^2 + 4m^2_p)(q^2 + 4m^2_\chi)\,,
\end{equation}
where $\sigma_{\chi p}$ is the DM-proton cross section at zero center-of-mass momentum, $\mu_{\chi p}$ is the reduced mass of $\chi$ and proton, and $s = m_{\chi}^{2} + m_{p}^{2} + 2Em_{\chi}$ is center-of-mass energy. The quantity $F_p$ is the proton form factor~\cite{Angeli:2004kvy}
\begin{equation}\label{eq:formfactor}
    F_{p}(q^2) = \left(\frac{1}{1+ q^2/\Lambda^2}\right)^2 \quad {\rm with}\quad \Lambda = 0.770 \, \rm{GeV}\,.
\end{equation}
At energies much higher than $m_p^2/2m_\chi$, DM-CR scatterings become inelastic, breaking the proton and producing additional gamma-rays from the pion decay~\cite{Alvey:2022pad}. We model this process via a simple semi-analytic approximation similar to Refs.~\cite{Guo:2020oum,Cyburt:2002uw,Hooper:2018bfw}: we assume the DM-CR inelastic cross section to follow the neutrino-nucleon one and rescale it to match the DM-CR cross section in the elastic regime. In this way, the inelastic cross section $(\sigma_\mathrm{inel})$ is totally defined by means of $\sigma_{\chi p}$ in Eq.~\eqref{eq:differentialcrosssection}. The timescale for energy loss from inelastic DM-CR collision is 
\begin{equation}\label{eq:inelastic_DM_p}
     \tau^\mathrm{inel}_{\chi p} = \left(\kappa\,\sigma_{\rm{inel}}\, \frac{\rho_{\chi}}{m_{\chi}} \right)^{-1}\,,
\end{equation}
where $\kappa$ is the inelasticity of the process, assumed to be 0.5 as for inelastic proton-proton collisions. Finally, to evaluate the gamma-ray production in inelastic DM-CR scattering, we assume from each collision a gamma-ray emissivity analogous to proton-proton collision (for details see the Supplemental Material~II).

\begin{figure*}[t!]
    \centering
    \includegraphics[width=0.95\textwidth]{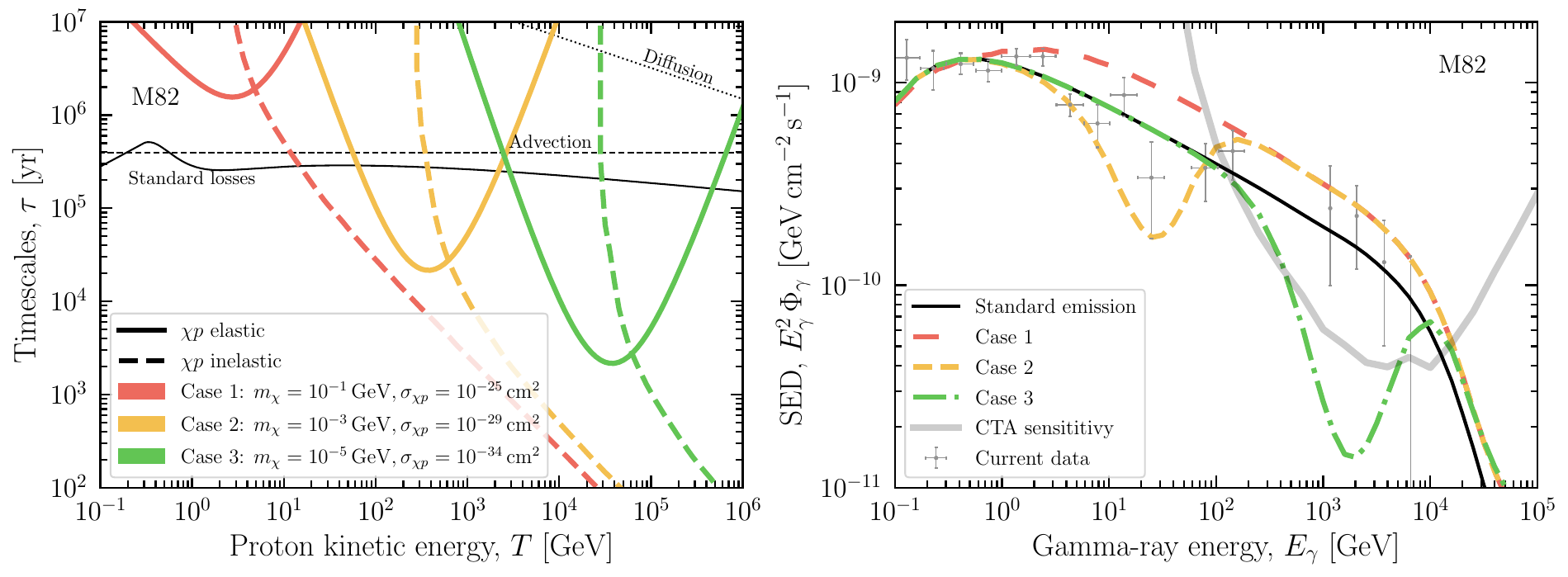}
    \caption{\textbf{Left panel.} Comparison between the proton timescales within M82 as a function of the proton kinetic energy $T$. The continuous, dashed and dotted black lines represent the standard losses, advection and diffusion timescales, respectively. The colored continuous (dashed) lines correspond to the elastic (inelastic) DM interactions for three different cases.
    \textbf{Right panel.} The expected gamma-ray fluxes from M82 compared to current data~\cite{Acciari:2009wq,Ajello:2020zna} and CTA sensitivity~\cite{CTAConsortium:2018tzg}. Analogously to the left panel, the black color line corresponds to the standard case (without DM-CR interactions), while the colored lines to the three different choices of $(m_{\chi},\,\sigma_{\chi p})$.}
    \label{fig:M82_timescale_fluxes}
\end{figure*}

The DM-CR scattering rate depends on the DM density distribution, which is pretty uncertain in the central cores of galaxies~\cite{Benito:2016kyp,Benito:2019ngh,Benito:2020lgu}. A benchmark parameterization is the Navarro-Frenk-White (NFW) distribution~\cite{Navarro:1995iw}
\begin{equation}\label{eq:NFW}
    \rho_\chi(r) = \frac{\rho_s}{r/r_s \left(1+r/r_s\right)^2}
\end{equation}
which is a function of the radial distance $r$ from the SBN center. The scale radius $r_s$ and the normalization $\rho_s$ can be expressed through the concentration parameter $c_{200} = r_{200}/r_s$ and the mass $M_{200}$ enclosed in a sphere of radius $r_{200}$, which is defined as the distance at which the mean DM density is 200 times the critical Universe density $\rho_{c}$.
These parameters are not measured, so we use the results of the simulations in Refs.~\cite{Werhahn:2021bal,Werhahn:2021gvl,Werhahn:2021jvy}, showing that $7 \leq c_{200} \leq 12$ and $10^{10} \leq M_{200}/\rm{M}_{\odot} \leq 10^{12}$. As benchmark cases in the following analysis we use $c_{200} = 7$ for both sources, $M_{200} = 10^{12}\, \rm{M}_{\odot}$ for M82 and $M_{200}= 3\times 10^{11}\, \rm{M}_{\odot}$ for NGC 253~\cite{Werhahn:2021bal,Werhahn:2021gvl,Werhahn:2021jvy}. In the Supplemental Material~IV (which includes Refs.~\cite{{Burkert:1995yz,Lin:2019yux}}) we quantify the impact of varying the halo parameters in the expected range, showing that it leads to an uncertainty of at most one order of magnitude in the constraints on DM-proton cross section $\sigma_{\chi p}$.

{\bf{Observable features in the gamma-ray spectrum.~---}}
The additional energy loss from elastic DM-CR interactions cause a suppression in the CR, and therefore in the gamma-ray spectrum, whereas the inelastic DM-CR production can replenish the gamma-ray spectrum at higher energies. These effects are visible in Fig.~\ref{fig:M82_timescale_fluxes} which represents the case of M82 source. The left panel shows the DM-CR energy-loss timescales ($\tau^\mathrm{el}_{\chi p}$ and $\tau^\mathrm{inel}_{\chi p}$), averaged within the SBN volume, in comparison with the standard timescales. At low CR energies, $T\ll E^p_\mathrm{dip}=m_p^2/(2m_\chi)$, $\tau^\mathrm{el}_{\chi p}\simeq 3 m_p^4 / 2 \rho_\chi \sigma_{\chi p} T^3$ rapidly decreases with the CR kinetic energy.  At high CR energies, elastic scattering becomes progressively unlikely compared with the inelastic one, so $\tau^\mathrm{el}_{\chi p}\simeq  128 m_\chi^6 T^3/\rho_\chi \sigma_{\chi p} \Lambda^8 \ln(T m_\chi / 2 m_p^2)$ increases with the CR kinetic energy. Elastic DM-CR scattering thus can cause a dip in the CR spectrum at an energy $E^p_\mathrm{dip}\simeq m_p^2/2m_\chi$, due to protons being pushed to lower energies. Above the dip, inelastic DM-CR scattering becomes the dominant source of CR energy loss. In each scattering the CR energy is reprocessed in gamma-rays, leading to a new calorimetric regime in which the gamma-ray spectrum again follows the CR injection power-law spectrum.

The right panel of Fig.~\ref{fig:M82_timescale_fluxes} shows the resulting gamma-ray spectrum, evidencing the dips corresponding to different masses $m_{\chi}$ due to elastic DM-CR scattering at an energy $E^\gamma_\mathrm{dip}\simeq 0.1 E^p_\mathrm{dip}$ -- since gamma-rays carry on average $10\%$ of the parent CR energy -- and the higher-energy power-law behavior of the gamma-rays from inelastic DM-CR scattering. The latter can exceed the gamma-rays produced in the standard proton-proton dominated regime (black line), in which the calorimetry is partial due to competition with advective escape. In the standard case without DM-CR interactions, only a fraction $\tau^\mathrm{eff}_\mathrm{loss} /(\tau^\mathrm{eff}_\mathrm{loss}+\tau_\mathrm{adv})\sim 40\%$ of the protons lose all of their energy to gamma-rays. However, we emphasize that the normalization of the gamma-ray spectrum after the dip also depends on the assumed inelasticity, which by rigor should be determined from the specific DM-quark coupling. Nevertheless, this has no significant impact on the bounds we derive, which essentially depend only on the behavior in the dip region and therefore on the elastic DM-CR scattering. Moreover, it is worth noticing that leptonic processes are completely subdominant in SBNi and cannot reduce the amplitude of the dip.

{\bf Statistical analysis.~---}
We analyze GeV-TeV data for both M82 and NGC 253. GeV data are obtained from the 10-year Fermi-LAT observation~\cite{Ajello:2020zna}. TeV data are taken for M82 from VERITAS~\cite{Acciari:2009wq} and for NGC 253 from H.E.S.S.~\cite{HESS:2018yqa}. All data-sets show a gamma-ray production up to TeV, with no hint of a break. Therefore, they strongly constrain DM-CR interactions.

To obtain these bounds, we follow Refs.~\cite{Ambrosone:2021aaw,Ambrosone:2022fip}, defining the likelihood as
\begin{equation}
    \mathcal{L}(m_\chi,\sigma_{\chi p},{\theta}) = e^{-\frac{1}{2}\sum_{i}\left(\frac{\mathrm{SED}_{i} - E_{i}^{2}\Phi_\gamma(E_{i} | m_\chi,\sigma_{\chi p},{\theta})}{\sigma_{i}}\right)^2}\,,
\end{equation}
where $\mathrm{SED}_{i}$ is the measured spectral energy distribution data, $E_{i}$ and $\sigma_{i}$ are respectively the centered energy bin value and the uncertainty on the data, and $i$ runs over the number of data points. Finally, $\Phi_\gamma(E_{i} | m_\chi,\sigma_{\chi p},{\theta})$ is the gamma-ray flux we compute, where $\theta$ represents the astrophysical nuisance parameters which are: $\dot{M}_{*}$, $\Gamma$, $R_{\mathrm{SBN}}$, $v_{\mathrm{wind}}$, $n_{\mathrm{ISM}}$, with  $n_{\mathrm{ISM}}$ being the interstellar gas density (the target for proton-proton collisions). For each of these parameters, we consider the realistic linear priors discussed in Ref.~\cite{Ambrosone:2022fip} to take into account the astrophysical uncertainties on the structural properties of M82 and NGC 253 (see for details Supplemental Material~III, which includes Refs.~\cite{Kennicutt:1997ng,Kennicutt:1998zb,Chevalier:1985pc,Kennicutt_2021}).

In order to obtain bounds on the DM-CR cross section, from the marginalized chi-squared $\chi^2(m_\chi,\sigma_{\chi p})=-2\ln\mathrm{max}_{\theta} \mathcal{L}(m_\chi,\sigma_{\chi p},\theta)$  we define the test statistic $\Delta\chi^2=\chi^2(m_\chi,\sigma_{\chi p})-\chi^2(m_\chi,0)$, comparing with the zero interaction case. We set bounds at $5\sigma$ confidence level by requiring $\Delta\chi^2=23.6$, since in the hypothesis of a DM signature the test statistic is distributed as a half-chi-squared variable.

\begin{figure}[t!]
    \centering
    \includegraphics[width=0.95\columnwidth]{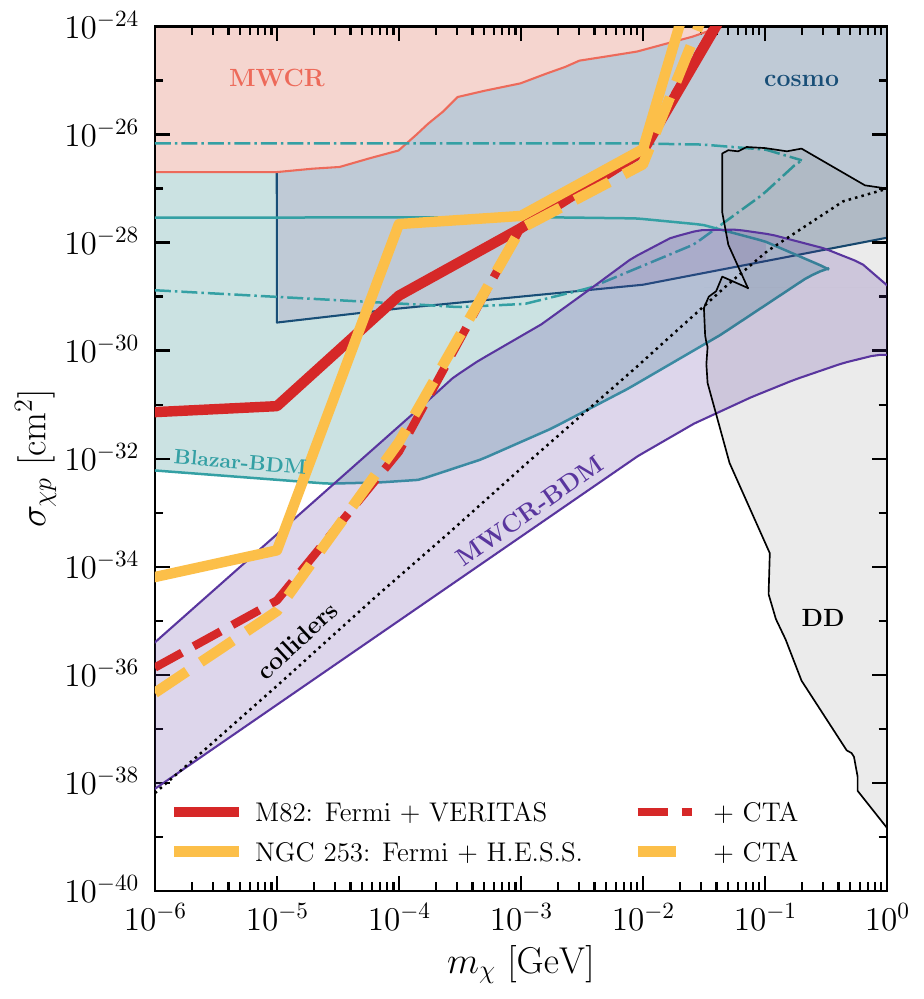}
    \caption{Constraints at 5$\sigma$ on DM-proton cross section placed by means of current (thick solid lines) and future CTA (thick dashed lines) data for M82 (red color) and NGC 253 (yellow color) galaxies. For comparison, the constraints from cosmological observations~\cite{Rogers:2021byl}, direct-detection experiments~\cite{PhysRevD.76.042007,PhysRevD.94.082006,CRESST:2015txj,Emken:2018run,EDELWEISS:2019vjv,CRESST:2017ues,Kouvaris:2016afs,PhysRevD.98.023005,Awe:2018fei,PhysRevLett.121.101801,Mahdawi:2018euy,LUX:2018akb,XENON:2019zpr,CRESST:2019jnq,CDEX:2019hzn,CDEX:2021cll,EDELWEISS:2022ktt}, colliders~\cite{Daci:2015hca}, Milky-Way Cosmic-Rays (MWCR)~\cite{Cappiello:2018hsu}, boosted dark matter from the blazar BL Lac (Blazar-BDM) with MiniBooNE (cyan dot-dashed lines) and XENON1T (cyan solid lines) detectors~\cite{Wang:2021jic}, and boosted dark matter due to MW cosmic-rays scatterings through a heavy mediator (MWCR-BDM)~\cite{Bondarenko:2019vrb} are reported.}
    \label{fig:constraints}
\end{figure}

{\bf Results and discussion.~---}
Fig. \ref{fig:constraints} summarizes the bounds we find on $\sigma_{\chi p}$ as a function of $m_{\chi}$, both for the case of M82 and NGC 253. The bounds flatten out for $m_\chi\lesssim 1$~keV, since lighter masses cause a dip at $E^\gamma_\mathrm{dip}\gtrsim 50$~TeV, where gamma-rays cannot be observed due to attenuation on extragalactic background light. NGC 253 leads to significantly better bounds at low masses due to the larger number of data points in the TeV region. Indeed, the main limitation from present-day data is the limited statistics in the $1-10$~TeV window. To quantify this, we perform a forecast analysis for the CTA telescope~\cite{CTAConsortium:2018tzg}, for both sources. CTA will dramatically improve the gamma-ray measurements in this energy region, as already shown in Ref.~\cite{Ambrosone:2022fip}. We generate 50 mock data samples (see the Supplemental Material~V for details), and we obtain the projected bounds for each sample. Fig.~\ref{fig:constraints} shows the mean values of these bounds. CTA will strengthen the constraints up to two orders of magnitude for NGC 253 and five orders of magnitude for M82 in the low-mass region. We emphasize that constraining DM-CR scattering using starburst galaxies has the additional advantage that different galaxies can be used to make the results more robust, and the bounds from different sources can be combined to provide more stringent exclusions on the DM properties.

Our bounds are complementary to the direct-detection of boosted DM, whose bounds exhibit a ceiling due to the atmosphere attenuation of the BDM flux.
Our bounds also look significantly stronger than the ones placed in Ref.~\cite{Cappiello:2018hsu} by searching for distortions of the Milky-Way CR spectrum due to DM-CR scattering, while for $m_{\chi} \lesssim \, 1 \, \rm{MeV}$ they are comparable with the ones derived from the non-observation of BDM particles from DM-CR interactions in blazars~\cite{Wang:2021jic}. However, the limits~\cite{Cappiello:2018hsu,Wang:2021jic} have been both obtained assuming an energy-independent DM-CR cross section, whereas we include the typical $\sigma_{\chi p}\propto E^2$ behavior due to a massive mediator, and a flat distribution for the DM recoil energy. Naively, since our bounds primarily come from CRs around $10$~TeV energies in the low-mass range, whereas $\sigma_{\chi p}$ is defined at a center-of-mass energy of order GeVs, they are stronger than the ones in Ref.~\cite{Cappiello:2018hsu} by about $10^8$ just because of the different cross section behavior. However, the difference in the recoil energy distribution also leads to a completely different shape for the bounds. For this reason, a comprehensive comparison would require a re-evaluation of their results, which is beyond the scope of this Letter.

Differently from the Milky Way~\cite{Cyburt:2002uw}, in SBNi the inelastic DM-CR scatterings are also less observationally interesting, since they just replace the proton-proton scatterings in making CRs lose their energy to gamma-rays. However, bounds based on inelastic scattering in the Milky Way are strongly dependent on how the differential cross section for gamma-ray production is modeled, which in turn requires a specific choice of the quark-DM coupling. Furthermore, these bounds are applicable only at large enough DM masses, in order that the cosmic-rays exceed the pion production threshold. Our bounds instead depend essentially on the elastic scattering, which requires no threshold condition, and therefore are robust against these uncertainties.

Concerning the blazar-BDM bounds, we also emphasize that they rely on the existence of a DM spike close to the central black hole. This is first of all impacted by the possibility of DM annihilation in the spike, as shown by Ref.~\cite{Wang:2021jic}. Furthermore, the steepness of the DM profile in the spike itself is subject of debate~\cite{Blandford:1999hh,Ullio:2001fb,PhysRevLett.83.1719,Gondolo:2000pn,Bertone:2001jv}, and these bounds may weaken considerably if the spike is less cuspy. On the other hand, our limits are pretty robust against astrophysical uncertainties, since they only rely on the existence of a CR power-law spectrum in SBNi.

In Fig.~\ref{fig:constraints}, the region above the dotted line is excluded by trackless jet searches at LHC, as pointed out by Ref.~\cite{Daci:2015hca}. However, this bound itself should possess a ceiling, since, if the particles interact too strongly, they do not reach the hadronic calorimeter. Since the size of the tracker and electromagnetic calorimeter is of the same order of magnitude as the hadronic calorimeter, we do not expect the collider limits to hold much more than an order of magnitude above the dotted line in Fig.~\ref{fig:constraints}.

Further, additional constraints could be placed by supernova observations~\cite{Knapen:2017xzo,Chang:2018rso,DeRocco:2019jti}, though to our knowledge no such bound exists in the literature for DM coupled to nucleons alone.

Finally, we briefly discuss the perspectives offered by neutrino astronomy. Neutrinos are in principle able to probe an energy range even higher than the gamma-ray one, since they travel unimpeded. Therefore, they could provide more stringent bounds in the low-mass region. The possibility of observing starburst galaxies as point sources at neutrino telescopes has been recently studied in Ref.~\cite{Ambrosone:2021aaw}. If the neutrino spectrum from one starburst galaxies is measured with a sufficiently good precision, precious knowledge will be gained on this mass region. We leave this possibility open for a future work. Furthermore, if the primary source of the diffuse neutrino flux is established to be hadronic production in SBNi, the absence of dips in the flux can also be used to constrain the DM-CR cross section. The present large uncertainties on the diffuse energy spectrum, and on the astrophysical origin of these neutrinos, make this possibility interesting only for future perspectives.

{\bf Conclusions.~---}
In this Letter, we have studied the phenomenology arising from scattering between high-energy protons and DM particles inside SBNi. We have shown that current data can exclude DM-CR interaction down to $\sigma_{\chi p}\lesssim10^{-34}\, \rm{cm}^{2}$ for $m_{\chi} \lesssim 10^{-6}\, \rm{GeV}$. We have also obtained projected bounds from the future CTA, showing that it will improve the DM-CR constraints down to $\sigma_{\chi p}\sim 10^{-37}-10^{-36}\, \rm{cm}^{2}$. Therefore, due to their nature of cosmic-rays reservoirs, starburst galaxies could play a significant role in investigating sub-GeV DM candidates, probing a region in between cosmological and collider bounds.

{\bf Acknowledgements.~---}
We thank Edoardo Vitagliano for fruitful discussion on supernovae limits.
AA, MC, AM and GM are supported by the research grant number 2017W4HA7S ``NAT-NET: Neutrino and Astroparticle Theory Network'' under the program PRIN 2017 funded by the Italian Ministero dell'Universit\`a e della Ricerca (MUR) and by the research project TAsP (Theoretical Astroparticle Physics) funded by the Istituto Nazionale di Fisica Nucleare (INFN).
DF is supported by the {\sc Villum Fonden} under project no.~29388.   This project has received funding from the European Union's Horizon 2020 research and innovation program under the Marie Sklodowska-Curie grant agreement No.~847523 ‘INTERACTIONS’.   This work used resources provided by the High Performance Computing Center at the University of Copenhagen.

\bibliography{references}

\begin{thebibliography}{118}%
\makeatletter
\providecommand \@ifxundefined [1]{%
 \@ifx{#1\undefined}
}%
\providecommand \@ifnum [1]{%
 \ifnum #1\expandafter \@firstoftwo
 \else \expandafter \@secondoftwo
 \fi
}%
\providecommand \@ifx [1]{%
 \ifx #1\expandafter \@firstoftwo
 \else \expandafter \@secondoftwo
 \fi
}%
\providecommand \natexlab [1]{#1}%
\providecommand \enquote  [1]{``#1''}%
\providecommand \bibnamefont  [1]{#1}%
\providecommand \bibfnamefont [1]{#1}%
\providecommand \citenamefont [1]{#1}%
\providecommand \href@noop [0]{\@secondoftwo}%
\providecommand \href [0]{\begingroup \@sanitize@url \@href}%
\providecommand \@href[1]{\@@startlink{#1}\@@href}%
\providecommand \@@href[1]{\endgroup#1\@@endlink}%
\providecommand \@sanitize@url [0]{\catcode `\\12\catcode `\$12\catcode
  `\&12\catcode `\#12\catcode `\^12\catcode `\_12\catcode `\%12\relax}%
\providecommand \@@startlink[1]{}%
\providecommand \@@endlink[0]{}%
\providecommand \url  [0]{\begingroup\@sanitize@url \@url }%
\providecommand \@url [1]{\endgroup\@href {#1}{\urlprefix }}%
\providecommand \urlprefix  [0]{URL }%
\providecommand \Eprint [0]{\href }%
\providecommand \doibase [0]{http://dx.doi.org/}%
\providecommand \selectlanguage [0]{\@gobble}%
\providecommand \bibinfo  [0]{\@secondoftwo}%
\providecommand \bibfield  [0]{\@secondoftwo}%
\providecommand \translation [1]{[#1]}%
\providecommand \BibitemOpen [0]{}%
\providecommand \bibitemStop [0]{}%
\providecommand \bibitemNoStop [0]{.\EOS\space}%
\providecommand \EOS [0]{\spacefactor3000\relax}%
\providecommand \BibitemShut  [1]{\csname bibitem#1\endcsname}%
\let\auto@bib@innerbib\@empty
\bibitem [{\citenamefont {Aghanim}\ \emph {et~al.}(2020)\citenamefont {Aghanim}
  \emph {et~al.}}]{Planck:2018vyg}%
  \BibitemOpen
  \bibfield  {author} {\bibinfo {author} {\bibfnamefont {N.}~\bibnamefont
  {Aghanim}} \emph {et~al.} (\bibinfo {collaboration} {Planck}),\ }\bibfield
  {title} {\enquote {\bibinfo {title} {{Planck 2018 results. VI. Cosmological
  parameters}},}\ }\href {\doibase 10.1051/0004-6361/201833910} {\bibfield
  {journal} {\bibinfo  {journal} {Astron. Astrophys.}\ }\textbf {\bibinfo
  {volume} {641}},\ \bibinfo {pages} {A6} (\bibinfo {year} {2020})},\ \bibinfo
  {note} {[Erratum: Astron.Astrophys. 652, C4 (2021)]},\ \Eprint
  {http://arxiv.org/abs/1807.06209} {arXiv:1807.06209 [astro-ph.CO]}
  \BibitemShut {NoStop}%
\bibitem [{\citenamefont {Bertone}\ and\ \citenamefont
  {Tait}(2018)}]{Bertone:2018krk}%
  \BibitemOpen
  \bibfield  {author} {\bibinfo {author} {\bibfnamefont {Gianfranco}\
  \bibnamefont {Bertone}}\ and\ \bibinfo {author} {\bibfnamefont {Tim}\
  \bibnamefont {Tait}, \bibfnamefont {M.~P.}},\ }\bibfield  {title} {\enquote
  {\bibinfo {title} {{A new era in the search for dark matter}},}\ }\href
  {\doibase 10.1038/s41586-018-0542-z} {\bibfield  {journal} {\bibinfo
  {journal} {Nature}\ }\textbf {\bibinfo {volume} {562}},\ \bibinfo {pages}
  {51--56} (\bibinfo {year} {2018})},\ \Eprint
  {http://arxiv.org/abs/1810.01668} {arXiv:1810.01668 [astro-ph.CO]}
  \BibitemShut {NoStop}%
\bibitem [{\citenamefont {Kahlhoefer}(2017)}]{Kahlhoefer:2017dnp}%
  \BibitemOpen
  \bibfield  {author} {\bibinfo {author} {\bibfnamefont {Felix}\ \bibnamefont
  {Kahlhoefer}},\ }\bibfield  {title} {\enquote {\bibinfo {title} {{Review of
  LHC Dark Matter Searches}},}\ }\href {\doibase 10.1142/S0217751X1730006X}
  {\bibfield  {journal} {\bibinfo  {journal} {Int. J. Mod. Phys. A}\ }\textbf
  {\bibinfo {volume} {32}},\ \bibinfo {pages} {1730006} (\bibinfo {year}
  {2017})},\ \Eprint {http://arxiv.org/abs/1702.02430} {arXiv:1702.02430
  [hep-ph]} \BibitemShut {NoStop}%
\bibitem [{\citenamefont {P\'erez de~los
  Heros}(2020)}]{PerezdelosHeros:2020qyt}%
  \BibitemOpen
  \bibfield  {author} {\bibinfo {author} {\bibfnamefont {Carlos}\ \bibnamefont
  {P\'erez de~los Heros}},\ }\bibfield  {title} {\enquote {\bibinfo {title}
  {{Status, Challenges and Directions in Indirect Dark Matter Searches}},}\
  }\href {\doibase 10.3390/sym12101648} {\bibfield  {journal} {\bibinfo
  {journal} {Symmetry}\ }\textbf {\bibinfo {volume} {12}},\ \bibinfo {pages}
  {1648} (\bibinfo {year} {2020})},\ \Eprint {http://arxiv.org/abs/2008.11561}
  {arXiv:2008.11561 [astro-ph.HE]} \BibitemShut {NoStop}%
\bibitem [{\citenamefont {Billard}\ \emph {et~al.}(2022)\citenamefont {Billard}
  \emph {et~al.}}]{Billard:2021uyg}%
  \BibitemOpen
  \bibfield  {author} {\bibinfo {author} {\bibfnamefont {Julien}\ \bibnamefont
  {Billard}} \emph {et~al.},\ }\bibfield  {title} {\enquote {\bibinfo {title}
  {{Direct detection of dark matter\textemdash{}APPEC committee report*}},}\
  }\href {\doibase 10.1088/1361-6633/ac5754} {\bibfield  {journal} {\bibinfo
  {journal} {Rept. Prog. Phys.}\ }\textbf {\bibinfo {volume} {85}},\ \bibinfo
  {pages} {056201} (\bibinfo {year} {2022})},\ \Eprint
  {http://arxiv.org/abs/2104.07634} {arXiv:2104.07634 [hep-ex]} \BibitemShut
  {NoStop}%
\bibitem [{\citenamefont {Iocco}\ \emph {et~al.}(2015)\citenamefont {Iocco},
  \citenamefont {Pato},\ and\ \citenamefont {Bertone}}]{Iocco:2015xga}%
  \BibitemOpen
  \bibfield  {author} {\bibinfo {author} {\bibfnamefont {Fabio}\ \bibnamefont
  {Iocco}}, \bibinfo {author} {\bibfnamefont {Miguel}\ \bibnamefont {Pato}}, \
  and\ \bibinfo {author} {\bibfnamefont {Gianfranco}\ \bibnamefont {Bertone}},\
  }\bibfield  {title} {\enquote {\bibinfo {title} {{Evidence for dark matter in
  the inner Milky Way}},}\ }\href {\doibase 10.1038/nphys3237} {\bibfield
  {journal} {\bibinfo  {journal} {Nature Phys.}\ }\textbf {\bibinfo {volume}
  {11}},\ \bibinfo {pages} {245--248} (\bibinfo {year} {2015})},\ \Eprint
  {http://arxiv.org/abs/1502.03821} {arXiv:1502.03821 [astro-ph.GA]}
  \BibitemShut {NoStop}%
\bibitem [{\citenamefont {Bertone}\ and\ \citenamefont
  {Hooper}(2018)}]{Bertone:2016nfn}%
  \BibitemOpen
  \bibfield  {author} {\bibinfo {author} {\bibfnamefont {Gianfranco}\
  \bibnamefont {Bertone}}\ and\ \bibinfo {author} {\bibfnamefont {Dan}\
  \bibnamefont {Hooper}},\ }\bibfield  {title} {\enquote {\bibinfo {title}
  {{History of dark matter}},}\ }\href {\doibase 10.1103/RevModPhys.90.045002}
  {\bibfield  {journal} {\bibinfo  {journal} {Rev. Mod. Phys.}\ }\textbf
  {\bibinfo {volume} {90}},\ \bibinfo {pages} {045002} (\bibinfo {year}
  {2018})},\ \Eprint {http://arxiv.org/abs/1605.04909} {arXiv:1605.04909
  [astro-ph.CO]} \BibitemShut {NoStop}%
\bibitem [{\citenamefont {Salucci}(2019)}]{Salucci:2018hqu}%
  \BibitemOpen
  \bibfield  {author} {\bibinfo {author} {\bibfnamefont {Paolo}\ \bibnamefont
  {Salucci}},\ }\bibfield  {title} {\enquote {\bibinfo {title} {{The
  distribution of dark matter in galaxies}},}\ }\href {\doibase
  10.1007/s00159-018-0113-1} {\bibfield  {journal} {\bibinfo  {journal}
  {Astron. Astrophys. Rev.}\ }\textbf {\bibinfo {volume} {27}},\ \bibinfo
  {pages} {2} (\bibinfo {year} {2019})},\ \Eprint
  {http://arxiv.org/abs/1811.08843} {arXiv:1811.08843 [astro-ph.GA]}
  \BibitemShut {NoStop}%
\bibitem [{\citenamefont {Werhahn}\ \emph
  {et~al.}(2021{\natexlab{a}})\citenamefont {Werhahn}, \citenamefont
  {Pfrommer}, \citenamefont {Girichidis},\ and\ \citenamefont
  {Winner}}]{Werhahn:2021bal}%
  \BibitemOpen
  \bibfield  {author} {\bibinfo {author} {\bibfnamefont {Maria}\ \bibnamefont
  {Werhahn}}, \bibinfo {author} {\bibfnamefont {Christoph}\ \bibnamefont
  {Pfrommer}}, \bibinfo {author} {\bibfnamefont {Philipp}\ \bibnamefont
  {Girichidis}}, \ and\ \bibinfo {author} {\bibfnamefont {Georg}\ \bibnamefont
  {Winner}},\ }\bibfield  {title} {\enquote {\bibinfo {title} {{Cosmic rays and
  non-thermal emission in simulated galaxies \textendash{} II.
  \ensuremath{\gamma}-ray maps, spectra, and the
  far-infrared\textendash{}\ensuremath{\gamma}-ray relation}},}\ }\href
  {\doibase 10.1093/mnras/stab1325} {\bibfield  {journal} {\bibinfo  {journal}
  {Mon. Not. Roy. Astron. Soc.}\ }\textbf {\bibinfo {volume} {505}},\ \bibinfo
  {pages} {3295--3313} (\bibinfo {year} {2021}{\natexlab{a}})},\ \Eprint
  {http://arxiv.org/abs/2105.11463} {arXiv:2105.11463 [astro-ph.HE]}
  \BibitemShut {NoStop}%
\bibitem [{\citenamefont {Werhahn}\ \emph
  {et~al.}(2021{\natexlab{b}})\citenamefont {Werhahn}, \citenamefont
  {Pfrommer}, \citenamefont {Girichidis}, \citenamefont {Puchwein},\ and\
  \citenamefont {Pakmor}}]{Werhahn:2021gvl}%
  \BibitemOpen
  \bibfield  {author} {\bibinfo {author} {\bibfnamefont {Maria}\ \bibnamefont
  {Werhahn}}, \bibinfo {author} {\bibfnamefont {Christoph}\ \bibnamefont
  {Pfrommer}}, \bibinfo {author} {\bibfnamefont {Philipp}\ \bibnamefont
  {Girichidis}}, \bibinfo {author} {\bibfnamefont {Ewald}\ \bibnamefont
  {Puchwein}}, \ and\ \bibinfo {author} {\bibfnamefont {R\"udiger}\
  \bibnamefont {Pakmor}},\ }\bibfield  {title} {\enquote {\bibinfo {title}
  {{Cosmic rays and non-thermal emission in simulated galaxies \ensuremath{-}
  I. Electron and proton spectra compared to Voyager-1 data}},}\ }\href
  {\doibase 10.1093/mnras/stab1324} {\bibfield  {journal} {\bibinfo  {journal}
  {Mon. Not. Roy. Astron. Soc.}\ }\textbf {\bibinfo {volume} {505}},\ \bibinfo
  {pages} {3273--3294} (\bibinfo {year} {2021}{\natexlab{b}})},\ \Eprint
  {http://arxiv.org/abs/2105.10509} {arXiv:2105.10509 [astro-ph.HE]}
  \BibitemShut {NoStop}%
\bibitem [{\citenamefont {Arina}(2018)}]{Arina:2018zcq}%
  \BibitemOpen
  \bibfield  {author} {\bibinfo {author} {\bibfnamefont {Chiara}\ \bibnamefont
  {Arina}},\ }\bibfield  {title} {\enquote {\bibinfo {title} {{Impact of
  cosmological and astrophysical constraints on dark matter simplified
  models}},}\ }\href {\doibase 10.3389/fspas.2018.00030} {\bibfield  {journal}
  {\bibinfo  {journal} {Front. Astron. Space Sci.}\ }\textbf {\bibinfo {volume}
  {5}},\ \bibinfo {pages} {30} (\bibinfo {year} {2018})},\ \Eprint
  {http://arxiv.org/abs/1805.04290} {arXiv:1805.04290 [hep-ph]} \BibitemShut
  {NoStop}%
\bibitem [{\citenamefont {Cotner}\ and\ \citenamefont
  {Kusenko}(2016)}]{Cotner:2016dhw}%
  \BibitemOpen
  \bibfield  {author} {\bibinfo {author} {\bibfnamefont {Eric}\ \bibnamefont
  {Cotner}}\ and\ \bibinfo {author} {\bibfnamefont {Alexander}\ \bibnamefont
  {Kusenko}},\ }\bibfield  {title} {\enquote {\bibinfo {title} {{Astrophysical
  constraints on dark-matter $Q$-balls in the presence of baryon-violating
  operators}},}\ }\href {\doibase 10.1103/PhysRevD.94.123006} {\bibfield
  {journal} {\bibinfo  {journal} {Phys. Rev. D}\ }\textbf {\bibinfo {volume}
  {94}},\ \bibinfo {pages} {123006} (\bibinfo {year} {2016})},\ \Eprint
  {http://arxiv.org/abs/1609.00970} {arXiv:1609.00970 [hep-ph]} \BibitemShut
  {NoStop}%
\bibitem [{\citenamefont {Green}(2022)}]{Green:2021jrr}%
  \BibitemOpen
  \bibfield  {author} {\bibinfo {author} {\bibfnamefont {Anne~M.}\ \bibnamefont
  {Green}},\ }\bibfield  {title} {\enquote {\bibinfo {title} {{Dark matter in
  astrophysics/cosmology}},}\ }\href {\doibase
  10.21468/SciPostPhysLectNotes.37} {\bibfield  {journal} {\bibinfo  {journal}
  {SciPost Phys. Lect. Notes}\ }\textbf {\bibinfo {volume} {37}},\ \bibinfo
  {pages} {1} (\bibinfo {year} {2022})},\ \Eprint
  {http://arxiv.org/abs/2109.05854} {arXiv:2109.05854 [hep-ph]} \BibitemShut
  {NoStop}%
\bibitem [{\citenamefont {Del~Popolo}(2007)}]{DelPopolo:2007dna}%
  \BibitemOpen
  \bibfield  {author} {\bibinfo {author} {\bibfnamefont {Antonino}\
  \bibnamefont {Del~Popolo}},\ }\bibfield  {title} {\enquote {\bibinfo {title}
  {{Dark matter and structure formation a review}},}\ }\href {\doibase
  10.1134/S1063772907030018} {\bibfield  {journal} {\bibinfo  {journal}
  {Astron. Rep.}\ }\textbf {\bibinfo {volume} {51}},\ \bibinfo {pages}
  {169--196} (\bibinfo {year} {2007})},\ \Eprint
  {http://arxiv.org/abs/0801.1091} {arXiv:0801.1091 [astro-ph]} \BibitemShut
  {NoStop}%
\bibitem [{\citenamefont {Ali-Ha\"\i{}moud}\ \emph {et~al.}(2015)\citenamefont
  {Ali-Ha\"\i{}moud}, \citenamefont {Chluba},\ and\ \citenamefont
  {Kamionkowski}}]{Ali-Haimoud:2015pwa}%
  \BibitemOpen
  \bibfield  {author} {\bibinfo {author} {\bibfnamefont {Yacine}\ \bibnamefont
  {Ali-Ha\"\i{}moud}}, \bibinfo {author} {\bibfnamefont {Jens}\ \bibnamefont
  {Chluba}}, \ and\ \bibinfo {author} {\bibfnamefont {Marc}\ \bibnamefont
  {Kamionkowski}},\ }\bibfield  {title} {\enquote {\bibinfo {title}
  {{Constraints on Dark Matter Interactions with Standard Model Particles from
  Cosmic Microwave Background Spectral Distortions}},}\ }\href {\doibase
  10.1103/PhysRevLett.115.071304} {\bibfield  {journal} {\bibinfo  {journal}
  {Phys. Rev. Lett.}\ }\textbf {\bibinfo {volume} {115}},\ \bibinfo {pages}
  {071304} (\bibinfo {year} {2015})},\ \Eprint
  {http://arxiv.org/abs/1506.04745} {arXiv:1506.04745 [astro-ph.CO]}
  \BibitemShut {NoStop}%
\bibitem [{\citenamefont {Gluscevic}\ and\ \citenamefont
  {Boddy}(2018)}]{Gluscevic:2017ywp}%
  \BibitemOpen
  \bibfield  {author} {\bibinfo {author} {\bibfnamefont {Vera}\ \bibnamefont
  {Gluscevic}}\ and\ \bibinfo {author} {\bibfnamefont {Kimberly~K.}\
  \bibnamefont {Boddy}},\ }\bibfield  {title} {\enquote {\bibinfo {title}
  {{Constraints on Scattering of keV\textendash{}TeV Dark Matter with Protons
  in the Early Universe}},}\ }\href {\doibase 10.1103/PhysRevLett.121.081301}
  {\bibfield  {journal} {\bibinfo  {journal} {Phys. Rev. Lett.}\ }\textbf
  {\bibinfo {volume} {121}},\ \bibinfo {pages} {081301} (\bibinfo {year}
  {2018})},\ \Eprint {http://arxiv.org/abs/1712.07133} {arXiv:1712.07133
  [astro-ph.CO]} \BibitemShut {NoStop}%
\bibitem [{\citenamefont {Xu}\ \emph {et~al.}(2018)\citenamefont {Xu},
  \citenamefont {Dvorkin},\ and\ \citenamefont {Chael}}]{Xu:2018efh}%
  \BibitemOpen
  \bibfield  {author} {\bibinfo {author} {\bibfnamefont {Weishuang~Linda}\
  \bibnamefont {Xu}}, \bibinfo {author} {\bibfnamefont {Cora}\ \bibnamefont
  {Dvorkin}}, \ and\ \bibinfo {author} {\bibfnamefont {Andrew}\ \bibnamefont
  {Chael}},\ }\bibfield  {title} {\enquote {\bibinfo {title} {{Probing sub-GeV
  Dark Matter-Baryon Scattering with Cosmological Observables}},}\ }\href
  {\doibase 10.1103/PhysRevD.97.103530} {\bibfield  {journal} {\bibinfo
  {journal} {Phys. Rev. D}\ }\textbf {\bibinfo {volume} {97}},\ \bibinfo
  {pages} {103530} (\bibinfo {year} {2018})},\ \Eprint
  {http://arxiv.org/abs/1802.06788} {arXiv:1802.06788 [astro-ph.CO]}
  \BibitemShut {NoStop}%
\bibitem [{\citenamefont {Slatyer}\ and\ \citenamefont
  {Wu}(2018)}]{PhysRevD.98.023013}%
  \BibitemOpen
  \bibfield  {author} {\bibinfo {author} {\bibfnamefont {Tracy~R.}\
  \bibnamefont {Slatyer}}\ and\ \bibinfo {author} {\bibfnamefont {Chih-Liang}\
  \bibnamefont {Wu}},\ }\bibfield  {title} {\enquote {\bibinfo {title}
  {Early-universe constraints on dark matter-baryon scattering and their
  implications for a global 21 cm signal},}\ }\href {\doibase
  10.1103/PhysRevD.98.023013} {\bibfield  {journal} {\bibinfo  {journal} {Phys.
  Rev. D}\ }\textbf {\bibinfo {volume} {98}},\ \bibinfo {pages} {023013}
  (\bibinfo {year} {2018})}\BibitemShut {NoStop}%
\bibitem [{\citenamefont {Nadler}\ \emph {et~al.}(2021)\citenamefont {Nadler}
  \emph {et~al.}}]{DES:2020fxi}%
  \BibitemOpen
  \bibfield  {author} {\bibinfo {author} {\bibfnamefont {E.~O.}\ \bibnamefont
  {Nadler}} \emph {et~al.} (\bibinfo {collaboration} {DES}),\ }\bibfield
  {title} {\enquote {\bibinfo {title} {{Milky Way Satellite Census. III.
  Constraints on Dark Matter Properties from Observations of Milky Way
  Satellite Galaxies}},}\ }\href {\doibase 10.1103/PhysRevLett.126.091101}
  {\bibfield  {journal} {\bibinfo  {journal} {Phys. Rev. Lett.}\ }\textbf
  {\bibinfo {volume} {126}},\ \bibinfo {pages} {091101} (\bibinfo {year}
  {2021})},\ \Eprint {http://arxiv.org/abs/2008.00022} {arXiv:2008.00022
  [astro-ph.CO]} \BibitemShut {NoStop}%
\bibitem [{\citenamefont {Rogers}\ \emph {et~al.}(2022)\citenamefont {Rogers},
  \citenamefont {Dvorkin},\ and\ \citenamefont {Peiris}}]{Rogers:2021byl}%
  \BibitemOpen
  \bibfield  {author} {\bibinfo {author} {\bibfnamefont {Keir~K.}\ \bibnamefont
  {Rogers}}, \bibinfo {author} {\bibfnamefont {Cora}\ \bibnamefont {Dvorkin}},
  \ and\ \bibinfo {author} {\bibfnamefont {Hiranya~V.}\ \bibnamefont
  {Peiris}},\ }\bibfield  {title} {\enquote {\bibinfo {title} {{Limits on the
  Light Dark Matter\textendash{}Proton Cross Section from Cosmic Large-Scale
  Structure}},}\ }\href {\doibase 10.1103/PhysRevLett.128.171301} {\bibfield
  {journal} {\bibinfo  {journal} {Phys. Rev. Lett.}\ }\textbf {\bibinfo
  {volume} {128}},\ \bibinfo {pages} {171301} (\bibinfo {year} {2022})},\
  \Eprint {http://arxiv.org/abs/2111.10386} {arXiv:2111.10386 [astro-ph.CO]}
  \BibitemShut {NoStop}%
\bibitem [{\citenamefont {Daci}\ \emph {et~al.}(2015)\citenamefont {Daci},
  \citenamefont {De~Bruyn}, \citenamefont {Lowette}, \citenamefont {Tytgat},\
  and\ \citenamefont {Zaldivar}}]{Daci:2015hca}%
  \BibitemOpen
  \bibfield  {author} {\bibinfo {author} {\bibfnamefont {N.}~\bibnamefont
  {Daci}}, \bibinfo {author} {\bibfnamefont {Isabelle}\ \bibnamefont
  {De~Bruyn}}, \bibinfo {author} {\bibfnamefont {S.}~\bibnamefont {Lowette}},
  \bibinfo {author} {\bibfnamefont {M.~H.~G.}\ \bibnamefont {Tytgat}}, \ and\
  \bibinfo {author} {\bibfnamefont {B.}~\bibnamefont {Zaldivar}},\ }\bibfield
  {title} {\enquote {\bibinfo {title} {{Simplified SIMPs and the LHC}},}\
  }\href {\doibase 10.1007/JHEP11(2015)108} {\bibfield  {journal} {\bibinfo
  {journal} {JHEP}\ }\textbf {\bibinfo {volume} {11}},\ \bibinfo {pages} {108}
  (\bibinfo {year} {2015})},\ \Eprint {http://arxiv.org/abs/1503.05505}
  {arXiv:1503.05505 [hep-ph]} \BibitemShut {NoStop}%
\bibitem [{\citenamefont {Cappiello}\ \emph {et~al.}(2019)\citenamefont
  {Cappiello}, \citenamefont {Ng},\ and\ \citenamefont
  {Beacom}}]{Cappiello:2018hsu}%
  \BibitemOpen
  \bibfield  {author} {\bibinfo {author} {\bibfnamefont {Christopher~V.}\
  \bibnamefont {Cappiello}}, \bibinfo {author} {\bibfnamefont {Kenny C.~Y.}\
  \bibnamefont {Ng}}, \ and\ \bibinfo {author} {\bibfnamefont {John~F.}\
  \bibnamefont {Beacom}},\ }\bibfield  {title} {\enquote {\bibinfo {title}
  {{Reverse Direct Detection: Cosmic Ray Scattering With Light Dark Matter}},}\
  }\href {\doibase 10.1103/PhysRevD.99.063004} {\bibfield  {journal} {\bibinfo
  {journal} {Phys. Rev. D}\ }\textbf {\bibinfo {volume} {99}},\ \bibinfo
  {pages} {063004} (\bibinfo {year} {2019})},\ \Eprint
  {http://arxiv.org/abs/1810.07705} {arXiv:1810.07705 [hep-ph]} \BibitemShut
  {NoStop}%
\bibitem [{\citenamefont {Bringmann}\ and\ \citenamefont
  {Pospelov}(2019)}]{Bringmann:2018cvk}%
  \BibitemOpen
  \bibfield  {author} {\bibinfo {author} {\bibfnamefont {Torsten}\ \bibnamefont
  {Bringmann}}\ and\ \bibinfo {author} {\bibfnamefont {Maxim}\ \bibnamefont
  {Pospelov}},\ }\bibfield  {title} {\enquote {\bibinfo {title} {{Novel direct
  detection constraints on light dark matter}},}\ }\href {\doibase
  10.1103/PhysRevLett.122.171801} {\bibfield  {journal} {\bibinfo  {journal}
  {Phys. Rev. Lett.}\ }\textbf {\bibinfo {volume} {122}},\ \bibinfo {pages}
  {171801} (\bibinfo {year} {2019})},\ \Eprint
  {http://arxiv.org/abs/1810.10543} {arXiv:1810.10543 [hep-ph]} \BibitemShut
  {NoStop}%
\bibitem [{\citenamefont {Ema}\ \emph {et~al.}(2019)\citenamefont {Ema},
  \citenamefont {Sala},\ and\ \citenamefont {Sato}}]{Ema:2018bih}%
  \BibitemOpen
  \bibfield  {author} {\bibinfo {author} {\bibfnamefont {Yohei}\ \bibnamefont
  {Ema}}, \bibinfo {author} {\bibfnamefont {Filippo}\ \bibnamefont {Sala}}, \
  and\ \bibinfo {author} {\bibfnamefont {Ryosuke}\ \bibnamefont {Sato}},\
  }\bibfield  {title} {\enquote {\bibinfo {title} {{Light Dark Matter at
  Neutrino Experiments}},}\ }\href {\doibase 10.1103/PhysRevLett.122.181802}
  {\bibfield  {journal} {\bibinfo  {journal} {Phys. Rev. Lett.}\ }\textbf
  {\bibinfo {volume} {122}},\ \bibinfo {pages} {181802} (\bibinfo {year}
  {2019})},\ \Eprint {http://arxiv.org/abs/1811.00520} {arXiv:1811.00520
  [hep-ph]} \BibitemShut {NoStop}%
\bibitem [{\citenamefont {Ema}\ \emph {et~al.}(2021)\citenamefont {Ema},
  \citenamefont {Sala},\ and\ \citenamefont {Sato}}]{Ema:2020ulo}%
  \BibitemOpen
  \bibfield  {author} {\bibinfo {author} {\bibfnamefont {Yohei}\ \bibnamefont
  {Ema}}, \bibinfo {author} {\bibfnamefont {Filippo}\ \bibnamefont {Sala}}, \
  and\ \bibinfo {author} {\bibfnamefont {Ryosuke}\ \bibnamefont {Sato}},\
  }\bibfield  {title} {\enquote {\bibinfo {title} {{Neutrino experiments probe
  hadrophilic light dark matter}},}\ }\href {\doibase
  10.21468/SciPostPhys.10.3.072} {\bibfield  {journal} {\bibinfo  {journal}
  {SciPost Phys.}\ }\textbf {\bibinfo {volume} {10}},\ \bibinfo {pages} {072}
  (\bibinfo {year} {2021})},\ \Eprint {http://arxiv.org/abs/2011.01939}
  {arXiv:2011.01939 [hep-ph]} \BibitemShut {NoStop}%
\bibitem [{\citenamefont {Wang}\ \emph {et~al.}(2022)\citenamefont {Wang},
  \citenamefont {Granelli},\ and\ \citenamefont {Ullio}}]{Wang:2021jic}%
  \BibitemOpen
  \bibfield  {author} {\bibinfo {author} {\bibfnamefont {Jin-Wei}\ \bibnamefont
  {Wang}}, \bibinfo {author} {\bibfnamefont {Alessandro}\ \bibnamefont
  {Granelli}}, \ and\ \bibinfo {author} {\bibfnamefont {Piero}\ \bibnamefont
  {Ullio}},\ }\bibfield  {title} {\enquote {\bibinfo {title} {{Direct Detection
  Constraints on Blazar-Boosted Dark Matter}},}\ }\href {\doibase
  10.1103/PhysRevLett.128.221104} {\bibfield  {journal} {\bibinfo  {journal}
  {Phys. Rev. Lett.}\ }\textbf {\bibinfo {volume} {128}},\ \bibinfo {pages}
  {221104} (\bibinfo {year} {2022})},\ \Eprint
  {http://arxiv.org/abs/2111.13644} {arXiv:2111.13644 [astro-ph.HE]}
  \BibitemShut {NoStop}%
\bibitem [{\citenamefont {Granelli}\ \emph {et~al.}(2022)\citenamefont
  {Granelli}, \citenamefont {Ullio},\ and\ \citenamefont
  {Wang}}]{Granelli:2022ysi}%
  \BibitemOpen
  \bibfield  {author} {\bibinfo {author} {\bibfnamefont {Alessandro}\
  \bibnamefont {Granelli}}, \bibinfo {author} {\bibfnamefont {Piero}\
  \bibnamefont {Ullio}}, \ and\ \bibinfo {author} {\bibfnamefont {Jin-Wei}\
  \bibnamefont {Wang}},\ }\bibfield  {title} {\enquote {\bibinfo {title}
  {{Blazar-boosted dark matter at Super-Kamiokande}},}\ }\href {\doibase
  10.1088/1475-7516/2022/07/013} {\bibfield  {journal} {\bibinfo  {journal}
  {JCAP}\ }\textbf {\bibinfo {volume} {07}},\ \bibinfo {pages} {013} (\bibinfo
  {year} {2022})},\ \Eprint {http://arxiv.org/abs/2202.07598} {arXiv:2202.07598
  [astro-ph.HE]} \BibitemShut {NoStop}%
\bibitem [{\citenamefont {Calabrese}\ \emph
  {et~al.}(2022{\natexlab{a}})\citenamefont {Calabrese}, \citenamefont
  {Chianese}, \citenamefont {Fiorillo},\ and\ \citenamefont
  {Saviano}}]{Calabrese:2021src}%
  \BibitemOpen
  \bibfield  {author} {\bibinfo {author} {\bibfnamefont {Roberta}\ \bibnamefont
  {Calabrese}}, \bibinfo {author} {\bibfnamefont {Marco}\ \bibnamefont
  {Chianese}}, \bibinfo {author} {\bibfnamefont {Damiano F.~G.}\ \bibnamefont
  {Fiorillo}}, \ and\ \bibinfo {author} {\bibfnamefont {Ninetta}\ \bibnamefont
  {Saviano}},\ }\bibfield  {title} {\enquote {\bibinfo {title} {{Direct
  detection of light dark matter from evaporating primordial black holes}},}\
  }\href {\doibase 10.1103/PhysRevD.105.L021302} {\bibfield  {journal}
  {\bibinfo  {journal} {Phys. Rev. D}\ }\textbf {\bibinfo {volume} {105}},\
  \bibinfo {pages} {L021302} (\bibinfo {year} {2022}{\natexlab{a}})},\ \Eprint
  {http://arxiv.org/abs/2107.13001} {arXiv:2107.13001 [hep-ph]} \BibitemShut
  {NoStop}%
\bibitem [{\citenamefont {Calabrese}\ \emph
  {et~al.}(2022{\natexlab{b}})\citenamefont {Calabrese}, \citenamefont
  {Chianese}, \citenamefont {Fiorillo},\ and\ \citenamefont
  {Saviano}}]{Calabrese:2022rfa}%
  \BibitemOpen
  \bibfield  {author} {\bibinfo {author} {\bibfnamefont {Roberta}\ \bibnamefont
  {Calabrese}}, \bibinfo {author} {\bibfnamefont {Marco}\ \bibnamefont
  {Chianese}}, \bibinfo {author} {\bibfnamefont {Damiano F.~G.}\ \bibnamefont
  {Fiorillo}}, \ and\ \bibinfo {author} {\bibfnamefont {Ninetta}\ \bibnamefont
  {Saviano}},\ }\bibfield  {title} {\enquote {\bibinfo {title} {{Electron
  scattering of light new particles from evaporating primordial black
  holes}},}\ }\href {\doibase 10.1103/PhysRevD.105.103024} {\bibfield
  {journal} {\bibinfo  {journal} {Phys. Rev. D}\ }\textbf {\bibinfo {volume}
  {105}},\ \bibinfo {pages} {103024} (\bibinfo {year} {2022}{\natexlab{b}})},\
  \Eprint {http://arxiv.org/abs/2203.17093} {arXiv:2203.17093 [hep-ph]}
  \BibitemShut {NoStop}%
\bibitem [{\citenamefont {Agashe}\ \emph {et~al.}(2014)\citenamefont {Agashe},
  \citenamefont {Cui}, \citenamefont {Necib},\ and\ \citenamefont
  {Thaler}}]{Agashe:2014yua}%
  \BibitemOpen
  \bibfield  {author} {\bibinfo {author} {\bibfnamefont {Kaustubh}\
  \bibnamefont {Agashe}}, \bibinfo {author} {\bibfnamefont {Yanou}\
  \bibnamefont {Cui}}, \bibinfo {author} {\bibfnamefont {Lina}\ \bibnamefont
  {Necib}}, \ and\ \bibinfo {author} {\bibfnamefont {Jesse}\ \bibnamefont
  {Thaler}},\ }\bibfield  {title} {\enquote {\bibinfo {title} {{(In)direct
  Detection of Boosted Dark Matter}},}\ }\href {\doibase
  10.1088/1475-7516/2014/10/062} {\bibfield  {journal} {\bibinfo  {journal}
  {JCAP}\ }\textbf {\bibinfo {volume} {10}},\ \bibinfo {pages} {062} (\bibinfo
  {year} {2014})},\ \Eprint {http://arxiv.org/abs/1405.7370} {arXiv:1405.7370
  [hep-ph]} \BibitemShut {NoStop}%
\bibitem [{\citenamefont {Giudice}\ \emph {et~al.}(2018)\citenamefont
  {Giudice}, \citenamefont {Kim}, \citenamefont {Park},\ and\ \citenamefont
  {Shin}}]{Giudice:2017zke}%
  \BibitemOpen
  \bibfield  {author} {\bibinfo {author} {\bibfnamefont {Gian~F.}\ \bibnamefont
  {Giudice}}, \bibinfo {author} {\bibfnamefont {Doojin}\ \bibnamefont {Kim}},
  \bibinfo {author} {\bibfnamefont {Jong-Chul}\ \bibnamefont {Park}}, \ and\
  \bibinfo {author} {\bibfnamefont {Seodong}\ \bibnamefont {Shin}},\ }\bibfield
   {title} {\enquote {\bibinfo {title} {{Inelastic Boosted Dark Matter at
  Direct Detection Experiments}},}\ }\href {\doibase
  10.1016/j.physletb.2018.03.043} {\bibfield  {journal} {\bibinfo  {journal}
  {Phys. Lett. B}\ }\textbf {\bibinfo {volume} {780}},\ \bibinfo {pages}
  {543--552} (\bibinfo {year} {2018})},\ \Eprint
  {http://arxiv.org/abs/1712.07126} {arXiv:1712.07126 [hep-ph]} \BibitemShut
  {NoStop}%
\bibitem [{\citenamefont {Cappiello}\ and\ \citenamefont
  {Beacom}(2019)}]{Cappiello:2019qsw}%
  \BibitemOpen
  \bibfield  {author} {\bibinfo {author} {\bibfnamefont {Christopher~V.}\
  \bibnamefont {Cappiello}}\ and\ \bibinfo {author} {\bibfnamefont {John~F.}\
  \bibnamefont {Beacom}},\ }\bibfield  {title} {\enquote {\bibinfo {title}
  {{Strong New Limits on Light Dark Matter from Neutrino Experiments}},}\
  }\href {\doibase 10.1103/PhysRevD.104.069901} {\bibfield  {journal} {\bibinfo
   {journal} {Phys. Rev. D}\ }\textbf {\bibinfo {volume} {100}},\ \bibinfo
  {pages} {103011} (\bibinfo {year} {2019})},\ \bibinfo {note} {[Erratum:
  Phys.Rev.D 104, 069901 (2021)]},\ \Eprint {http://arxiv.org/abs/1906.11283}
  {arXiv:1906.11283 [hep-ph]} \BibitemShut {NoStop}%
\bibitem [{\citenamefont {Alvey}\ \emph {et~al.}(2019)\citenamefont {Alvey},
  \citenamefont {Campos}, \citenamefont {Fairbairn},\ and\ \citenamefont
  {You}}]{Alvey:2019zaa}%
  \BibitemOpen
  \bibfield  {author} {\bibinfo {author} {\bibfnamefont {James}\ \bibnamefont
  {Alvey}}, \bibinfo {author} {\bibfnamefont {Miguel}\ \bibnamefont {Campos}},
  \bibinfo {author} {\bibfnamefont {Malcolm}\ \bibnamefont {Fairbairn}}, \ and\
  \bibinfo {author} {\bibfnamefont {Tevong}\ \bibnamefont {You}},\ }\bibfield
  {title} {\enquote {\bibinfo {title} {{Detecting Light Dark Matter via
  Inelastic Cosmic Ray Collisions}},}\ }\href {\doibase
  10.1103/PhysRevLett.123.261802} {\bibfield  {journal} {\bibinfo  {journal}
  {Phys. Rev. Lett.}\ }\textbf {\bibinfo {volume} {123}},\ \bibinfo {pages}
  {261802} (\bibinfo {year} {2019})},\ \Eprint
  {http://arxiv.org/abs/1905.05776} {arXiv:1905.05776 [hep-ph]} \BibitemShut
  {NoStop}%
\bibitem [{\citenamefont {Dent}\ \emph {et~al.}(2020)\citenamefont {Dent},
  \citenamefont {Dutta}, \citenamefont {Newstead},\ and\ \citenamefont
  {Shoemaker}}]{Dent:2019krz}%
  \BibitemOpen
  \bibfield  {author} {\bibinfo {author} {\bibfnamefont {James~B.}\
  \bibnamefont {Dent}}, \bibinfo {author} {\bibfnamefont {Bhaskar}\
  \bibnamefont {Dutta}}, \bibinfo {author} {\bibfnamefont {Jayden~L.}\
  \bibnamefont {Newstead}}, \ and\ \bibinfo {author} {\bibfnamefont {Ian~M.}\
  \bibnamefont {Shoemaker}},\ }\bibfield  {title} {\enquote {\bibinfo {title}
  {{Bounds on Cosmic Ray-Boosted Dark Matter in Simplified Models and its
  Corresponding Neutrino-Floor}},}\ }\href {\doibase
  10.1103/PhysRevD.101.116007} {\bibfield  {journal} {\bibinfo  {journal}
  {Phys. Rev. D}\ }\textbf {\bibinfo {volume} {101}},\ \bibinfo {pages}
  {116007} (\bibinfo {year} {2020})},\ \Eprint
  {http://arxiv.org/abs/1907.03782} {arXiv:1907.03782 [hep-ph]} \BibitemShut
  {NoStop}%
\bibitem [{\citenamefont {Berger}\ \emph {et~al.}(2021)\citenamefont {Berger},
  \citenamefont {Cui}, \citenamefont {Graham}, \citenamefont {Necib},
  \citenamefont {Petrillo}, \citenamefont {Stocks}, \citenamefont {Tsai},\ and\
  \citenamefont {Zhao}}]{Berger:2019ttc}%
  \BibitemOpen
  \bibfield  {author} {\bibinfo {author} {\bibfnamefont {Joshua}\ \bibnamefont
  {Berger}}, \bibinfo {author} {\bibfnamefont {Yanou}\ \bibnamefont {Cui}},
  \bibinfo {author} {\bibfnamefont {Mathew}\ \bibnamefont {Graham}}, \bibinfo
  {author} {\bibfnamefont {Lina}\ \bibnamefont {Necib}}, \bibinfo {author}
  {\bibfnamefont {Gianluca}\ \bibnamefont {Petrillo}}, \bibinfo {author}
  {\bibfnamefont {Dane}\ \bibnamefont {Stocks}}, \bibinfo {author}
  {\bibfnamefont {Yun-Tse}\ \bibnamefont {Tsai}}, \ and\ \bibinfo {author}
  {\bibfnamefont {Yue}\ \bibnamefont {Zhao}},\ }\bibfield  {title} {\enquote
  {\bibinfo {title} {{Prospects for detecting boosted dark matter in DUNE
  through hadronic interactions}},}\ }\href {\doibase
  10.1103/PhysRevD.103.095012} {\bibfield  {journal} {\bibinfo  {journal}
  {Phys. Rev. D}\ }\textbf {\bibinfo {volume} {103}},\ \bibinfo {pages}
  {095012} (\bibinfo {year} {2021})},\ \Eprint
  {http://arxiv.org/abs/1912.05558} {arXiv:1912.05558 [hep-ph]} \BibitemShut
  {NoStop}%
\bibitem [{\citenamefont {Wang}\ \emph {et~al.}(2020)\citenamefont {Wang},
  \citenamefont {Wu}, \citenamefont {Yang}, \citenamefont {Zhou},\ and\
  \citenamefont {Zhu}}]{Wang:2019jtk}%
  \BibitemOpen
  \bibfield  {author} {\bibinfo {author} {\bibfnamefont {Wenyu}\ \bibnamefont
  {Wang}}, \bibinfo {author} {\bibfnamefont {Lei}\ \bibnamefont {Wu}}, \bibinfo
  {author} {\bibfnamefont {Jin~Min}\ \bibnamefont {Yang}}, \bibinfo {author}
  {\bibfnamefont {Hang}\ \bibnamefont {Zhou}}, \ and\ \bibinfo {author}
  {\bibfnamefont {Bin}\ \bibnamefont {Zhu}},\ }\bibfield  {title} {\enquote
  {\bibinfo {title} {{Cosmic ray boosted sub-GeV gravitationally interacting
  dark matter in direct detection}},}\ }\href {\doibase
  10.1007/JHEP12(2020)072} {\bibfield  {journal} {\bibinfo  {journal} {JHEP}\
  }\textbf {\bibinfo {volume} {12}},\ \bibinfo {pages} {072} (\bibinfo {year}
  {2020})},\ \bibinfo {note} {[Erratum: JHEP 02, 052 (2021)]},\ \Eprint
  {http://arxiv.org/abs/1912.09904} {arXiv:1912.09904 [hep-ph]} \BibitemShut
  {NoStop}%
\bibitem [{\citenamefont {Guo}\ \emph {et~al.}(2020{\natexlab{a}})\citenamefont
  {Guo}, \citenamefont {Tsai},\ and\ \citenamefont {Wu}}]{Guo:2020drq}%
  \BibitemOpen
  \bibfield  {author} {\bibinfo {author} {\bibfnamefont {Gang}\ \bibnamefont
  {Guo}}, \bibinfo {author} {\bibfnamefont {Yue-Lin~Sming}\ \bibnamefont
  {Tsai}}, \ and\ \bibinfo {author} {\bibfnamefont {Meng-Ru}\ \bibnamefont
  {Wu}},\ }\bibfield  {title} {\enquote {\bibinfo {title} {{Probing cosmic-ray
  accelerated light dark matter with IceCube}},}\ }\href {\doibase
  10.1088/1475-7516/2020/10/049} {\bibfield  {journal} {\bibinfo  {journal}
  {JCAP}\ }\textbf {\bibinfo {volume} {10}},\ \bibinfo {pages} {049} (\bibinfo
  {year} {2020}{\natexlab{a}})},\ \Eprint {http://arxiv.org/abs/2004.03161}
  {arXiv:2004.03161 [astro-ph.HE]} \BibitemShut {NoStop}%
\bibitem [{\citenamefont {Jho}\ \emph {et~al.}(2020)\citenamefont {Jho},
  \citenamefont {Park}, \citenamefont {Park},\ and\ \citenamefont
  {Tseng}}]{Jho:2020sku}%
  \BibitemOpen
  \bibfield  {author} {\bibinfo {author} {\bibfnamefont {Yongsoo}\ \bibnamefont
  {Jho}}, \bibinfo {author} {\bibfnamefont {Jong-Chul}\ \bibnamefont {Park}},
  \bibinfo {author} {\bibfnamefont {Seong~Chan}\ \bibnamefont {Park}}, \ and\
  \bibinfo {author} {\bibfnamefont {Po-Yan}\ \bibnamefont {Tseng}},\ }\bibfield
   {title} {\enquote {\bibinfo {title} {{Leptonic New Force and Cosmic-ray
  Boosted Dark Matter for the XENON1T Excess}},}\ }\href {\doibase
  10.1016/j.physletb.2020.135863} {\bibfield  {journal} {\bibinfo  {journal}
  {Phys. Lett. B}\ }\textbf {\bibinfo {volume} {811}},\ \bibinfo {pages}
  {135863} (\bibinfo {year} {2020})},\ \Eprint
  {http://arxiv.org/abs/2006.13910} {arXiv:2006.13910 [hep-ph]} \BibitemShut
  {NoStop}%
\bibitem [{\citenamefont {Guo}\ \emph {et~al.}(2020{\natexlab{b}})\citenamefont
  {Guo}, \citenamefont {Tsai}, \citenamefont {Wu},\ and\ \citenamefont
  {Yuan}}]{Guo:2020oum}%
  \BibitemOpen
  \bibfield  {author} {\bibinfo {author} {\bibfnamefont {Gang}\ \bibnamefont
  {Guo}}, \bibinfo {author} {\bibfnamefont {Yue-Lin~Sming}\ \bibnamefont
  {Tsai}}, \bibinfo {author} {\bibfnamefont {Meng-Ru}\ \bibnamefont {Wu}}, \
  and\ \bibinfo {author} {\bibfnamefont {Qiang}\ \bibnamefont {Yuan}},\
  }\bibfield  {title} {\enquote {\bibinfo {title} {{Elastic and Inelastic
  Scattering of Cosmic-Rays on Sub-GeV Dark Matter}},}\ }\href {\doibase
  10.1103/PhysRevD.102.103004} {\bibfield  {journal} {\bibinfo  {journal}
  {Phys. Rev. D}\ }\textbf {\bibinfo {volume} {102}},\ \bibinfo {pages}
  {103004} (\bibinfo {year} {2020}{\natexlab{b}})},\ \Eprint
  {http://arxiv.org/abs/2008.12137} {arXiv:2008.12137 [astro-ph.HE]}
  \BibitemShut {NoStop}%
\bibitem [{\citenamefont {Dent}\ \emph {et~al.}(2021)\citenamefont {Dent},
  \citenamefont {Dutta}, \citenamefont {Newstead}, \citenamefont {Shoemaker},\
  and\ \citenamefont {Arellano}}]{Dent:2020syp}%
  \BibitemOpen
  \bibfield  {author} {\bibinfo {author} {\bibfnamefont {James~B.}\
  \bibnamefont {Dent}}, \bibinfo {author} {\bibfnamefont {Bhaskar}\
  \bibnamefont {Dutta}}, \bibinfo {author} {\bibfnamefont {Jayden~L.}\
  \bibnamefont {Newstead}}, \bibinfo {author} {\bibfnamefont {Ian~M.}\
  \bibnamefont {Shoemaker}}, \ and\ \bibinfo {author} {\bibfnamefont
  {Natalia~Tapia}\ \bibnamefont {Arellano}},\ }\bibfield  {title} {\enquote
  {\bibinfo {title} {{Present and future status of light dark matter models
  from cosmic-ray electron upscattering}},}\ }\href {\doibase
  10.1103/PhysRevD.103.095015} {\bibfield  {journal} {\bibinfo  {journal}
  {Phys. Rev. D}\ }\textbf {\bibinfo {volume} {103}},\ \bibinfo {pages}
  {095015} (\bibinfo {year} {2021})},\ \Eprint
  {http://arxiv.org/abs/2010.09749} {arXiv:2010.09749 [hep-ph]} \BibitemShut
  {NoStop}%
\bibitem [{\citenamefont {Bell}\ \emph {et~al.}(2021)\citenamefont {Bell},
  \citenamefont {Dent}, \citenamefont {Dutta}, \citenamefont {Ghosh},
  \citenamefont {Kumar}, \citenamefont {Newstead},\ and\ \citenamefont
  {Shoemaker}}]{Bell:2021xff}%
  \BibitemOpen
  \bibfield  {author} {\bibinfo {author} {\bibfnamefont {Nicole~F.}\
  \bibnamefont {Bell}}, \bibinfo {author} {\bibfnamefont {James~B.}\
  \bibnamefont {Dent}}, \bibinfo {author} {\bibfnamefont {Bhaskar}\
  \bibnamefont {Dutta}}, \bibinfo {author} {\bibfnamefont {Sumit}\ \bibnamefont
  {Ghosh}}, \bibinfo {author} {\bibfnamefont {Jason}\ \bibnamefont {Kumar}},
  \bibinfo {author} {\bibfnamefont {Jayden~L.}\ \bibnamefont {Newstead}}, \
  and\ \bibinfo {author} {\bibfnamefont {Ian~M.}\ \bibnamefont {Shoemaker}},\
  }\bibfield  {title} {\enquote {\bibinfo {title} {{Cosmic-ray upscattered
  inelastic dark matter}},}\ }\href {\doibase 10.1103/PhysRevD.104.076020}
  {\bibfield  {journal} {\bibinfo  {journal} {Phys. Rev. D}\ }\textbf {\bibinfo
  {volume} {104}},\ \bibinfo {pages} {076020} (\bibinfo {year} {2021})},\
  \Eprint {http://arxiv.org/abs/2108.00583} {arXiv:2108.00583 [hep-ph]}
  \BibitemShut {NoStop}%
\bibitem [{\citenamefont {Feng}\ \emph {et~al.}(2022)\citenamefont {Feng},
  \citenamefont {Kang}, \citenamefont {Lu}, \citenamefont {Tsai},\ and\
  \citenamefont {Zhang}}]{Feng:2021hyz}%
  \BibitemOpen
  \bibfield  {author} {\bibinfo {author} {\bibfnamefont {Jie-Cheng}\
  \bibnamefont {Feng}}, \bibinfo {author} {\bibfnamefont {Xian-Wei}\
  \bibnamefont {Kang}}, \bibinfo {author} {\bibfnamefont {Chih-Ting}\
  \bibnamefont {Lu}}, \bibinfo {author} {\bibfnamefont {Yue-Lin~Sming}\
  \bibnamefont {Tsai}}, \ and\ \bibinfo {author} {\bibfnamefont {Feng-Shou}\
  \bibnamefont {Zhang}},\ }\bibfield  {title} {\enquote {\bibinfo {title}
  {{Revising inelastic dark matter direct detection by including the cosmic ray
  acceleration}},}\ }\href {\doibase 10.1007/JHEP04(2022)080} {\bibfield
  {journal} {\bibinfo  {journal} {JHEP}\ }\textbf {\bibinfo {volume} {04}},\
  \bibinfo {pages} {080} (\bibinfo {year} {2022})},\ \Eprint
  {http://arxiv.org/abs/2110.08863} {arXiv:2110.08863 [hep-ph]} \BibitemShut
  {NoStop}%
\bibitem [{\citenamefont {Das}\ and\ \citenamefont {Sen}(2021)}]{Das:2021lcr}%
  \BibitemOpen
  \bibfield  {author} {\bibinfo {author} {\bibfnamefont {Anirban}\ \bibnamefont
  {Das}}\ and\ \bibinfo {author} {\bibfnamefont {Manibrata}\ \bibnamefont
  {Sen}},\ }\bibfield  {title} {\enquote {\bibinfo {title} {{Boosted dark
  matter from diffuse supernova neutrinos}},}\ }\href {\doibase
  10.1103/PhysRevD.104.075029} {\bibfield  {journal} {\bibinfo  {journal}
  {Phys. Rev. D}\ }\textbf {\bibinfo {volume} {104}},\ \bibinfo {pages}
  {075029} (\bibinfo {year} {2021})},\ \Eprint
  {http://arxiv.org/abs/2104.00027} {arXiv:2104.00027 [hep-ph]} \BibitemShut
  {NoStop}%
\bibitem [{\citenamefont {Xia}\ \emph {et~al.}(2022{\natexlab{a}})\citenamefont
  {Xia}, \citenamefont {Xu},\ and\ \citenamefont {Zhou}}]{Xia:2021vbz}%
  \BibitemOpen
  \bibfield  {author} {\bibinfo {author} {\bibfnamefont {Chen}\ \bibnamefont
  {Xia}}, \bibinfo {author} {\bibfnamefont {Yan-Hao}\ \bibnamefont {Xu}}, \
  and\ \bibinfo {author} {\bibfnamefont {Yu-Feng}\ \bibnamefont {Zhou}},\
  }\bibfield  {title} {\enquote {\bibinfo {title} {{Production and attenuation
  of cosmic-ray boosted dark matter}},}\ }\href {\doibase
  10.1088/1475-7516/2022/02/028} {\bibfield  {journal} {\bibinfo  {journal}
  {JCAP}\ }\textbf {\bibinfo {volume} {02}},\ \bibinfo {pages} {028} (\bibinfo
  {year} {2022}{\natexlab{a}})},\ \Eprint {http://arxiv.org/abs/2111.05559}
  {arXiv:2111.05559 [hep-ph]} \BibitemShut {NoStop}%
\bibitem [{\citenamefont {Xia}\ \emph {et~al.}(2022{\natexlab{b}})\citenamefont
  {Xia}, \citenamefont {Xu},\ and\ \citenamefont {Zhou}}]{Xia:2022tid}%
  \BibitemOpen
  \bibfield  {author} {\bibinfo {author} {\bibfnamefont {Chen}\ \bibnamefont
  {Xia}}, \bibinfo {author} {\bibfnamefont {Yan-Hao}\ \bibnamefont {Xu}}, \
  and\ \bibinfo {author} {\bibfnamefont {Yu-Feng}\ \bibnamefont {Zhou}},\
  }\bibfield  {title} {\enquote {\bibinfo {title} {{Azimuthal asymmetry in
  cosmic-ray boosted dark matter flux}},}\ }\href@noop {} {\  (\bibinfo {year}
  {2022}{\natexlab{b}})},\ \Eprint {http://arxiv.org/abs/2206.11454}
  {arXiv:2206.11454 [hep-ph]} \BibitemShut {NoStop}%
\bibitem [{\citenamefont {Kachulis}\ \emph {et~al.}(2018)\citenamefont
  {Kachulis} \emph {et~al.}}]{Super-Kamiokande:2017dch}%
  \BibitemOpen
  \bibfield  {author} {\bibinfo {author} {\bibfnamefont {C.}~\bibnamefont
  {Kachulis}} \emph {et~al.} (\bibinfo {collaboration} {Super-Kamiokande}),\
  }\bibfield  {title} {\enquote {\bibinfo {title} {{Search for Boosted Dark
  Matter Interacting With Electrons in Super-Kamiokande}},}\ }\href {\doibase
  10.1103/PhysRevLett.120.221301} {\bibfield  {journal} {\bibinfo  {journal}
  {Phys. Rev. Lett.}\ }\textbf {\bibinfo {volume} {120}},\ \bibinfo {pages}
  {221301} (\bibinfo {year} {2018})},\ \Eprint
  {http://arxiv.org/abs/1711.05278} {arXiv:1711.05278 [hep-ex]} \BibitemShut
  {NoStop}%
\bibitem [{\citenamefont {Bondarenko}\ \emph {et~al.}(2020)\citenamefont
  {Bondarenko}, \citenamefont {Boyarsky}, \citenamefont {Bringmann},
  \citenamefont {Hufnagel}, \citenamefont {Schmidt-Hoberg},\ and\ \citenamefont
  {Sokolenko}}]{Bondarenko:2019vrb}%
  \BibitemOpen
  \bibfield  {author} {\bibinfo {author} {\bibfnamefont {Kyrylo}\ \bibnamefont
  {Bondarenko}}, \bibinfo {author} {\bibfnamefont {Alexey}\ \bibnamefont
  {Boyarsky}}, \bibinfo {author} {\bibfnamefont {Torsten}\ \bibnamefont
  {Bringmann}}, \bibinfo {author} {\bibfnamefont {Marco}\ \bibnamefont
  {Hufnagel}}, \bibinfo {author} {\bibfnamefont {Kai}\ \bibnamefont
  {Schmidt-Hoberg}}, \ and\ \bibinfo {author} {\bibfnamefont {Anastasia}\
  \bibnamefont {Sokolenko}},\ }\bibfield  {title} {\enquote {\bibinfo {title}
  {{Direct detection and complementary constraints for sub-GeV dark matter}},}\
  }\href {\doibase 10.1007/JHEP03(2020)118} {\bibfield  {journal} {\bibinfo
  {journal} {JHEP}\ }\textbf {\bibinfo {volume} {03}},\ \bibinfo {pages} {118}
  (\bibinfo {year} {2020})},\ \Eprint {http://arxiv.org/abs/1909.08632}
  {arXiv:1909.08632 [hep-ph]} \BibitemShut {NoStop}%
\bibitem [{\citenamefont {Andriamirado}\ \emph {et~al.}(2021)\citenamefont
  {Andriamirado} \emph {et~al.}}]{PROSPECT:2021awi}%
  \BibitemOpen
  \bibfield  {author} {\bibinfo {author} {\bibfnamefont {M.}~\bibnamefont
  {Andriamirado}} \emph {et~al.} (\bibinfo {collaboration} {PROSPECT, (PROSPECT
  Collaboration)*}),\ }\bibfield  {title} {\enquote {\bibinfo {title} {{Limits
  on sub-GeV dark matter from the PROSPECT reactor antineutrino experiment}},}\
  }\href {\doibase 10.1103/PhysRevD.104.012009} {\bibfield  {journal} {\bibinfo
   {journal} {Phys. Rev. D}\ }\textbf {\bibinfo {volume} {104}},\ \bibinfo
  {pages} {012009} (\bibinfo {year} {2021})},\ \Eprint
  {http://arxiv.org/abs/2104.11219} {arXiv:2104.11219 [hep-ex]} \BibitemShut
  {NoStop}%
\bibitem [{\citenamefont {Cui}\ \emph {et~al.}(2022)\citenamefont {Cui} \emph
  {et~al.}}]{PandaX-II:2021kai}%
  \BibitemOpen
  \bibfield  {author} {\bibinfo {author} {\bibfnamefont {Xiangyi}\ \bibnamefont
  {Cui}} \emph {et~al.} (\bibinfo {collaboration} {PandaX-II}),\ }\bibfield
  {title} {\enquote {\bibinfo {title} {{Search for Cosmic-Ray Boosted Sub-GeV
  Dark Matter at the PandaX-II Experiment}},}\ }\href {\doibase
  10.1103/PhysRevLett.128.171801} {\bibfield  {journal} {\bibinfo  {journal}
  {Phys. Rev. Lett.}\ }\textbf {\bibinfo {volume} {128}},\ \bibinfo {pages}
  {171801} (\bibinfo {year} {2022})},\ \Eprint
  {http://arxiv.org/abs/2112.08957} {arXiv:2112.08957 [hep-ex]} \BibitemShut
  {NoStop}%
\bibitem [{\citenamefont {Xu}\ \emph {et~al.}(2022)\citenamefont {Xu} \emph
  {et~al.}}]{CDEX:2022fig}%
  \BibitemOpen
  \bibfield  {author} {\bibinfo {author} {\bibfnamefont {R.}~\bibnamefont {Xu}}
  \emph {et~al.} (\bibinfo {collaboration} {CDEX}),\ }\bibfield  {title}
  {\enquote {\bibinfo {title} {{Constraints on sub-GeV dark matter boosted by
  cosmic rays from the CDEX-10 experiment at the China Jinping Underground
  Laboratory}},}\ }\href {\doibase 10.1103/PhysRevD.106.052008} {\bibfield
  {journal} {\bibinfo  {journal} {Phys. Rev. D}\ }\textbf {\bibinfo {volume}
  {106}},\ \bibinfo {pages} {052008} (\bibinfo {year} {2022})},\ \Eprint
  {http://arxiv.org/abs/2201.01704} {arXiv:2201.01704 [hep-ex]} \BibitemShut
  {NoStop}%
\bibitem [{\citenamefont {Maity}\ and\ \citenamefont
  {Laha}(2022)}]{Maity:2022exk}%
  \BibitemOpen
  \bibfield  {author} {\bibinfo {author} {\bibfnamefont {Tarak~Nath}\
  \bibnamefont {Maity}}\ and\ \bibinfo {author} {\bibfnamefont {Ranjan}\
  \bibnamefont {Laha}},\ }\bibfield  {title} {\enquote {\bibinfo {title}
  {{Cosmic-ray boosted dark matter in Xe-based direct detection
  experiments}},}\ }\href@noop {} {\  (\bibinfo {year} {2022})},\ \Eprint
  {http://arxiv.org/abs/2210.01815} {arXiv:2210.01815 [hep-ph]} \BibitemShut
  {NoStop}%
\bibitem [{\citenamefont {Ferrer}\ \emph {et~al.}(2022)\citenamefont {Ferrer},
  \citenamefont {Herrera},\ and\ \citenamefont {Ibarra}}]{Ferrer:2022kei}%
  \BibitemOpen
  \bibfield  {author} {\bibinfo {author} {\bibfnamefont {Francesc}\
  \bibnamefont {Ferrer}}, \bibinfo {author} {\bibfnamefont {Gonzalo}\
  \bibnamefont {Herrera}}, \ and\ \bibinfo {author} {\bibfnamefont {Alejandro}\
  \bibnamefont {Ibarra}},\ }\bibfield  {title} {\enquote {\bibinfo {title}
  {{New constraints on the dark matter-neutrino and dark matter-photon
  scattering cross sections from TXS 0506+056}},}\ }\href@noop {} {\  (\bibinfo
  {year} {2022})},\ \Eprint {http://arxiv.org/abs/2209.06339} {arXiv:2209.06339
  [hep-ph]} \BibitemShut {NoStop}%
\bibitem [{\citenamefont {Gorchtein}\ \emph {et~al.}(2011)\citenamefont
  {Gorchtein}, \citenamefont {Profumo},\ and\ \citenamefont
  {Ubaldi}}]{PhysRevD.84.069903}%
  \BibitemOpen
  \bibfield  {author} {\bibinfo {author} {\bibfnamefont {Mikhail}\ \bibnamefont
  {Gorchtein}}, \bibinfo {author} {\bibfnamefont {Stefano}\ \bibnamefont
  {Profumo}}, \ and\ \bibinfo {author} {\bibfnamefont {Lorenzo}\ \bibnamefont
  {Ubaldi}},\ }\bibfield  {title} {\enquote {\bibinfo {title} {Erratum: Probing
  dark matter with active galactic nuclei jets [phys. rev. d 82, 083514
  (2010)]},}\ }\href {\doibase 10.1103/PhysRevD.84.069903} {\bibfield
  {journal} {\bibinfo  {journal} {Phys. Rev. D}\ }\textbf {\bibinfo {volume}
  {84}},\ \bibinfo {pages} {069903} (\bibinfo {year} {2011})}\BibitemShut
  {NoStop}%
\bibitem [{\citenamefont {Cerme\~no}\ \emph {et~al.}(2022)\citenamefont
  {Cerme\~no}, \citenamefont {Degrande},\ and\ \citenamefont
  {Mantani}}]{Cermeno:2022rni}%
  \BibitemOpen
  \bibfield  {author} {\bibinfo {author} {\bibfnamefont {Marina}\ \bibnamefont
  {Cerme\~no}}, \bibinfo {author} {\bibfnamefont {C\'eline}\ \bibnamefont
  {Degrande}}, \ and\ \bibinfo {author} {\bibfnamefont {Luca}\ \bibnamefont
  {Mantani}},\ }\bibfield  {title} {\enquote {\bibinfo {title} {{Signatures of
  leptophilic t-channel dark matter from active galactic nuclei}},}\ }\href
  {\doibase 10.1103/PhysRevD.105.083019} {\bibfield  {journal} {\bibinfo
  {journal} {Phys. Rev. D}\ }\textbf {\bibinfo {volume} {105}},\ \bibinfo
  {pages} {083019} (\bibinfo {year} {2022})},\ \Eprint
  {http://arxiv.org/abs/2201.07247} {arXiv:2201.07247 [hep-ph]} \BibitemShut
  {NoStop}%
\bibitem [{\citenamefont {Peretti}\ \emph {et~al.}(2019)\citenamefont
  {Peretti}, \citenamefont {Blasi}, \citenamefont {Aharonian},\ and\
  \citenamefont {Morlino}}]{Peretti:2018tmo}%
  \BibitemOpen
  \bibfield  {author} {\bibinfo {author} {\bibfnamefont {Enrico}\ \bibnamefont
  {Peretti}}, \bibinfo {author} {\bibfnamefont {Pasquale}\ \bibnamefont
  {Blasi}}, \bibinfo {author} {\bibfnamefont {Felix}\ \bibnamefont
  {Aharonian}}, \ and\ \bibinfo {author} {\bibfnamefont {Giovanni}\
  \bibnamefont {Morlino}},\ }\bibfield  {title} {\enquote {\bibinfo {title}
  {{Cosmic ray transport and radiative processes in nuclei of starburst
  galaxies}},}\ }\href {\doibase 10.1093/mnras/stz1161} {\bibfield  {journal}
  {\bibinfo  {journal} {Mon. Not. Roy. Astron. Soc.}\ }\textbf {\bibinfo
  {volume} {487}},\ \bibinfo {pages} {168--180} (\bibinfo {year} {2019})},\
  \Eprint {http://arxiv.org/abs/1812.01996} {arXiv:1812.01996 [astro-ph.HE]}
  \BibitemShut {NoStop}%
\bibitem [{\citenamefont {Ambrosone}\ \emph
  {et~al.}(2021{\natexlab{a}})\citenamefont {Ambrosone}, \citenamefont
  {Chianese}, \citenamefont {Fiorillo}, \citenamefont {Marinelli},
  \citenamefont {Miele},\ and\ \citenamefont {Pisanti}}]{Ambrosone:2020evo}%
  \BibitemOpen
  \bibfield  {author} {\bibinfo {author} {\bibfnamefont {Antonio}\ \bibnamefont
  {Ambrosone}}, \bibinfo {author} {\bibfnamefont {Marco}\ \bibnamefont
  {Chianese}}, \bibinfo {author} {\bibfnamefont {Damiano F.~G.}\ \bibnamefont
  {Fiorillo}}, \bibinfo {author} {\bibfnamefont {Antonio}\ \bibnamefont
  {Marinelli}}, \bibinfo {author} {\bibfnamefont {Gennaro}\ \bibnamefont
  {Miele}}, \ and\ \bibinfo {author} {\bibfnamefont {Ofelia}\ \bibnamefont
  {Pisanti}},\ }\bibfield  {title} {\enquote {\bibinfo {title} {{Starburst
  galaxies strike back: a multi-messenger analysis with Fermi-LAT and IceCube
  data}},}\ }\href {\doibase 10.1093/mnras/stab659} {\bibfield  {journal}
  {\bibinfo  {journal} {Mon. Not. Roy. Astron. Soc.}\ }\textbf {\bibinfo
  {volume} {503}},\ \bibinfo {pages} {4032} (\bibinfo {year}
  {2021}{\natexlab{a}})},\ \Eprint {http://arxiv.org/abs/2011.02483}
  {arXiv:2011.02483 [astro-ph.HE]} \BibitemShut {NoStop}%
\bibitem [{\citenamefont {Ambrosone}\ \emph
  {et~al.}(2021{\natexlab{b}})\citenamefont {Ambrosone}, \citenamefont
  {Chianese}, \citenamefont {Fiorillo}, \citenamefont {Marinelli},\ and\
  \citenamefont {Miele}}]{Ambrosone:2021aaw}%
  \BibitemOpen
  \bibfield  {author} {\bibinfo {author} {\bibfnamefont {Antonio}\ \bibnamefont
  {Ambrosone}}, \bibinfo {author} {\bibfnamefont {Marco}\ \bibnamefont
  {Chianese}}, \bibinfo {author} {\bibfnamefont {Damiano F.~G.}\ \bibnamefont
  {Fiorillo}}, \bibinfo {author} {\bibfnamefont {Antonio}\ \bibnamefont
  {Marinelli}}, \ and\ \bibinfo {author} {\bibfnamefont {Gennaro}\ \bibnamefont
  {Miele}},\ }\bibfield  {title} {\enquote {\bibinfo {title} {{Could Nearby
  Star-forming Galaxies Light Up the Pointlike Neutrino Sky?}}}\ }\href
  {\doibase 10.3847/2041-8213/ac25ff} {\bibfield  {journal} {\bibinfo
  {journal} {Astrophys. J. Lett.}\ }\textbf {\bibinfo {volume} {919}},\
  \bibinfo {pages} {L32} (\bibinfo {year} {2021}{\natexlab{b}})},\ \Eprint
  {http://arxiv.org/abs/2106.13248} {arXiv:2106.13248 [astro-ph.HE]}
  \BibitemShut {NoStop}%
\bibitem [{\citenamefont {Peretti}\ \emph {et~al.}(2020)\citenamefont
  {Peretti}, \citenamefont {Blasi}, \citenamefont {Aharonian}, \citenamefont
  {Morlino},\ and\ \citenamefont {Cristofari}}]{Peretti:2019vsj}%
  \BibitemOpen
  \bibfield  {author} {\bibinfo {author} {\bibfnamefont {Enrico}\ \bibnamefont
  {Peretti}}, \bibinfo {author} {\bibfnamefont {Pasquale}\ \bibnamefont
  {Blasi}}, \bibinfo {author} {\bibfnamefont {Felix}\ \bibnamefont
  {Aharonian}}, \bibinfo {author} {\bibfnamefont {Giovanni}\ \bibnamefont
  {Morlino}}, \ and\ \bibinfo {author} {\bibfnamefont {Pierre}\ \bibnamefont
  {Cristofari}},\ }\bibfield  {title} {\enquote {\bibinfo {title}
  {{Contribution of starburst nuclei to the diffuse gamma-ray and neutrino
  flux}},}\ }\href {\doibase 10.1093/mnras/staa698} {\bibfield  {journal}
  {\bibinfo  {journal} {Mon. Not. Roy. Astron. Soc.}\ }\textbf {\bibinfo
  {volume} {493}} (\bibinfo {year} {2020}),\ 10.1093/mnras/staa698},\ \Eprint
  {http://arxiv.org/abs/1911.06163} {arXiv:1911.06163 [astro-ph.HE]}
  \BibitemShut {NoStop}%
\bibitem [{\citenamefont {Kornecki}\ \emph {et~al.}(2020)\citenamefont
  {Kornecki}, \citenamefont {Pellizza}, \citenamefont {del Palacio},
  \citenamefont {M\"uller}, \citenamefont {Albacete-Colombo},\ and\
  \citenamefont {Romero}}]{Kornecki:2020riv}%
  \BibitemOpen
  \bibfield  {author} {\bibinfo {author} {\bibfnamefont {P.}~\bibnamefont
  {Kornecki}}, \bibinfo {author} {\bibfnamefont {L.~J.}\ \bibnamefont
  {Pellizza}}, \bibinfo {author} {\bibfnamefont {S.}~\bibnamefont {del
  Palacio}}, \bibinfo {author} {\bibfnamefont {A.~L.}\ \bibnamefont
  {M\"uller}}, \bibinfo {author} {\bibfnamefont {J.~F.}\ \bibnamefont
  {Albacete-Colombo}}, \ and\ \bibinfo {author} {\bibfnamefont {G.~E.}\
  \bibnamefont {Romero}},\ }\bibfield  {title} {\enquote {\bibinfo {title}
  {{$\gamma$-ray/infrared luminosity correlation of star-forming galaxies}},}\
  }\href {\doibase 10.1051/0004-6361/202038428} {\bibfield  {journal} {\bibinfo
   {journal} {Astron. Astrophys.}\ }\textbf {\bibinfo {volume} {641}},\
  \bibinfo {pages} {A147} (\bibinfo {year} {2020})},\ \Eprint
  {http://arxiv.org/abs/2007.07430} {arXiv:2007.07430 [astro-ph.HE]}
  \BibitemShut {NoStop}%
\bibitem [{\citenamefont {Kornecki}\ \emph {et~al.}(2022)\citenamefont
  {Kornecki}, \citenamefont {Peretti}, \citenamefont {del Palacio},
  \citenamefont {Benaglia},\ and\ \citenamefont {Pellizza}}]{Kornecki:2021xiy}%
  \BibitemOpen
  \bibfield  {author} {\bibinfo {author} {\bibfnamefont {P.}~\bibnamefont
  {Kornecki}}, \bibinfo {author} {\bibfnamefont {E.}~\bibnamefont {Peretti}},
  \bibinfo {author} {\bibfnamefont {S.}~\bibnamefont {del Palacio}}, \bibinfo
  {author} {\bibfnamefont {P.}~\bibnamefont {Benaglia}}, \ and\ \bibinfo
  {author} {\bibfnamefont {L.~J.}\ \bibnamefont {Pellizza}},\ }\bibfield
  {title} {\enquote {\bibinfo {title} {{Exploring the physics behind the
  non-thermal emission from star-forming galaxies detected in
  \ensuremath{\gamma} rays}},}\ }\href {\doibase 10.1051/0004-6361/202141295}
  {\bibfield  {journal} {\bibinfo  {journal} {Astron. Astrophys.}\ }\textbf
  {\bibinfo {volume} {657}},\ \bibinfo {pages} {A49} (\bibinfo {year}
  {2022})},\ \Eprint {http://arxiv.org/abs/2107.00823} {arXiv:2107.00823
  [astro-ph.HE]} \BibitemShut {NoStop}%
\bibitem [{\citenamefont {Ambrosone}\ \emph {et~al.}(2022)\citenamefont
  {Ambrosone}, \citenamefont {Chianese}, \citenamefont {Fiorillo},
  \citenamefont {Marinelli},\ and\ \citenamefont {Miele}}]{Ambrosone:2022fip}%
  \BibitemOpen
  \bibfield  {author} {\bibinfo {author} {\bibfnamefont {Antonio}\ \bibnamefont
  {Ambrosone}}, \bibinfo {author} {\bibfnamefont {Marco}\ \bibnamefont
  {Chianese}}, \bibinfo {author} {\bibfnamefont {Damiano F.~G.}\ \bibnamefont
  {Fiorillo}}, \bibinfo {author} {\bibfnamefont {Antonio}\ \bibnamefont
  {Marinelli}}, \ and\ \bibinfo {author} {\bibfnamefont {Gennaro}\ \bibnamefont
  {Miele}},\ }\bibfield  {title} {\enquote {\bibinfo {title} {{Observable
  signatures of cosmic rays transport in Starburst Galaxies on gamma-ray and
  neutrino observations}},}\ }\href {\doibase 10.1093/mnras/stac2133}
  {\bibfield  {journal} {\bibinfo  {journal} {Mon. Not. Roy. Astron. Soc.}\
  }\textbf {\bibinfo {volume} {515}},\ \bibinfo {pages} {5389--5399} (\bibinfo
  {year} {2022})},\ \Eprint {http://arxiv.org/abs/2203.03642} {arXiv:2203.03642
  [astro-ph.HE]} \BibitemShut {NoStop}%
\bibitem [{\citenamefont {Acharya}\ \emph {et~al.}(2018)\citenamefont {Acharya}
  \emph {et~al.}}]{CTAConsortium:2018tzg}%
  \BibitemOpen
  \bibfield  {author} {\bibinfo {author} {\bibfnamefont {B.~S.}\ \bibnamefont
  {Acharya}} \emph {et~al.} (\bibinfo {collaboration} {CTA Consortium}),\
  }\href {\doibase 10.1142/10986} {\emph {\bibinfo {title} {{Science with the
  Cherenkov Telescope Array}}}}\ (\bibinfo  {publisher} {WSP},\ \bibinfo {year}
  {2018})\ \Eprint {http://arxiv.org/abs/1709.07997} {arXiv:1709.07997
  [astro-ph.IM]} \BibitemShut {NoStop}%
\bibitem [{\citenamefont {Gaisser}\ \emph {et~al.}(2013)\citenamefont
  {Gaisser}, \citenamefont {Stanev},\ and\ \citenamefont
  {Tilav}}]{Gaisser:2013bla}%
  \BibitemOpen
  \bibfield  {author} {\bibinfo {author} {\bibfnamefont {Thomas~K.}\
  \bibnamefont {Gaisser}}, \bibinfo {author} {\bibfnamefont {Todor}\
  \bibnamefont {Stanev}}, \ and\ \bibinfo {author} {\bibfnamefont {Serap}\
  \bibnamefont {Tilav}},\ }\bibfield  {title} {\enquote {\bibinfo {title}
  {{Cosmic Ray Energy Spectrum from Measurements of Air Showers}},}\ }\href
  {\doibase 10.1007/s11467-013-0319-7} {\bibfield  {journal} {\bibinfo
  {journal} {Front. Phys. (Beijing)}\ }\textbf {\bibinfo {volume} {8}},\
  \bibinfo {pages} {748--758} (\bibinfo {year} {2013})},\ \Eprint
  {http://arxiv.org/abs/1303.3565} {arXiv:1303.3565 [astro-ph.HE]} \BibitemShut
  {NoStop}%
\bibitem [{\citenamefont {Blasi}\ and\ \citenamefont
  {Amato}(2012)}]{Blasi_2012}%
  \BibitemOpen
  \bibfield  {author} {\bibinfo {author} {\bibfnamefont {Pasquale}\
  \bibnamefont {Blasi}}\ and\ \bibinfo {author} {\bibfnamefont {Elena}\
  \bibnamefont {Amato}},\ }\bibfield  {title} {\enquote {\bibinfo {title}
  {Diffusive propagation of cosmic rays from supernova remnants in the galaxy.
  i: spectrum and chemical composition},}\ }\href {\doibase
  10.1088/1475-7516/2012/01/010} {\bibfield  {journal} {\bibinfo  {journal}
  {Journal of Cosmology and Astroparticle Physics}\ }\textbf {\bibinfo {volume}
  {2012}},\ \bibinfo {pages} {010--010} (\bibinfo {year} {2012})}\BibitemShut
  {NoStop}%
\bibitem [{\citenamefont {Evoli}\ \emph {et~al.}(2008)\citenamefont {Evoli},
  \citenamefont {Gaggero}, \citenamefont {Grasso},\ and\ \citenamefont
  {Maccione}}]{Evoli:2008dv}%
  \BibitemOpen
  \bibfield  {author} {\bibinfo {author} {\bibfnamefont {Carmelo}\ \bibnamefont
  {Evoli}}, \bibinfo {author} {\bibfnamefont {Daniele}\ \bibnamefont
  {Gaggero}}, \bibinfo {author} {\bibfnamefont {Dario}\ \bibnamefont {Grasso}},
  \ and\ \bibinfo {author} {\bibfnamefont {Luca}\ \bibnamefont {Maccione}},\
  }\bibfield  {title} {\enquote {\bibinfo {title} {{Cosmic-Ray Nuclei,
  Antiprotons and Gamma-rays in the Galaxy: a New Diffusion Model}},}\ }\href
  {\doibase 10.1088/1475-7516/2008/10/018} {\bibfield  {journal} {\bibinfo
  {journal} {JCAP}\ }\textbf {\bibinfo {volume} {10}},\ \bibinfo {pages} {018}
  (\bibinfo {year} {2008})},\ \bibinfo {note} {[Erratum: JCAP 04, E01
  (2016)]},\ \Eprint {http://arxiv.org/abs/0807.4730} {arXiv:0807.4730
  [astro-ph]} \BibitemShut {NoStop}%
\bibitem [{\citenamefont {Joshi}\ \emph {et~al.}(2014)\citenamefont {Joshi},
  \citenamefont {Winter},\ and\ \citenamefont {Gupta}}]{Joshi:2013aua}%
  \BibitemOpen
  \bibfield  {author} {\bibinfo {author} {\bibfnamefont {Jagdish~C.}\
  \bibnamefont {Joshi}}, \bibinfo {author} {\bibfnamefont {Walter}\
  \bibnamefont {Winter}}, \ and\ \bibinfo {author} {\bibfnamefont {Nayantara}\
  \bibnamefont {Gupta}},\ }\bibfield  {title} {\enquote {\bibinfo {title} {{How
  Many of the Observed Neutrino Events Can Be Described by Cosmic Ray
  Interactions in the Milky Way?}}}\ }\href {\doibase 10.1093/mnras/stu189}
  {\bibfield  {journal} {\bibinfo  {journal} {Mon. Not. Roy. Astron. Soc.}\
  }\textbf {\bibinfo {volume} {439}},\ \bibinfo {pages} {3414--3419} (\bibinfo
  {year} {2014})},\ \bibinfo {note} {[Erratum: Mon.Not.Roy.Astron.Soc. 446, 892
  (2014)]},\ \Eprint {http://arxiv.org/abs/1310.5123} {arXiv:1310.5123
  [astro-ph.HE]} \BibitemShut {NoStop}%
\bibitem [{\citenamefont {Luque}\ \emph {et~al.}(2023)\citenamefont {Luque},
  \citenamefont {Gaggero}, \citenamefont {Grasso}, \citenamefont {Fornieri},
  \citenamefont {Egberts}, \citenamefont {Steppa},\ and\ \citenamefont
  {Evoli}}]{Luque:2022buq}%
  \BibitemOpen
  \bibfield  {author} {\bibinfo {author} {\bibfnamefont {Pedro De la~Torre}\
  \bibnamefont {Luque}}, \bibinfo {author} {\bibfnamefont {Daniele}\
  \bibnamefont {Gaggero}}, \bibinfo {author} {\bibfnamefont {Dario}\
  \bibnamefont {Grasso}}, \bibinfo {author} {\bibfnamefont {Ottavio}\
  \bibnamefont {Fornieri}}, \bibinfo {author} {\bibfnamefont {Kathrin}\
  \bibnamefont {Egberts}}, \bibinfo {author} {\bibfnamefont {Constantin}\
  \bibnamefont {Steppa}}, \ and\ \bibinfo {author} {\bibfnamefont {Carmelo}\
  \bibnamefont {Evoli}},\ }\bibfield  {title} {\enquote {\bibinfo {title}
  {{Galactic diffuse gamma rays meet the PeV frontier}},}\ }\href {\doibase
  10.1051/0004-6361/202243714} {\bibfield  {journal} {\bibinfo  {journal}
  {Astron. Astrophys.}\ }\textbf {\bibinfo {volume} {672}},\ \bibinfo {pages}
  {A58} (\bibinfo {year} {2023})},\ \Eprint {http://arxiv.org/abs/2203.15759}
  {arXiv:2203.15759 [astro-ph.HE]} \BibitemShut {NoStop}%
\bibitem [{\citenamefont {Kafexhiu}\ \emph {et~al.}(2014)\citenamefont
  {Kafexhiu}, \citenamefont {Aharonian}, \citenamefont {Taylor},\ and\
  \citenamefont {Vila}}]{Kafexhiu:2014cua}%
  \BibitemOpen
  \bibfield  {author} {\bibinfo {author} {\bibfnamefont {Ervin}\ \bibnamefont
  {Kafexhiu}}, \bibinfo {author} {\bibfnamefont {Felix}\ \bibnamefont
  {Aharonian}}, \bibinfo {author} {\bibfnamefont {Andrew~M.}\ \bibnamefont
  {Taylor}}, \ and\ \bibinfo {author} {\bibfnamefont {Gabriela~S.}\
  \bibnamefont {Vila}},\ }\bibfield  {title} {\enquote {\bibinfo {title}
  {{Parametrization of gamma-ray production cross-sections for pp interactions
  in a broad proton energy range from the kinematic threshold to PeV
  energies}},}\ }\href {\doibase 10.1103/PhysRevD.90.123014} {\bibfield
  {journal} {\bibinfo  {journal} {Phys. Rev. D}\ }\textbf {\bibinfo {volume}
  {90}},\ \bibinfo {pages} {123014} (\bibinfo {year} {2014})},\ \Eprint
  {http://arxiv.org/abs/1406.7369} {arXiv:1406.7369 [astro-ph.HE]} \BibitemShut
  {NoStop}%
\bibitem [{\citenamefont {Yoast-Hull}\ \emph {et~al.}(2013)\citenamefont
  {Yoast-Hull}, \citenamefont {Everett}, \citenamefont {Gallagher},\ and\
  \citenamefont {Zweibel}}]{Yoast-Hull:2013wwa}%
  \BibitemOpen
  \bibfield  {author} {\bibinfo {author} {\bibfnamefont {Tova~M.}\ \bibnamefont
  {Yoast-Hull}}, \bibinfo {author} {\bibfnamefont {John~E.}\ \bibnamefont
  {Everett}}, \bibinfo {author} {\bibfnamefont {J.S.}\ \bibnamefont
  {Gallagher}}, \ and\ \bibinfo {author} {\bibfnamefont {Ellen~G.}\
  \bibnamefont {Zweibel}},\ }\bibfield  {title} {\enquote {\bibinfo {title}
  {{Winds, Clumps, and Interacting Cosmic Rays in M82}},}\ }\href {\doibase
  10.1088/0004-637X/768/1/53} {\bibfield  {journal} {\bibinfo  {journal}
  {Astrophys. J.}\ }\textbf {\bibinfo {volume} {768}},\ \bibinfo {pages} {53}
  (\bibinfo {year} {2013})},\ \Eprint {http://arxiv.org/abs/1303.4305}
  {arXiv:1303.4305 [astro-ph.HE]} \BibitemShut {NoStop}%
\bibitem [{\citenamefont {Lacki}\ \emph {et~al.}(2011)\citenamefont {Lacki},
  \citenamefont {Thompson}, \citenamefont {Quataert}, \citenamefont {Loeb},\
  and\ \citenamefont {Waxman}}]{Lacki:2010vs}%
  \BibitemOpen
  \bibfield  {author} {\bibinfo {author} {\bibfnamefont {Brian~C.}\
  \bibnamefont {Lacki}}, \bibinfo {author} {\bibfnamefont {Todd~A.}\
  \bibnamefont {Thompson}}, \bibinfo {author} {\bibfnamefont {Eliot}\
  \bibnamefont {Quataert}}, \bibinfo {author} {\bibfnamefont {Abraham}\
  \bibnamefont {Loeb}}, \ and\ \bibinfo {author} {\bibfnamefont {Eli}\
  \bibnamefont {Waxman}},\ }\bibfield  {title} {\enquote {\bibinfo {title} {{On
  The GeV \& TeV Detections of the Starburst Galaxies M82 \& NGC 253}},}\
  }\href {\doibase 10.1088/0004-637X/734/2/107} {\bibfield  {journal} {\bibinfo
   {journal} {Astrophys. J.}\ }\textbf {\bibinfo {volume} {734}},\ \bibinfo
  {pages} {107} (\bibinfo {year} {2011})},\ \Eprint
  {http://arxiv.org/abs/1003.3257} {arXiv:1003.3257 [astro-ph.HE]} \BibitemShut
  {NoStop}%
\bibitem [{\citenamefont {Lacki}\ and\ \citenamefont
  {Beck}(2013)}]{Lacki:2013ry}%
  \BibitemOpen
  \bibfield  {author} {\bibinfo {author} {\bibfnamefont {Brian~C.}\
  \bibnamefont {Lacki}}\ and\ \bibinfo {author} {\bibfnamefont {Rainer}\
  \bibnamefont {Beck}},\ }\bibfield  {title} {\enquote {\bibinfo {title} {{The
  Equipartition Magnetic Field Formula in Starburst Galaxies: Accounting for
  Pionic Secondaries and Strong Energy Losses}},}\ }\href {\doibase
  10.1093/mnras/stt122} {\bibfield  {journal} {\bibinfo  {journal} {Mon. Not.
  Roy. Astron. Soc.}\ }\textbf {\bibinfo {volume} {430}},\ \bibinfo {pages}
  {3171} (\bibinfo {year} {2013})},\ \Eprint {http://arxiv.org/abs/1301.5391}
  {arXiv:1301.5391 [astro-ph.CO]} \BibitemShut {NoStop}%
\bibitem [{\citenamefont {Peretti}\ \emph {et~al.}(2022)\citenamefont
  {Peretti}, \citenamefont {Morlino}, \citenamefont {Blasi},\ and\
  \citenamefont {Cristofari}}]{Peretti:2021yhc}%
  \BibitemOpen
  \bibfield  {author} {\bibinfo {author} {\bibfnamefont {Enrico}\ \bibnamefont
  {Peretti}}, \bibinfo {author} {\bibfnamefont {Giovanni}\ \bibnamefont
  {Morlino}}, \bibinfo {author} {\bibfnamefont {Pasquale}\ \bibnamefont
  {Blasi}}, \ and\ \bibinfo {author} {\bibfnamefont {Pierre}\ \bibnamefont
  {Cristofari}},\ }\bibfield  {title} {\enquote {\bibinfo {title} {{Particle
  acceleration and multimessenger emission from starburst-driven galactic
  winds}},}\ }\href {\doibase 10.1093/mnras/stac084} {\bibfield  {journal}
  {\bibinfo  {journal} {Mon. Not. Roy. Astron. Soc.}\ }\textbf {\bibinfo
  {volume} {511}},\ \bibinfo {pages} {1336--1348} (\bibinfo {year} {2022})},\
  \Eprint {http://arxiv.org/abs/2104.10978} {arXiv:2104.10978 [astro-ph.HE]}
  \BibitemShut {NoStop}%
\bibitem [{\citenamefont {Bell}(1978)}]{10.1093/mnras/182.3.443}%
  \BibitemOpen
  \bibfield  {author} {\bibinfo {author} {\bibfnamefont {A.~R.}\ \bibnamefont
  {Bell}},\ }\bibfield  {title} {\enquote {\bibinfo {title} {{The acceleration
  of cosmic rays in shock fronts – II}},}\ }\href {\doibase
  10.1093/mnras/182.3.443} {\bibfield  {journal} {\bibinfo  {journal} {Monthly
  Notices of the Royal Astronomical Society}\ }\textbf {\bibinfo {volume}
  {182}},\ \bibinfo {pages} {443--455} (\bibinfo {year} {1978})},\ \Eprint
  {http://arxiv.org/abs/https://academic.oup.com/mnras/article-pdf/182/3/443/3856040/mnras182-0443.pdf}
  {https://academic.oup.com/mnras/article-pdf/182/3/443/3856040/mnras182-0443.pdf}
  \BibitemShut {NoStop}%
\bibitem [{\citenamefont {Kelner}\ \emph {et~al.}(2006)\citenamefont {Kelner},
  \citenamefont {Aharonian},\ and\ \citenamefont {Bugayov}}]{Kelner:2006tc}%
  \BibitemOpen
  \bibfield  {author} {\bibinfo {author} {\bibfnamefont {S.R.}\ \bibnamefont
  {Kelner}}, \bibinfo {author} {\bibfnamefont {Felex~A.}\ \bibnamefont
  {Aharonian}}, \ and\ \bibinfo {author} {\bibfnamefont {V.V.}\ \bibnamefont
  {Bugayov}},\ }\bibfield  {title} {\enquote {\bibinfo {title} {{Energy spectra
  of gamma-rays, electrons and neutrinos produced at proton-proton interactions
  in the very high energy regime}},}\ }\href {\doibase
  10.1103/PhysRevD.74.034018} {\bibfield  {journal} {\bibinfo  {journal} {Phys.
  Rev. D}\ }\textbf {\bibinfo {volume} {74}},\ \bibinfo {pages} {034018}
  (\bibinfo {year} {2006})},\ \bibinfo {note} {[Erratum: Phys.Rev.D 79, 039901
  (2009)]},\ \Eprint {http://arxiv.org/abs/astro-ph/0606058}
  {arXiv:astro-ph/0606058} \BibitemShut {NoStop}%
\bibitem [{\citenamefont {Franceschini}\ and\ \citenamefont
  {Rodighiero}(2017)}]{Franceschini:2017iwq}%
  \BibitemOpen
  \bibfield  {author} {\bibinfo {author} {\bibfnamefont {Alberto}\ \bibnamefont
  {Franceschini}}\ and\ \bibinfo {author} {\bibfnamefont {Giulia}\ \bibnamefont
  {Rodighiero}},\ }\bibfield  {title} {\enquote {\bibinfo {title} {{The
  extragalactic background light revisited and the cosmic photon-photon
  opacity}},}\ }\href {\doibase 10.1051/0004-6361/201629684} {\bibfield
  {journal} {\bibinfo  {journal} {Astron. Astrophys.}\ }\textbf {\bibinfo
  {volume} {603}},\ \bibinfo {pages} {A34} (\bibinfo {year} {2017})},\ \Eprint
  {http://arxiv.org/abs/1705.10256} {arXiv:1705.10256 [astro-ph.HE]}
  \BibitemShut {NoStop}%
\bibitem [{\citenamefont {Angeli}(2004)}]{Angeli:2004kvy}%
  \BibitemOpen
  \bibfield  {author} {\bibinfo {author} {\bibfnamefont {I.}~\bibnamefont
  {Angeli}},\ }\bibfield  {title} {\enquote {\bibinfo {title} {{A consistent
  set of nuclear rms charge radii: properties of the radius surface R(N,Z)}},}\
  }\href {\doibase 10.1016/j.adt.2004.04.002} {\bibfield  {journal} {\bibinfo
  {journal} {Atom. Data Nucl. Data Tabl.}\ }\textbf {\bibinfo {volume} {87}},\
  \bibinfo {pages} {185--206} (\bibinfo {year} {2004})}\BibitemShut {NoStop}%
\bibitem [{\citenamefont {Alvey}\ \emph {et~al.}(2022)\citenamefont {Alvey},
  \citenamefont {Bringmann},\ and\ \citenamefont {Kolesova}}]{Alvey:2022pad}%
  \BibitemOpen
  \bibfield  {author} {\bibinfo {author} {\bibfnamefont {James}\ \bibnamefont
  {Alvey}}, \bibinfo {author} {\bibfnamefont {Torsten}\ \bibnamefont
  {Bringmann}}, \ and\ \bibinfo {author} {\bibfnamefont {Helena}\ \bibnamefont
  {Kolesova}},\ }\bibfield  {title} {\enquote {\bibinfo {title} {{No room to
  hide: implications of cosmic-ray upscattering for GeV-scale dark matter}},}\
  }\href@noop {} {\  (\bibinfo {year} {2022})},\ \Eprint
  {http://arxiv.org/abs/2209.03360} {arXiv:2209.03360 [hep-ph]} \BibitemShut
  {NoStop}%
\bibitem [{\citenamefont {Cyburt}\ \emph {et~al.}(2002)\citenamefont {Cyburt},
  \citenamefont {Fields}, \citenamefont {Pavlidou},\ and\ \citenamefont
  {Wandelt}}]{Cyburt:2002uw}%
  \BibitemOpen
  \bibfield  {author} {\bibinfo {author} {\bibfnamefont {Richard~H.}\
  \bibnamefont {Cyburt}}, \bibinfo {author} {\bibfnamefont {Brian~D.}\
  \bibnamefont {Fields}}, \bibinfo {author} {\bibfnamefont {Vasiliki}\
  \bibnamefont {Pavlidou}}, \ and\ \bibinfo {author} {\bibfnamefont
  {Benjamin~D.}\ \bibnamefont {Wandelt}},\ }\bibfield  {title} {\enquote
  {\bibinfo {title} {{Constraining strong baryon dark matter interactions with
  primordial nucleosynthesis and cosmic rays}},}\ }\href {\doibase
  10.1103/PhysRevD.65.123503} {\bibfield  {journal} {\bibinfo  {journal} {Phys.
  Rev. D}\ }\textbf {\bibinfo {volume} {65}},\ \bibinfo {pages} {123503}
  (\bibinfo {year} {2002})},\ \Eprint {http://arxiv.org/abs/astro-ph/0203240}
  {arXiv:astro-ph/0203240} \BibitemShut {NoStop}%
\bibitem [{\citenamefont {Hooper}\ and\ \citenamefont
  {McDermott}(2018)}]{Hooper:2018bfw}%
  \BibitemOpen
  \bibfield  {author} {\bibinfo {author} {\bibfnamefont {Dan}\ \bibnamefont
  {Hooper}}\ and\ \bibinfo {author} {\bibfnamefont {Samuel~D.}\ \bibnamefont
  {McDermott}},\ }\bibfield  {title} {\enquote {\bibinfo {title} {{Robust
  Constraints and Novel Gamma-Ray Signatures of Dark Matter That Interacts
  Strongly With Nucleons}},}\ }\href {\doibase 10.1103/PhysRevD.97.115006}
  {\bibfield  {journal} {\bibinfo  {journal} {Phys. Rev. D}\ }\textbf {\bibinfo
  {volume} {97}},\ \bibinfo {pages} {115006} (\bibinfo {year} {2018})},\
  \Eprint {http://arxiv.org/abs/1802.03025} {arXiv:1802.03025 [hep-ph]}
  \BibitemShut {NoStop}%
\bibitem [{\citenamefont {Acciari}\ \emph {et~al.}(2009)\citenamefont {Acciari}
  \emph {et~al.}}]{Acciari:2009wq}%
  \BibitemOpen
  \bibfield  {author} {\bibinfo {author} {\bibfnamefont {V.~A.}\ \bibnamefont
  {Acciari}} \emph {et~al.},\ }\bibfield  {title} {\enquote {\bibinfo {title}
  {{A connection between star formation activity and cosmic rays in the
  starburst galaxy M 82}},}\ }\href {\doibase 10.1038/nature08557} {\bibfield
  {journal} {\bibinfo  {journal} {Nature}\ }\textbf {\bibinfo {volume} {462}},\
  \bibinfo {pages} {770--772} (\bibinfo {year} {2009})},\ \Eprint
  {http://arxiv.org/abs/0911.0873} {arXiv:0911.0873 [astro-ph.CO]} \BibitemShut
  {NoStop}%
\bibitem [{\citenamefont {Ajello}\ \emph {et~al.}(2020)\citenamefont {Ajello},
  \citenamefont {Di~Mauro}, \citenamefont {Paliya},\ and\ \citenamefont
  {Garrappa}}]{Ajello:2020zna}%
  \BibitemOpen
  \bibfield  {author} {\bibinfo {author} {\bibfnamefont {M.}~\bibnamefont
  {Ajello}}, \bibinfo {author} {\bibfnamefont {M.}~\bibnamefont {Di~Mauro}},
  \bibinfo {author} {\bibfnamefont {V.S.}\ \bibnamefont {Paliya}}, \ and\
  \bibinfo {author} {\bibfnamefont {S.}~\bibnamefont {Garrappa}},\ }\bibfield
  {title} {\enquote {\bibinfo {title} {{The $\gamma$-ray Emission of
  Star-Forming Galaxies}},}\ }\href {\doibase 10.3847/1538-4357/ab86a6}
  {\bibfield  {journal} {\bibinfo  {journal} {Astrophys. J.}\ }\textbf
  {\bibinfo {volume} {894}},\ \bibinfo {pages} {88} (\bibinfo {year} {2020})},\
  \Eprint {http://arxiv.org/abs/2003.05493} {arXiv:2003.05493 [astro-ph.GA]}
  \BibitemShut {NoStop}%
\bibitem [{\citenamefont {Benito}\ \emph {et~al.}(2017)\citenamefont {Benito},
  \citenamefont {Bernal}, \citenamefont {Bozorgnia}, \citenamefont {Calore},\
  and\ \citenamefont {Iocco}}]{Benito:2016kyp}%
  \BibitemOpen
  \bibfield  {author} {\bibinfo {author} {\bibfnamefont {Maria}\ \bibnamefont
  {Benito}}, \bibinfo {author} {\bibfnamefont {Nicolas}\ \bibnamefont
  {Bernal}}, \bibinfo {author} {\bibfnamefont {Nassim}\ \bibnamefont
  {Bozorgnia}}, \bibinfo {author} {\bibfnamefont {Francesca}\ \bibnamefont
  {Calore}}, \ and\ \bibinfo {author} {\bibfnamefont {Fabio}\ \bibnamefont
  {Iocco}},\ }\bibfield  {title} {\enquote {\bibinfo {title} {{Particle Dark
  Matter Constraints: the Effect of Galactic Uncertainties}},}\ }\href
  {\doibase 10.1088/1475-7516/2017/02/007} {\bibfield  {journal} {\bibinfo
  {journal} {JCAP}\ }\textbf {\bibinfo {volume} {02}},\ \bibinfo {pages} {007}
  (\bibinfo {year} {2017})},\ \bibinfo {note} {[Erratum: JCAP 06, E01
  (2018)]},\ \Eprint {http://arxiv.org/abs/1612.02010} {arXiv:1612.02010
  [hep-ph]} \BibitemShut {NoStop}%
\bibitem [{\citenamefont {Benito}\ \emph {et~al.}(2019)\citenamefont {Benito},
  \citenamefont {Cuoco},\ and\ \citenamefont {Iocco}}]{Benito:2019ngh}%
  \BibitemOpen
  \bibfield  {author} {\bibinfo {author} {\bibfnamefont {Maria}\ \bibnamefont
  {Benito}}, \bibinfo {author} {\bibfnamefont {Alessandro}\ \bibnamefont
  {Cuoco}}, \ and\ \bibinfo {author} {\bibfnamefont {Fabio}\ \bibnamefont
  {Iocco}},\ }\bibfield  {title} {\enquote {\bibinfo {title} {{Handling the
  Uncertainties in the Galactic Dark Matter Distribution for Particle Dark
  Matter Searches}},}\ }\href {\doibase 10.1088/1475-7516/2019/03/033}
  {\bibfield  {journal} {\bibinfo  {journal} {JCAP}\ }\textbf {\bibinfo
  {volume} {03}},\ \bibinfo {pages} {033} (\bibinfo {year} {2019})},\ \Eprint
  {http://arxiv.org/abs/1901.02460} {arXiv:1901.02460 [astro-ph.GA]}
  \BibitemShut {NoStop}%
\bibitem [{\citenamefont {Benito}\ \emph {et~al.}(2021)\citenamefont {Benito},
  \citenamefont {Iocco},\ and\ \citenamefont {Cuoco}}]{Benito:2020lgu}%
  \BibitemOpen
  \bibfield  {author} {\bibinfo {author} {\bibfnamefont {Mar\'\i{}a}\
  \bibnamefont {Benito}}, \bibinfo {author} {\bibfnamefont {Fabio}\
  \bibnamefont {Iocco}}, \ and\ \bibinfo {author} {\bibfnamefont {Alessandro}\
  \bibnamefont {Cuoco}},\ }\bibfield  {title} {\enquote {\bibinfo {title}
  {{Uncertainties in the Galactic Dark Matter distribution: An update}},}\
  }\href {\doibase 10.1016/j.dark.2021.100826} {\bibfield  {journal} {\bibinfo
  {journal} {Phys. Dark Univ.}\ }\textbf {\bibinfo {volume} {32}},\ \bibinfo
  {pages} {100826} (\bibinfo {year} {2021})},\ \Eprint
  {http://arxiv.org/abs/2009.13523} {arXiv:2009.13523 [astro-ph.GA]}
  \BibitemShut {NoStop}%
\bibitem [{\citenamefont {Navarro}\ \emph {et~al.}(1996)\citenamefont
  {Navarro}, \citenamefont {Frenk},\ and\ \citenamefont
  {White}}]{Navarro:1995iw}%
  \BibitemOpen
  \bibfield  {author} {\bibinfo {author} {\bibfnamefont {Julio~F.}\
  \bibnamefont {Navarro}}, \bibinfo {author} {\bibfnamefont {Carlos~S.}\
  \bibnamefont {Frenk}}, \ and\ \bibinfo {author} {\bibfnamefont {Simon D.~M.}\
  \bibnamefont {White}},\ }\bibfield  {title} {\enquote {\bibinfo {title} {{The
  Structure of cold dark matter halos}},}\ }\href {\doibase 10.1086/177173}
  {\bibfield  {journal} {\bibinfo  {journal} {Astrophys. J.}\ }\textbf
  {\bibinfo {volume} {462}},\ \bibinfo {pages} {563--575} (\bibinfo {year}
  {1996})},\ \Eprint {http://arxiv.org/abs/astro-ph/9508025}
  {arXiv:astro-ph/9508025} \BibitemShut {NoStop}%
\bibitem [{\citenamefont {Werhahn}\ \emph
  {et~al.}(2021{\natexlab{c}})\citenamefont {Werhahn}, \citenamefont
  {Pfrommer},\ and\ \citenamefont {Girichidis}}]{Werhahn:2021jvy}%
  \BibitemOpen
  \bibfield  {author} {\bibinfo {author} {\bibfnamefont {Maria}\ \bibnamefont
  {Werhahn}}, \bibinfo {author} {\bibfnamefont {Christoph}\ \bibnamefont
  {Pfrommer}}, \ and\ \bibinfo {author} {\bibfnamefont {Philipp}\ \bibnamefont
  {Girichidis}},\ }\bibfield  {title} {\enquote {\bibinfo {title} {{Cosmic rays
  and non-thermal emission in simulated galaxies \textendash{} III. Probing
  cosmic-ray calorimetry with radio spectra and the FIR\textendash{}radio
  correlation}},}\ }\href {\doibase 10.1093/mnras/stab2535} {\bibfield
  {journal} {\bibinfo  {journal} {Mon. Not. Roy. Astron. Soc.}\ }\textbf
  {\bibinfo {volume} {508}},\ \bibinfo {pages} {4072--4095} (\bibinfo {year}
  {2021}{\natexlab{c}})},\ \Eprint {http://arxiv.org/abs/2105.12134}
  {arXiv:2105.12134 [astro-ph.GA]} \BibitemShut {NoStop}%
\bibitem [{\citenamefont {Burkert}(1995)}]{Burkert:1995yz}%
  \BibitemOpen
  \bibfield  {author} {\bibinfo {author} {\bibfnamefont {A.}~\bibnamefont
  {Burkert}},\ }\bibfield  {title} {\enquote {\bibinfo {title} {{The Structure
  of dark matter halos in dwarf galaxies}},}\ }\href {\doibase 10.1086/309560}
  {\bibfield  {journal} {\bibinfo  {journal} {Astrophys. J. Lett.}\ }\textbf
  {\bibinfo {volume} {447}},\ \bibinfo {pages} {L25} (\bibinfo {year}
  {1995})},\ \Eprint {http://arxiv.org/abs/astro-ph/9504041}
  {arXiv:astro-ph/9504041} \BibitemShut {NoStop}%
\bibitem [{\citenamefont {Lin}\ and\ \citenamefont {Li}(2019)}]{Lin:2019yux}%
  \BibitemOpen
  \bibfield  {author} {\bibinfo {author} {\bibfnamefont {Hai-Nan}\ \bibnamefont
  {Lin}}\ and\ \bibinfo {author} {\bibfnamefont {Xin}\ \bibnamefont {Li}},\
  }\bibfield  {title} {\enquote {\bibinfo {title} {{The Dark Matter Profiles in
  the Milky Way}},}\ }\href {\doibase 10.1093/mnras/stz1698} {\bibfield
  {journal} {\bibinfo  {journal} {Mon. Not. Roy. Astron. Soc.}\ }\textbf
  {\bibinfo {volume} {487}},\ \bibinfo {pages} {5679--5684} (\bibinfo {year}
  {2019})},\ \Eprint {http://arxiv.org/abs/1906.08419} {arXiv:1906.08419
  [astro-ph.GA]} \BibitemShut {NoStop}%
\bibitem [{\citenamefont {Abdalla}\ \emph {et~al.}(2018)\citenamefont {Abdalla}
  \emph {et~al.}}]{HESS:2018yqa}%
  \BibitemOpen
  \bibfield  {author} {\bibinfo {author} {\bibfnamefont {H.}~\bibnamefont
  {Abdalla}} \emph {et~al.} (\bibinfo {collaboration} {H.E.S.S.}),\ }\bibfield
  {title} {\enquote {\bibinfo {title} {{The starburst galaxy NGC 253 revisited
  by H.E.S.S. and Fermi-LAT}},}\ }\href {\doibase 10.1051/0004-6361/201833202}
  {\bibfield  {journal} {\bibinfo  {journal} {Astron. Astrophys.}\ }\textbf
  {\bibinfo {volume} {617}},\ \bibinfo {pages} {A73} (\bibinfo {year}
  {2018})},\ \Eprint {http://arxiv.org/abs/1806.03866} {arXiv:1806.03866
  [astro-ph.HE]} \BibitemShut {NoStop}%
\bibitem [{\citenamefont {Kennicutt}(1998{\natexlab{a}})}]{Kennicutt:1997ng}%
  \BibitemOpen
  \bibfield  {author} {\bibinfo {author} {\bibfnamefont {Jr.}\ \bibnamefont
  {Kennicutt}, \bibfnamefont {Robert~C.}},\ }\bibfield  {title} {\enquote
  {\bibinfo {title} {{The Global Schmidt law in star forming galaxies}},}\
  }\href {\doibase 10.1086/305588} {\bibfield  {journal} {\bibinfo  {journal}
  {Astrophys. J.}\ }\textbf {\bibinfo {volume} {498}},\ \bibinfo {pages} {541}
  (\bibinfo {year} {1998}{\natexlab{a}})},\ \Eprint
  {http://arxiv.org/abs/astro-ph/9712213} {arXiv:astro-ph/9712213} \BibitemShut
  {NoStop}%
\bibitem [{\citenamefont {Kennicutt}(1998{\natexlab{b}})}]{Kennicutt:1998zb}%
  \BibitemOpen
  \bibfield  {author} {\bibinfo {author} {\bibfnamefont {Jr.}\ \bibnamefont
  {Kennicutt}, \bibfnamefont {Robert~C.}},\ }\bibfield  {title} {\enquote
  {\bibinfo {title} {{Star formation in galaxies along the Hubble sequence}},}\
  }\href {\doibase 10.1146/annurev.astro.36.1.189} {\bibfield  {journal}
  {\bibinfo  {journal} {Ann. Rev. Astron. Astrophys.}\ }\textbf {\bibinfo
  {volume} {36}},\ \bibinfo {pages} {189--231} (\bibinfo {year}
  {1998}{\natexlab{b}})},\ \Eprint {http://arxiv.org/abs/astro-ph/9807187}
  {arXiv:astro-ph/9807187} \BibitemShut {NoStop}%
\bibitem [{\citenamefont {Chevalier}\ and\ \citenamefont
  {Clegg}(1985)}]{Chevalier:1985pc}%
  \BibitemOpen
  \bibfield  {author} {\bibinfo {author} {\bibfnamefont {R.A.}\ \bibnamefont
  {Chevalier}}\ and\ \bibinfo {author} {\bibfnamefont {Andrew~W.}\ \bibnamefont
  {Clegg}},\ }\bibfield  {title} {\enquote {\bibinfo {title} {{Wind from a
  starburst galaxy nucleus}},}\ }\href {\doibase 10.1038/317044a0} {\bibfield
  {journal} {\bibinfo  {journal} {Nature}\ }\textbf {\bibinfo {volume} {317}},\
  \bibinfo {pages} {44} (\bibinfo {year} {1985})}\BibitemShut {NoStop}%
\bibitem [{\citenamefont {Kennicutt}\ and\ \citenamefont
  {De~Los~Reyes}(2021)}]{Kennicutt_2021}%
  \BibitemOpen
  \bibfield  {author} {\bibinfo {author} {\bibfnamefont {Robert~C.}\
  \bibnamefont {Kennicutt}}\ and\ \bibinfo {author} {\bibfnamefont {Mithi
  A.~C.}\ \bibnamefont {De~Los~Reyes}},\ }\bibfield  {title} {\enquote
  {\bibinfo {title} {Revisiting the integrated star formation law. ii.
  starbursts and the combined global schmidt law},}\ }\href {\doibase
  10.3847/1538-4357/abd3a2} {\bibfield  {journal} {\bibinfo  {journal} {The
  Astrophysical Journal}\ }\textbf {\bibinfo {volume} {908}},\ \bibinfo {pages}
  {61} (\bibinfo {year} {2021})}\BibitemShut {NoStop}%
\bibitem [{\citenamefont {Erickcek}\ \emph {et~al.}(2007)\citenamefont
  {Erickcek}, \citenamefont {Steinhardt}, \citenamefont {McCammon},\ and\
  \citenamefont {McGuire}}]{PhysRevD.76.042007}%
  \BibitemOpen
  \bibfield  {author} {\bibinfo {author} {\bibfnamefont {Adrienne~L.}\
  \bibnamefont {Erickcek}}, \bibinfo {author} {\bibfnamefont {Paul~J.}\
  \bibnamefont {Steinhardt}}, \bibinfo {author} {\bibfnamefont {Dan}\
  \bibnamefont {McCammon}}, \ and\ \bibinfo {author} {\bibfnamefont
  {Patrick~C.}\ \bibnamefont {McGuire}},\ }\bibfield  {title} {\enquote
  {\bibinfo {title} {Constraints on the interactions between dark matter and
  baryons from the x-ray quantum calorimetry experiment},}\ }\href {\doibase
  10.1103/PhysRevD.76.042007} {\bibfield  {journal} {\bibinfo  {journal} {Phys.
  Rev. D}\ }\textbf {\bibinfo {volume} {76}},\ \bibinfo {pages} {042007}
  (\bibinfo {year} {2007})}\BibitemShut {NoStop}%
\bibitem [{\citenamefont {Aguilar-Arevalo}\ \emph {et~al.}(2016)\citenamefont
  {Aguilar-Arevalo}, \citenamefont {Amidei}, \citenamefont {Bertou},
  \citenamefont {Butner}, \citenamefont {Cancelo}, \citenamefont {Casta\~neda
  V\'azquez}, \citenamefont {Cervantes~Vergara}, \citenamefont {Chavarria},
  \citenamefont {Chavez}, \citenamefont {de~Mello~Neto}, \citenamefont
  {D'Olivo}, \citenamefont {Estrada}, \citenamefont {Fernandez~Moroni},
  \citenamefont {Ga\"{\i}or}, \citenamefont {Guardincerri}, \citenamefont
  {Hern\'andez~Torres}, \citenamefont {Izraelevitch}, \citenamefont {Kavner},
  \citenamefont {Kilminster}, \citenamefont {Lawson}, \citenamefont
  {Letessier-Selvon}, \citenamefont {Liao}, \citenamefont {Mello},
  \citenamefont {Molina}, \citenamefont {Pe\~na}, \citenamefont {Privitera},
  \citenamefont {Ramanathan}, \citenamefont {Sarkis}, \citenamefont {Schwarz},
  \citenamefont {Sengul}, \citenamefont {Settimo}, \citenamefont {Sofo~Haro},
  \citenamefont {Thomas}, \citenamefont {Tiffenberg}, \citenamefont
  {Tiouchichine}, \citenamefont {Torres~Machado}, \citenamefont {Trillaud},
  \citenamefont {You},\ and\ \citenamefont {Zhou}}]{PhysRevD.94.082006}%
  \BibitemOpen
  \bibfield  {author} {\bibinfo {author} {\bibfnamefont {A.}~\bibnamefont
  {Aguilar-Arevalo}}, \bibinfo {author} {\bibfnamefont {D.}~\bibnamefont
  {Amidei}}, \bibinfo {author} {\bibfnamefont {X.}~\bibnamefont {Bertou}},
  \bibinfo {author} {\bibfnamefont {M.}~\bibnamefont {Butner}}, \bibinfo
  {author} {\bibfnamefont {G.}~\bibnamefont {Cancelo}}, \bibinfo {author}
  {\bibfnamefont {A.}~\bibnamefont {Casta\~neda V\'azquez}}, \bibinfo {author}
  {\bibfnamefont {B.~A.}\ \bibnamefont {Cervantes~Vergara}}, \bibinfo {author}
  {\bibfnamefont {A.~E.}\ \bibnamefont {Chavarria}}, \bibinfo {author}
  {\bibfnamefont {C.~R.}\ \bibnamefont {Chavez}}, \bibinfo {author}
  {\bibfnamefont {J.~R.~T.}\ \bibnamefont {de~Mello~Neto}}, \bibinfo {author}
  {\bibfnamefont {J.~C.}\ \bibnamefont {D'Olivo}}, \bibinfo {author}
  {\bibfnamefont {J.}~\bibnamefont {Estrada}}, \bibinfo {author} {\bibfnamefont
  {G.}~\bibnamefont {Fernandez~Moroni}}, \bibinfo {author} {\bibfnamefont
  {R.}~\bibnamefont {Ga\"{\i}or}}, \bibinfo {author} {\bibfnamefont
  {Y.}~\bibnamefont {Guardincerri}}, \bibinfo {author} {\bibfnamefont {K.~P.}\
  \bibnamefont {Hern\'andez~Torres}}, \bibinfo {author} {\bibfnamefont
  {F.}~\bibnamefont {Izraelevitch}}, \bibinfo {author} {\bibfnamefont
  {A.}~\bibnamefont {Kavner}}, \bibinfo {author} {\bibfnamefont
  {B.}~\bibnamefont {Kilminster}}, \bibinfo {author} {\bibfnamefont
  {I.}~\bibnamefont {Lawson}}, \bibinfo {author} {\bibfnamefont
  {A.}~\bibnamefont {Letessier-Selvon}}, \bibinfo {author} {\bibfnamefont
  {J.}~\bibnamefont {Liao}}, \bibinfo {author} {\bibfnamefont {V.~B.~B.}\
  \bibnamefont {Mello}}, \bibinfo {author} {\bibfnamefont {J.}~\bibnamefont
  {Molina}}, \bibinfo {author} {\bibfnamefont {J.~R.}\ \bibnamefont {Pe\~na}},
  \bibinfo {author} {\bibfnamefont {P.}~\bibnamefont {Privitera}}, \bibinfo
  {author} {\bibfnamefont {K.}~\bibnamefont {Ramanathan}}, \bibinfo {author}
  {\bibfnamefont {Y.}~\bibnamefont {Sarkis}}, \bibinfo {author} {\bibfnamefont
  {T.}~\bibnamefont {Schwarz}}, \bibinfo {author} {\bibfnamefont
  {C.}~\bibnamefont {Sengul}}, \bibinfo {author} {\bibfnamefont
  {M.}~\bibnamefont {Settimo}}, \bibinfo {author} {\bibfnamefont
  {M.}~\bibnamefont {Sofo~Haro}}, \bibinfo {author} {\bibfnamefont
  {R.}~\bibnamefont {Thomas}}, \bibinfo {author} {\bibfnamefont
  {J.}~\bibnamefont {Tiffenberg}}, \bibinfo {author} {\bibfnamefont
  {E.}~\bibnamefont {Tiouchichine}}, \bibinfo {author} {\bibfnamefont
  {D.}~\bibnamefont {Torres~Machado}}, \bibinfo {author} {\bibfnamefont
  {F.}~\bibnamefont {Trillaud}}, \bibinfo {author} {\bibfnamefont
  {X.}~\bibnamefont {You}}, \ and\ \bibinfo {author} {\bibfnamefont
  {J.}~\bibnamefont {Zhou}} (\bibinfo {collaboration} {DAMIC Collaboration}),\
  }\bibfield  {title} {\enquote {\bibinfo {title} {Search for low-mass wimps in
  a 0.6 kg day exposure of the damic experiment at snolab},}\ }\href {\doibase
  10.1103/PhysRevD.94.082006} {\bibfield  {journal} {\bibinfo  {journal} {Phys.
  Rev. D}\ }\textbf {\bibinfo {volume} {94}},\ \bibinfo {pages} {082006}
  (\bibinfo {year} {2016})}\BibitemShut {NoStop}%
\bibitem [{\citenamefont {Angloher}\ \emph {et~al.}(2016)\citenamefont
  {Angloher} \emph {et~al.}}]{CRESST:2015txj}%
  \BibitemOpen
  \bibfield  {author} {\bibinfo {author} {\bibfnamefont {G.}~\bibnamefont
  {Angloher}} \emph {et~al.} (\bibinfo {collaboration} {CRESST}),\ }\bibfield
  {title} {\enquote {\bibinfo {title} {{Results on light dark matter particles
  with a low-threshold CRESST-II detector}},}\ }\href {\doibase
  10.1140/epjc/s10052-016-3877-3} {\bibfield  {journal} {\bibinfo  {journal}
  {Eur. Phys. J. C}\ }\textbf {\bibinfo {volume} {76}},\ \bibinfo {pages} {25}
  (\bibinfo {year} {2016})},\ \Eprint {http://arxiv.org/abs/1509.01515}
  {arXiv:1509.01515 [astro-ph.CO]} \BibitemShut {NoStop}%
\bibitem [{\citenamefont {Emken}\ and\ \citenamefont
  {Kouvaris}(2018)}]{Emken:2018run}%
  \BibitemOpen
  \bibfield  {author} {\bibinfo {author} {\bibfnamefont {Timon}\ \bibnamefont
  {Emken}}\ and\ \bibinfo {author} {\bibfnamefont {Chris}\ \bibnamefont
  {Kouvaris}},\ }\bibfield  {title} {\enquote {\bibinfo {title} {{How blind are
  underground and surface detectors to strongly interacting Dark Matter?}}}\
  }\href {\doibase 10.1103/PhysRevD.97.115047} {\bibfield  {journal} {\bibinfo
  {journal} {Phys. Rev. D}\ }\textbf {\bibinfo {volume} {97}},\ \bibinfo
  {pages} {115047} (\bibinfo {year} {2018})},\ \Eprint
  {http://arxiv.org/abs/1802.04764} {arXiv:1802.04764 [hep-ph]} \BibitemShut
  {NoStop}%
\bibitem [{\citenamefont {Armengaud}\ \emph {et~al.}(2019)\citenamefont
  {Armengaud} \emph {et~al.}}]{EDELWEISS:2019vjv}%
  \BibitemOpen
  \bibfield  {author} {\bibinfo {author} {\bibfnamefont {E.}~\bibnamefont
  {Armengaud}} \emph {et~al.} (\bibinfo {collaboration} {EDELWEISS}),\
  }\bibfield  {title} {\enquote {\bibinfo {title} {{Searching for low-mass dark
  matter particles with a massive Ge bolometer operated above-ground}},}\
  }\href {\doibase 10.1103/PhysRevD.99.082003} {\bibfield  {journal} {\bibinfo
  {journal} {Phys. Rev. D}\ }\textbf {\bibinfo {volume} {99}},\ \bibinfo
  {pages} {082003} (\bibinfo {year} {2019})},\ \Eprint
  {http://arxiv.org/abs/1901.03588} {arXiv:1901.03588 [astro-ph.GA]}
  \BibitemShut {NoStop}%
\bibitem [{\citenamefont {Angloher}\ \emph {et~al.}(2017)\citenamefont
  {Angloher} \emph {et~al.}}]{CRESST:2017ues}%
  \BibitemOpen
  \bibfield  {author} {\bibinfo {author} {\bibfnamefont {G.}~\bibnamefont
  {Angloher}} \emph {et~al.} (\bibinfo {collaboration} {CRESST}),\ }\bibfield
  {title} {\enquote {\bibinfo {title} {{Results on MeV-scale dark matter from a
  gram-scale cryogenic calorimeter operated above ground}},}\ }\href {\doibase
  10.1140/epjc/s10052-017-5223-9} {\bibfield  {journal} {\bibinfo  {journal}
  {Eur. Phys. J. C}\ }\textbf {\bibinfo {volume} {77}},\ \bibinfo {pages} {637}
  (\bibinfo {year} {2017})},\ \Eprint {http://arxiv.org/abs/1707.06749}
  {arXiv:1707.06749 [astro-ph.CO]} \BibitemShut {NoStop}%
\bibitem [{\citenamefont {Kouvaris}\ and\ \citenamefont
  {Pradler}(2017)}]{Kouvaris:2016afs}%
  \BibitemOpen
  \bibfield  {author} {\bibinfo {author} {\bibfnamefont {Chris}\ \bibnamefont
  {Kouvaris}}\ and\ \bibinfo {author} {\bibfnamefont {Josef}\ \bibnamefont
  {Pradler}},\ }\bibfield  {title} {\enquote {\bibinfo {title} {{Probing
  sub-GeV Dark Matter with conventional detectors}},}\ }\href {\doibase
  10.1103/PhysRevLett.118.031803} {\bibfield  {journal} {\bibinfo  {journal}
  {Phys. Rev. Lett.}\ }\textbf {\bibinfo {volume} {118}},\ \bibinfo {pages}
  {031803} (\bibinfo {year} {2017})},\ \Eprint
  {http://arxiv.org/abs/1607.01789} {arXiv:1607.01789 [hep-ph]} \BibitemShut
  {NoStop}%
\bibitem [{\citenamefont {Collar}(2018)}]{PhysRevD.98.023005}%
  \BibitemOpen
  \bibfield  {author} {\bibinfo {author} {\bibfnamefont {J.~I.}\ \bibnamefont
  {Collar}},\ }\bibfield  {title} {\enquote {\bibinfo {title} {Search for a
  nonrelativistic component in the spectrum of cosmic rays at earth},}\ }\href
  {\doibase 10.1103/PhysRevD.98.023005} {\bibfield  {journal} {\bibinfo
  {journal} {Phys. Rev. D}\ }\textbf {\bibinfo {volume} {98}},\ \bibinfo
  {pages} {023005} (\bibinfo {year} {2018})}\BibitemShut {NoStop}%
\bibitem [{\citenamefont {Awe}\ \emph {et~al.}(2018)\citenamefont {Awe},
  \citenamefont {Barbeau}, \citenamefont {Collar}, \citenamefont {Hedges},\
  and\ \citenamefont {Li}}]{Awe:2018fei}%
  \BibitemOpen
  \bibfield  {author} {\bibinfo {author} {\bibfnamefont {C.}~\bibnamefont
  {Awe}}, \bibinfo {author} {\bibfnamefont {P.~S.}\ \bibnamefont {Barbeau}},
  \bibinfo {author} {\bibfnamefont {J.~I.}\ \bibnamefont {Collar}}, \bibinfo
  {author} {\bibfnamefont {S.}~\bibnamefont {Hedges}}, \ and\ \bibinfo {author}
  {\bibfnamefont {L.}~\bibnamefont {Li}},\ }\bibfield  {title} {\enquote
  {\bibinfo {title} {{Liquid scintillator response to proton recoils in the
  10\textendash{}100 keV range}},}\ }\href {\doibase
  10.1103/PhysRevC.98.045802} {\bibfield  {journal} {\bibinfo  {journal} {Phys.
  Rev. C}\ }\textbf {\bibinfo {volume} {98}},\ \bibinfo {pages} {045802}
  (\bibinfo {year} {2018})},\ \Eprint {http://arxiv.org/abs/1804.06457}
  {arXiv:1804.06457 [physics.ins-det]} \BibitemShut {NoStop}%
\bibitem [{\citenamefont {Dolan}\ \emph {et~al.}(2018)\citenamefont {Dolan},
  \citenamefont {Kahlhoefer},\ and\ \citenamefont
  {McCabe}}]{PhysRevLett.121.101801}%
  \BibitemOpen
  \bibfield  {author} {\bibinfo {author} {\bibfnamefont {Matthew~J.}\
  \bibnamefont {Dolan}}, \bibinfo {author} {\bibfnamefont {Felix}\ \bibnamefont
  {Kahlhoefer}}, \ and\ \bibinfo {author} {\bibfnamefont {Christopher}\
  \bibnamefont {McCabe}},\ }\bibfield  {title} {\enquote {\bibinfo {title}
  {Directly detecting sub-gev dark matter with electrons from nuclear
  scattering},}\ }\href {\doibase 10.1103/PhysRevLett.121.101801} {\bibfield
  {journal} {\bibinfo  {journal} {Phys. Rev. Lett.}\ }\textbf {\bibinfo
  {volume} {121}},\ \bibinfo {pages} {101801} (\bibinfo {year}
  {2018})}\BibitemShut {NoStop}%
\bibitem [{\citenamefont {Mahdawi}\ and\ \citenamefont
  {Farrar}(2018)}]{Mahdawi:2018euy}%
  \BibitemOpen
  \bibfield  {author} {\bibinfo {author} {\bibfnamefont {M.~Shafi}\
  \bibnamefont {Mahdawi}}\ and\ \bibinfo {author} {\bibfnamefont {Glennys~R.}\
  \bibnamefont {Farrar}},\ }\bibfield  {title} {\enquote {\bibinfo {title}
  {{Constraints on Dark Matter with a moderately large and velocity-dependent
  DM-nucleon cross-section}},}\ }\href {\doibase 10.1088/1475-7516/2018/10/007}
  {\bibfield  {journal} {\bibinfo  {journal} {JCAP}\ }\textbf {\bibinfo
  {volume} {10}},\ \bibinfo {pages} {007} (\bibinfo {year} {2018})},\ \Eprint
  {http://arxiv.org/abs/1804.03073} {arXiv:1804.03073 [hep-ph]} \BibitemShut
  {NoStop}%
\bibitem [{\citenamefont {Akerib}\ \emph {et~al.}(2019)\citenamefont {Akerib}
  \emph {et~al.}}]{LUX:2018akb}%
  \BibitemOpen
  \bibfield  {author} {\bibinfo {author} {\bibfnamefont {D.~S.}\ \bibnamefont
  {Akerib}} \emph {et~al.} (\bibinfo {collaboration} {LUX}),\ }\bibfield
  {title} {\enquote {\bibinfo {title} {{Results of a Search for Sub-GeV Dark
  Matter Using 2013 LUX Data}},}\ }\href {\doibase
  10.1103/PhysRevLett.122.131301} {\bibfield  {journal} {\bibinfo  {journal}
  {Phys. Rev. Lett.}\ }\textbf {\bibinfo {volume} {122}},\ \bibinfo {pages}
  {131301} (\bibinfo {year} {2019})},\ \Eprint
  {http://arxiv.org/abs/1811.11241} {arXiv:1811.11241 [astro-ph.CO]}
  \BibitemShut {NoStop}%
\bibitem [{\citenamefont {Aprile}\ \emph {et~al.}(2019)\citenamefont {Aprile}
  \emph {et~al.}}]{XENON:2019zpr}%
  \BibitemOpen
  \bibfield  {author} {\bibinfo {author} {\bibfnamefont {E.}~\bibnamefont
  {Aprile}} \emph {et~al.} (\bibinfo {collaboration} {XENON}),\ }\bibfield
  {title} {\enquote {\bibinfo {title} {{Search for Light Dark Matter
  Interactions Enhanced by the Migdal Effect or Bremsstrahlung in XENON1T}},}\
  }\href {\doibase 10.1103/PhysRevLett.123.241803} {\bibfield  {journal}
  {\bibinfo  {journal} {Phys. Rev. Lett.}\ }\textbf {\bibinfo {volume} {123}},\
  \bibinfo {pages} {241803} (\bibinfo {year} {2019})},\ \Eprint
  {http://arxiv.org/abs/1907.12771} {arXiv:1907.12771 [hep-ex]} \BibitemShut
  {NoStop}%
\bibitem [{\citenamefont {Abdelhameed}\ \emph {et~al.}(2019)\citenamefont
  {Abdelhameed} \emph {et~al.}}]{CRESST:2019jnq}%
  \BibitemOpen
  \bibfield  {author} {\bibinfo {author} {\bibfnamefont {A.~H.}\ \bibnamefont
  {Abdelhameed}} \emph {et~al.} (\bibinfo {collaboration} {CRESST}),\
  }\bibfield  {title} {\enquote {\bibinfo {title} {{First results from the
  CRESST-III low-mass dark matter program}},}\ }\href {\doibase
  10.1103/PhysRevD.100.102002} {\bibfield  {journal} {\bibinfo  {journal}
  {Phys. Rev. D}\ }\textbf {\bibinfo {volume} {100}},\ \bibinfo {pages}
  {102002} (\bibinfo {year} {2019})},\ \Eprint
  {http://arxiv.org/abs/1904.00498} {arXiv:1904.00498 [astro-ph.CO]}
  \BibitemShut {NoStop}%
\bibitem [{\citenamefont {Liu}\ \emph {et~al.}(2019)\citenamefont {Liu} \emph
  {et~al.}}]{CDEX:2019hzn}%
  \BibitemOpen
  \bibfield  {author} {\bibinfo {author} {\bibfnamefont {Z.~Z.}\ \bibnamefont
  {Liu}} \emph {et~al.} (\bibinfo {collaboration} {CDEX}),\ }\bibfield  {title}
  {\enquote {\bibinfo {title} {{Constraints on Spin-Independent Nucleus
  Scattering with sub-GeV Weakly Interacting Massive Particle Dark Matter from
  the CDEX-1B Experiment at the China Jinping Underground Laboratory}},}\
  }\href {\doibase 10.1103/PhysRevLett.123.161301} {\bibfield  {journal}
  {\bibinfo  {journal} {Phys. Rev. Lett.}\ }\textbf {\bibinfo {volume} {123}},\
  \bibinfo {pages} {161301} (\bibinfo {year} {2019})},\ \Eprint
  {http://arxiv.org/abs/1905.00354} {arXiv:1905.00354 [hep-ex]} \BibitemShut
  {NoStop}%
\bibitem [{\citenamefont {Liu}\ \emph {et~al.}(2022)\citenamefont {Liu} \emph
  {et~al.}}]{CDEX:2021cll}%
  \BibitemOpen
  \bibfield  {author} {\bibinfo {author} {\bibfnamefont {Z.~Z.}\ \bibnamefont
  {Liu}} \emph {et~al.} (\bibinfo {collaboration} {CDEX}),\ }\bibfield  {title}
  {\enquote {\bibinfo {title} {{Studies of the Earth shielding effect to direct
  dark matter searches at the China Jinping Underground Laboratory}},}\ }\href
  {\doibase 10.1103/PhysRevD.105.052005} {\bibfield  {journal} {\bibinfo
  {journal} {Phys. Rev. D}\ }\textbf {\bibinfo {volume} {105}},\ \bibinfo
  {pages} {052005} (\bibinfo {year} {2022})},\ \Eprint
  {http://arxiv.org/abs/2111.11243} {arXiv:2111.11243 [hep-ex]} \BibitemShut
  {NoStop}%
\bibitem [{\citenamefont {Armengaud}\ \emph {et~al.}(2022)\citenamefont
  {Armengaud} \emph {et~al.}}]{EDELWEISS:2022ktt}%
  \BibitemOpen
  \bibfield  {author} {\bibinfo {author} {\bibfnamefont {E.}~\bibnamefont
  {Armengaud}} \emph {et~al.} (\bibinfo {collaboration} {EDELWEISS}),\
  }\bibfield  {title} {\enquote {\bibinfo {title} {{Search for sub-GeV dark
  matter via the Migdal effect with an EDELWEISS germanium detector with NbSi
  transition-edge sensors}},}\ }\href {\doibase 10.1103/PhysRevD.106.062004}
  {\bibfield  {journal} {\bibinfo  {journal} {Phys. Rev. D}\ }\textbf {\bibinfo
  {volume} {106}},\ \bibinfo {pages} {062004} (\bibinfo {year} {2022})},\
  \Eprint {http://arxiv.org/abs/2203.03993} {arXiv:2203.03993 [astro-ph.GA]}
  \BibitemShut {NoStop}%
\bibitem [{\citenamefont {Blandford}(1999)}]{Blandford:1999hh}%
  \BibitemOpen
  \bibfield  {author} {\bibinfo {author} {\bibfnamefont {R.~D.}\ \bibnamefont
  {Blandford}},\ }\bibfield  {title} {\enquote {\bibinfo {title} {{Origin and
  evolution of massive black holes in galactic nuclei}},}\ }\href@noop {}
  {\bibfield  {journal} {\bibinfo  {journal} {ASP Conf. Ser.}\ }\textbf
  {\bibinfo {volume} {182}},\ \bibinfo {pages} {87} (\bibinfo {year} {1999})},\
  \Eprint {http://arxiv.org/abs/astro-ph/9906025} {arXiv:astro-ph/9906025}
  \BibitemShut {NoStop}%
\bibitem [{\citenamefont {Ullio}\ \emph {et~al.}(2001)\citenamefont {Ullio},
  \citenamefont {Zhao},\ and\ \citenamefont {Kamionkowski}}]{Ullio:2001fb}%
  \BibitemOpen
  \bibfield  {author} {\bibinfo {author} {\bibfnamefont {Piero}\ \bibnamefont
  {Ullio}}, \bibinfo {author} {\bibfnamefont {HongSheng}\ \bibnamefont {Zhao}},
  \ and\ \bibinfo {author} {\bibfnamefont {Marc}\ \bibnamefont
  {Kamionkowski}},\ }\bibfield  {title} {\enquote {\bibinfo {title} {{A Dark
  matter spike at the galactic center?}}}\ }\href {\doibase
  10.1103/PhysRevD.64.043504} {\bibfield  {journal} {\bibinfo  {journal} {Phys.
  Rev. D}\ }\textbf {\bibinfo {volume} {64}},\ \bibinfo {pages} {043504}
  (\bibinfo {year} {2001})},\ \Eprint {http://arxiv.org/abs/astro-ph/0101481}
  {arXiv:astro-ph/0101481} \BibitemShut {NoStop}%
\bibitem [{\citenamefont {Gondolo}\ and\ \citenamefont
  {Silk}(1999)}]{PhysRevLett.83.1719}%
  \BibitemOpen
  \bibfield  {author} {\bibinfo {author} {\bibfnamefont {Paolo}\ \bibnamefont
  {Gondolo}}\ and\ \bibinfo {author} {\bibfnamefont {Joseph}\ \bibnamefont
  {Silk}},\ }\bibfield  {title} {\enquote {\bibinfo {title} {Dark matter
  annihilation at the galactic center},}\ }\href {\doibase
  10.1103/PhysRevLett.83.1719} {\bibfield  {journal} {\bibinfo  {journal}
  {Phys. Rev. Lett.}\ }\textbf {\bibinfo {volume} {83}},\ \bibinfo {pages}
  {1719--1722} (\bibinfo {year} {1999})}\BibitemShut {NoStop}%
\bibitem [{\citenamefont {Gondolo}(2000)}]{Gondolo:2000pn}%
  \BibitemOpen
  \bibfield  {author} {\bibinfo {author} {\bibfnamefont {Paolo}\ \bibnamefont
  {Gondolo}},\ }\bibfield  {title} {\enquote {\bibinfo {title} {{Either
  neutralino dark matter or cuspy dark halos}},}\ }\href {\doibase
  10.1016/S0370-2693(00)01218-1} {\bibfield  {journal} {\bibinfo  {journal}
  {Phys. Lett. B}\ }\textbf {\bibinfo {volume} {494}},\ \bibinfo {pages}
  {181--186} (\bibinfo {year} {2000})},\ \Eprint
  {http://arxiv.org/abs/hep-ph/0002226} {arXiv:hep-ph/0002226} \BibitemShut
  {NoStop}%
\bibitem [{\citenamefont {Bertone}\ \emph {et~al.}(2001)\citenamefont
  {Bertone}, \citenamefont {Sigl},\ and\ \citenamefont
  {Silk}}]{Bertone:2001jv}%
  \BibitemOpen
  \bibfield  {author} {\bibinfo {author} {\bibfnamefont {G.}~\bibnamefont
  {Bertone}}, \bibinfo {author} {\bibfnamefont {G.}~\bibnamefont {Sigl}}, \
  and\ \bibinfo {author} {\bibfnamefont {J.}~\bibnamefont {Silk}},\ }\bibfield
  {title} {\enquote {\bibinfo {title} {{Astrophysical limits on massive dark
  matter}},}\ }\href {\doibase 10.1046/j.1365-8711.2001.04650.x} {\bibfield
  {journal} {\bibinfo  {journal} {Mon. Not. Roy. Astron. Soc.}\ }\textbf
  {\bibinfo {volume} {326}},\ \bibinfo {pages} {799--804} (\bibinfo {year}
  {2001})},\ \Eprint {http://arxiv.org/abs/astro-ph/0101134}
  {arXiv:astro-ph/0101134} \BibitemShut {NoStop}%
\bibitem [{\citenamefont {Knapen}\ \emph {et~al.}(2017)\citenamefont {Knapen},
  \citenamefont {Lin},\ and\ \citenamefont {Zurek}}]{Knapen:2017xzo}%
  \BibitemOpen
  \bibfield  {author} {\bibinfo {author} {\bibfnamefont {Simon}\ \bibnamefont
  {Knapen}}, \bibinfo {author} {\bibfnamefont {Tongyan}\ \bibnamefont {Lin}}, \
  and\ \bibinfo {author} {\bibfnamefont {Kathryn~M.}\ \bibnamefont {Zurek}},\
  }\bibfield  {title} {\enquote {\bibinfo {title} {{Light Dark Matter: Models
  and Constraints}},}\ }\href {\doibase 10.1103/PhysRevD.96.115021} {\bibfield
  {journal} {\bibinfo  {journal} {Phys. Rev. D}\ }\textbf {\bibinfo {volume}
  {96}},\ \bibinfo {pages} {115021} (\bibinfo {year} {2017})},\ \Eprint
  {http://arxiv.org/abs/1709.07882} {arXiv:1709.07882 [hep-ph]} \BibitemShut
  {NoStop}%
\bibitem [{\citenamefont {Chang}\ \emph {et~al.}(2018)\citenamefont {Chang},
  \citenamefont {Essig},\ and\ \citenamefont {McDermott}}]{Chang:2018rso}%
  \BibitemOpen
  \bibfield  {author} {\bibinfo {author} {\bibfnamefont {Jae~Hyeok}\
  \bibnamefont {Chang}}, \bibinfo {author} {\bibfnamefont {Rouven}\
  \bibnamefont {Essig}}, \ and\ \bibinfo {author} {\bibfnamefont {Samuel~D.}\
  \bibnamefont {McDermott}},\ }\bibfield  {title} {\enquote {\bibinfo {title}
  {{Supernova 1987A Constraints on Sub-GeV Dark Sectors, Millicharged
  Particles, the QCD Axion, and an Axion-like Particle}},}\ }\href {\doibase
  10.1007/JHEP09(2018)051} {\bibfield  {journal} {\bibinfo  {journal} {JHEP}\
  }\textbf {\bibinfo {volume} {09}},\ \bibinfo {pages} {051} (\bibinfo {year}
  {2018})},\ \Eprint {http://arxiv.org/abs/1803.00993} {arXiv:1803.00993
  [hep-ph]} \BibitemShut {NoStop}%
\bibitem [{\citenamefont {DeRocco}\ \emph {et~al.}(2019)\citenamefont
  {DeRocco}, \citenamefont {Graham}, \citenamefont {Kasen}, \citenamefont
  {Marques-Tavares},\ and\ \citenamefont {Rajendran}}]{DeRocco:2019jti}%
  \BibitemOpen
  \bibfield  {author} {\bibinfo {author} {\bibfnamefont {William}\ \bibnamefont
  {DeRocco}}, \bibinfo {author} {\bibfnamefont {Peter~W.}\ \bibnamefont
  {Graham}}, \bibinfo {author} {\bibfnamefont {Daniel}\ \bibnamefont {Kasen}},
  \bibinfo {author} {\bibfnamefont {Gustavo}\ \bibnamefont {Marques-Tavares}},
  \ and\ \bibinfo {author} {\bibfnamefont {Surjeet}\ \bibnamefont
  {Rajendran}},\ }\bibfield  {title} {\enquote {\bibinfo {title} {{Supernova
  signals of light dark matter}},}\ }\href {\doibase
  10.1103/PhysRevD.100.075018} {\bibfield  {journal} {\bibinfo  {journal}
  {Phys. Rev. D}\ }\textbf {\bibinfo {volume} {100}},\ \bibinfo {pages}
  {075018} (\bibinfo {year} {2019})},\ \Eprint
  {http://arxiv.org/abs/1905.09284} {arXiv:1905.09284 [hep-ph]} \BibitemShut
  {NoStop}%
\end{thebibliography}%

\clearpage
\newpage
\maketitle
\onecolumngrid

\begin{center}
\textbf{\large Supplemental Material for \\ \documenttitle} \\ 
\vspace{0.05in}
{Antonio Ambrosone, \ Marco Chianese, \ Damiano F.G. Fiorillo, \ Antonio Marinelli, and Gennaro Miele}
\end{center}
\onecolumngrid
\setcounter{equation}{0}
\setcounter{figure}{0}
\setcounter{table}{0}
\setcounter{section}{0}
\setcounter{page}{1}
\makeatletter
\renewcommand{\theequation}{S\arabic{equation}}
\renewcommand{\thefigure}{S\arabic{figure}}
\renewcommand{\thetable}{S\arabic{figure}}

The Supplemental Material is organized as follows. In Sec.~\ref{app:leaky} we focus on the leaky-box model approximation and its applicability to nearby star-forming and starburst galaxies, justifying the use of Eq.~(1). In Sec.~\ref{app:flux}, we detail our calculations for the gamma-ray spectrum from the SBNi. In Sec.~\ref{app:linear_priors} we discuss the linear priors used for each source for our analysis. In Sec.~\ref{app:halo} we scrutinize how our constraints on $\sigma_{\chi p}$ are affected by assuming different DM profiles as well as varying the halo parameters. In Sec.~\ref{app:cta}, we provide the details of the generation of CTA mock data and the forecast analysis we have performed. Finally, in Sec.~\ref{sec:heavier_CR} we assess the impact of heavier cosmic-ray species over the DM bounds.

\section{The Leaky-box Model Approximation}\label{app:leaky}
The leaky-box model provides a complete description of the complete CR transport equation for regions where the diffusion coefficient is not radial dependent and for sources where we do not expect any time-dependent CR distribution function. SBNi respect both of these conditions. Firstly, the steady state assumptions given by the leaky-box model is justified because the timescales involved in the non-thermal emissions $(\tau_{\rm loss} \sim 10^{5-6}\,\rm{yr})$ are lower than the duration of the starburst activity, which is typical of the order of $10^{7-8}\, \rm{yr}$~\cite{Peretti:2021yhc} (for other quantitative details see also Ref.~\cite{Lacki:2013ry}). Secondly, the light curves of these sources do not show any time variation~\cite{Ajello:2020zna}. Thirdly, SBNi have a tipycal dimension of $\sim 200 \, \rm{pc}$ thereby making negligible any radial variation for the diffusion timescale. Therefore, the CR transport equation can well be approximated by the leaky-box model equation~\cite{Cappiello:2018hsu,Werhahn:2021bal,Werhahn:2021gvl,Werhahn:2021jvy}
\begin{equation}\label{eq:leaky_energy}
    \frac{n_{\rm{CR}}(E)}{\tau_{\rm{esc}}} -\frac{{\rm d}}{{\rm d}E}\bigg(\frac{E}{\tau_{\rm{loss}}}n_{\rm{CR}}(E)\bigg) = Q_{{\rm CR},E}(E)\,,
\end{equation}
where $n_{\rm CR}(E)$ and $Q_{{\rm CR},E}(E)$ are the energy CR distribution inside the SBN and the injection rate into the energy space, respectively. The quantity $\tau_{\rm{esc}}=(\tau_{\rm{adv}}^{-1} + \tau_{\rm{diff}}^{-1})^{-1}$ is the time required for CRs to escape the SBN~\cite{Peretti:2018tmo,Ambrosone:2022fip,Werhahn:2021bal,Werhahn:2021gvl,Werhahn:2021jvy}. The quantities in energy space are linked to the ones in momentum space by the simple relationships
\begin{equation}\label{eq:energydistribution}
n_{\rm{CR}}(E) = 4\pi p^{2} f_{\rm{CR}}(p)\frac{{\rm d}p}{{\rm d}E} \quad {\rm and} \quad Q_{{\rm CR},E}(E) = 4\pi p^{2} Q_{\rm CR}(p)\frac{{\rm d}p}{{\rm d}E}\,,
\end{equation}
where $f_{\rm{CR}}(p)$ and $Q_{\rm CR}(p)$ appear in Eq.~(1). By inserting these expressions into Eq.~\eqref{eq:leaky_energy}, we can obtain the complete leaky-box model equation into the momentum space. In the regime where the escape time dominates over cooling phenomena, {\it i.e.} $\tau_{\rm{esc}} < \tau_{\rm{loss}}$, we have that
\begin{equation}
    f_{\rm{CR}}(p) = \tau_{\rm{esc}} \cdot Q_{\rm{CR}}(p)\,.
\end{equation}
On the other hand, in the energy loss regime we have 
\begin{equation}
   f_{\rm{CR}}(p) =  -\frac{\tau_{\rm{loss}}}{p\,E^{2}}\int p^{2}\, Q_{\rm{CR}}(p) \,{\rm d}p\,.
\end{equation}
For a power-law injection source $Q_{\rm{CR}}(p) \propto p^{-\alpha}$ with $\alpha = \Gamma +2$, we obtain
\begin{equation}\label{eq:solution_leaky_losses}
   f_{\rm{CR}}(p) =\frac{\tau_{\rm{loss}}\, p^{2}}{(p^2+m_{p}^2)}\, \frac{Q_{\rm{CR}}(p)}{(\alpha -3)} \simeq \tau_{\rm{loss}} \frac{Q_{\rm{CR}}(p)}{\Gamma -1} = \tau^{\rm eff}_{\rm loss} \cdot Q_{\rm{CR}}(p)\,,
\end{equation}
where the last passage into Eq.~\eqref{eq:solution_leaky_losses} holds for $p>m_{p}$ (the proton mass), which is the case we mostly focus on. This equation shows that scaling the energy-loss timescale by a factor of $(\Gamma-1)^{-1}$ gives a slightly more precise normalization for spectra with $\Gamma \ne 2$ than the simplest approximation that assumes $f_{\rm{CR}}(p) = \tau_{\rm{loss}} \cdot Q_{\rm{CR}}(p)$ even though the qualitative results do not change, as already emphasized by Ref.~\cite{Kornecki:2021xiy}. Finally, we stress that it is important to calculate the CR transport equation into the momentum space, because from a theoretical point of view the CRs are injected with a momentum power-law spectrum~\cite{Lacki:2013ry,10.1093/mnras/182.3.443}, even though many authors simply assume a power-law spectrum in the energy space.

\section{Gamma-Ray Production }\label{app:flux}

\begin{figure*}[t!]
    \centering
    \includegraphics[width=0.95\textwidth]{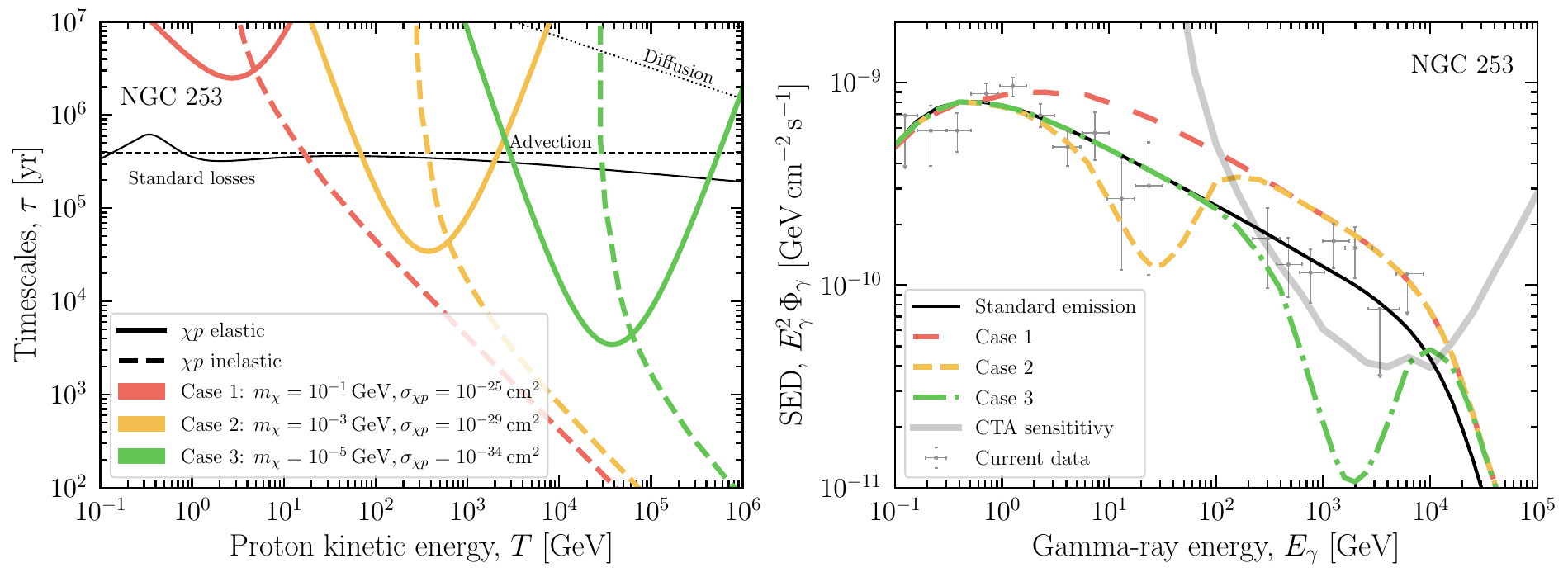}
    \caption{\textbf{Left panel.} Comparison between the proton timescales within NGC 253 as a function of the proton kinetic energy $T$. The continuous, dashed and dotted black lines represent the standard losses, advection and diffusion timescales, respectively. The colored continuous (dashed) lines correspond to the elastic (inelastic) for three different cases.
    \textbf{Right panel.} The expected gamma-ray fluxes from NGC 253 compared to current data~\cite{HESS:2018yqa,Ajello:2020zna} and CTA sensitivity~\cite{CTAConsortium:2018tzg}. Analogously to the left panel, the black color line corresponds to the standard case (without DM-CR interactions), while the colored lines to the three different choices of $(m_{\chi},\,\sigma_{\chi p})$.}
    \label{fig:NGC 253_timescale_fluxes}
\end{figure*}

We follow the modelling put forward by~\cite{Peretti:2018tmo,Peretti:2019vsj} which was later adopted by Refs.~\cite{Ambrosone:2020evo,Ambrosone:2021aaw,Ambrosone:2022fip}. Gamma-rays are mainly produced by the pion decays, bremsstrahlung and inverse Compton scatterings. Differently from the standard scenario, the interaction between DM particles and protons make the loss timescales dependent on the position $r$ within the SBN. Therefore, we compute the distribution $f_{\rm{CR}}(p,r)$ in Eq.~(1) for each $r$. We analytically estimate the pion production rate $q_{\pi}$ by assuming that each pion carry a fixed fraction $k_{\pi} = 17\%$ of the parent high-energy protons~\cite{Kelner:2006tc}. This brings us to
\begin{equation}
    q^{pp}_{\pi} (E_{\pi},r) = \frac{n_{\rm{ISM}}}{k_{\pi}} \sigma_{pp}\,\bigg(m_{p} +\frac{E_{\pi}}{k_{\pi}}\bigg)\,n_{p}\bigg(m_{p}+\frac{E_{\pi}}{k_{\pi}},r\bigg)\,,
\end{equation}
where $n_{\rm{ISM}}$ is the interstellar medium density and $n_{p}$ is defined through Eq.~\eqref{eq:energydistribution}. Regarding the gamma-ray emissions from inelastic DM-proton collisions, we consider the pion production to follow the proton-proton collisions
\begin{equation}
\label{eq:pion_production_inelastic_DM}
q^{\chi p}_{\pi} (E_{\pi},r) = \frac{\rho_{\chi}(r)}{m_{\chi} \, k_{\pi}}\, \sigma_{\text{inel}}\bigg(m_{p} +\frac{E_{\pi}}{k_{\pi}}\bigg)\,n_{p}\bigg(m_{p}+\frac{E_{\pi}}{k_{\pi}},r\bigg)\,.
\end{equation}
The emissivity of photons coming from neutral pion decay is 
\begin{equation}\label{eq:pi}
    Q_{\pi} (E,r) =2 \int_{E +m_{\pi}^2/4E}^{\infty} \frac{q^{pp}_{\pi}(E_{\pi},r) + q^{\chi p}_{\pi}(E_{\pi},r)}{\sqrt{E_{\pi}^2-m_{\pi}^2}}\, {\rm d}E_{\pi}\,.
\end{equation}
We also consider the primary and secondary bremsstrahlung processes, which lead to
\begin{equation}\label{eq:brem}
    Q_{\rm{brem}} (E,r) = \frac{n_{\rm{ISM}}  \,\sigma_{\rm{brem}}}{E}\int_{E}^{\infty} n_{e}(E_{e},r){\rm d}E_{e}\,,
\end{equation}
where $\sigma_{\rm{brem}} \simeq 3.4 \times 10^{-26}\, \rm{cm}^{-2}$ and $n_{e}$ is electron energy distribution inside the SBN defined through Eq.~\eqref{eq:energydistribution} as for protons. 
Finally, the inverse Compton scattering processes depend on the background photon distribution density which act as a target. Similar to Refs.~\cite{Ambrosone:2020evo,Ambrosone:2021aaw,Ambrosone:2022fip}, we implement it considering as a monochromatic spectrum peaking at $\epsilon_{\rm{peak}} = 0.1\, \rm{eV}$ with energy density $U_{\rm{rad}}$. Hence, we have
\begin{equation}\label{eq:compton}
    Q_{\rm{Compton}}(E,r) = \frac{3\sigma_{\rm{T}}}{4}\, \frac{U_{\rm rad}}{\epsilon_{\rm peak}^2} \int_{p_{\rm min}(E,\epsilon_{\rm{peak}})}^{\infty} f_{e}(p,r)\left(\frac{m_{e}}{E_{e}}\right)^2 G(q,\Lambda) 4 \pi p^2 {\rm d}p\,,
\end{equation}
where $p_{\rm{min}}$ and $G(q,\Lambda)$ are defined in Ref. \cite{Peretti:2018tmo}. The total gamma-ray spectrum at Earth can be then computed as
\begin{equation}
\label{eq:flux_at_earth}
    \Phi_{\gamma}(E,z) = \frac{{\rm Abs}(E(1+z))\, e^{-\tau_{\gamma\, \gamma} (E,z)}}{4\pi\, d_{c}(z)^2} \int_{V_{\rm SBN}} Q_{\rm{tot}}(E(1+z),r)\,{\rm d}V\,,
\end{equation}
where $Q_{\rm{tot}}(E,r)$ corresponds to the sum over all the contributions in Eqs.~\eqref{eq:pi}, \eqref{eq:brem} and \eqref{eq:compton}, $d_{c}(z)$ is the comoving distance between the source and the Earth, and the integral is performed over the SBN volume of. The quantity $\rm{Abs}(E)$ accounts for the internal gamma-ray absorption, which suppresses the flux above $1\, \rm{TeV}$: it is computed by averaging over all the possible lines of sight as in Ref.~\cite{Ambrosone:2020evo}. Moreover, $\tau_{\gamma \, \gamma}$ represents the optical depth for the CMB and EBL absorption, the latter modeled according to Ref.~\cite{Franceschini:2017iwq}. It is important to notice that, as long as the DM-proton interaction timescales are not competitive with respect to the others $(\tau^{\text{inel}}_{\chi p}, \, \tau^{\text{el}}_{\chi p}\gg \tau_{\rm{loss}},\tau_{\rm{adv}}) $, the integration in Eq.~\eqref{eq:flux_at_earth} results to be just a multiplication of the total production rate and the SBN volume. In the opposite scenario, when elastic $\chi$p collisions dominate $(\tau^{\text{el}}_{\chi p}\ll \tau_{\rm{loss}},\tau_{\rm{adv}})$, we have that protons mostly escape from the SBN and the gamma-ray spectrum becomes
\begin{equation}\label{eq:approximation_flux_DM}
    \Phi_{\gamma} \propto \int Q_{\pi}(r)\, {\rm d}V \propto \int \frac{Q_{p}(p,r)\, \tau^{\text{el,eff}}_{\chi p}(r)\, {\rm d}V}{V} \propto \int \frac{\rho_{\chi}^{-1}(r)\,{\rm d}V}{V}\,,
\end{equation}
where the first passage is due to the fact that the dominant contribution comes from the pion decay. Eq.~\eqref{eq:approximation_flux_DM} shows that the gamma-ray spectrum only depend on the average DM density inside the SBN rather than the actual DM profile. In the regime where $\tau_{\text{inel},\chi p}$ dominates over the others, the SBN starts being totally calorimetric. In this scenario the pion production, which is dominated by $q^{\chi p}_{\pi}(E_{\pi},r)$ (see Eq.~\eqref{eq:pion_production_inelastic_DM}), is just directly proportional to $Q_{p}$, therefore independent of $\rho_{\chi}$ and $m_{\chi}$. As a result, the gamma-ray spectrum is independent on $\rho_{\chi}$ as well as the structural parameters e.g. $R_{\mathrm{SBN}}$, $v_{\mathrm{wind}}$, $n_{\mathrm{ISM}}$ ($\Phi_{\gamma} \propto Q(p)$). All of this is crucial because it demonstrate that, the constraints for light DM particles from SBNi do not depend on the details of the DM profile. We end this section by showing in Fig.~\ref{fig:NGC 253_timescale_fluxes}, in the same way as for M82, the comparison between the standard timescales and elastic and inelastic $\chi$p timescales (left panel) for three different cases for $(\sigma_{\chi p}, m_{\chi})$. The corresponding fluxes are shown in the right panel. The spectral features are similar to the ones of M82.

\section{Linear Priors for the Sources}\label{app:linear_priors}

In Tab.~\ref{tab:linear_prior} we report the linear priors on the astrophysical nuisance parameters which are taken from the recent analysis on the SBN gamma-ray emission in Ref.~\cite{Ambrosone:2022fip}. For $\dot{M}^{*}$, we follow Refs.~\cite{Ambrosone:2021aaw,Ambrosone:2022fip} imposing the star-formation rates to be within a factor 3 with respect to the ones inferred through IR and UV observations~\cite{Kornecki:2020riv}. For the dimension of the nuclei $R_{\mathrm{SBN}}$, we account for a variation of a factor 2 with respect to the value expected for SBN circumnuclear region~\cite{Peretti:2018tmo,Peretti:2019vsj,Kennicutt:1997ng,Kennicutt:1998zb}. 
The range for the wind velocity is fixed according to the recent analyses~\cite{Peretti:2018tmo,Peretti:2019vsj,Peretti:2021yhc,Kornecki:2021xiy,Chevalier:1985pc}. Finally, for the gas density $n_{\mathrm{ISM}}$, the two sources require different priors according to the empirical Kennicut relation~\cite{Kennicutt:1997ng,Kennicutt:1998zb,Kennicutt_2021}.
\begin{table}[h!]
    \centering
    \begin{tabular}{c|c|c|c|c|c}
     Source & $\dot{M}_{*}\, [\mathrm{M}_{\odot}\, \mathrm{yr}^{-1}]$ & $\Gamma$ & $n_{\mathrm{ISM}}\, [\mathrm{cm}^{-3}]$ & $R_{\mathrm{SBN}}\, [\mathrm{pc}]$ & $v_{\mathrm{wind}}\, [\mathrm{km}\, \mathrm{s}^{-1}]$  \\ \hline
      M82  & $[3,30]$ & $[1,3]$ & $[100,400]$ & $[100,400]$ & $[200,1000]$ \\
      NGC 253 & $[1.4, 17]$ & $[1,3]$ & $[70,280]$ &  $[100,400]$ & $[200,1000]$ \\
    \end{tabular}
    \caption{Linear priors on the astrophysical nuisance parameters for the two sources adopted in the present analysis.}
    \label{tab:linear_prior}
\end{table}

Here, we also discuss the fact that the structural parameters might have a radial dependence (even though it should be a slight dependence due to the reduced dimension of the nuclei with respect to the dimension of the galaxies). The parameters whose radial dependence might impact our results are the following: DM profile, CR density and gas density. Regarding the DM profile, we discuss in the following section, the implications of different profiles. 
Regarding the CR density as well as the gas density, they could only impact our results in case their distribution correlate with the DM density, generating particular regions where the timescales of interaction becomes effectively lower than the ones we consider in the main text. However, since we cannot resolve such regions, what matters is the average of these parameters over the zone of interest. The values we chose are certainly of the correct order of magnitude, and furthermore the normalization and spectral shape of the flux is chosen to well fit the gamma-ray data, thus making our bounds very robust against these uncertainties.

\section{Constraints Dependence on the Dark Matter Profile}\label{app:halo}

Along with the NFW profile, we also consider the Burkert profile~\cite{Burkert:1995yz,Lin:2019yux}
\begin{equation}\label{bukertprofile}
    \rho_\chi (r) = \frac{\rho_{0}\,r_{0}^{3}}{(r+r_{0})(r^{2}+r_{0}^2)}\,,
\end{equation}
where the quantities $\rho_{0}$ and $r_{0}$ are defined by means of the parameters $c_{200}$ and $M_{200}$ in an analogous way as for the NFW profile. The main difference between the Burkert and NFW profiles is the fact the that the former is not divergent at $r = 0$, leading to a different average density within the SBN.
\begin{figure*}[t!]
    \centering
    \includegraphics[width=0.95\textwidth]{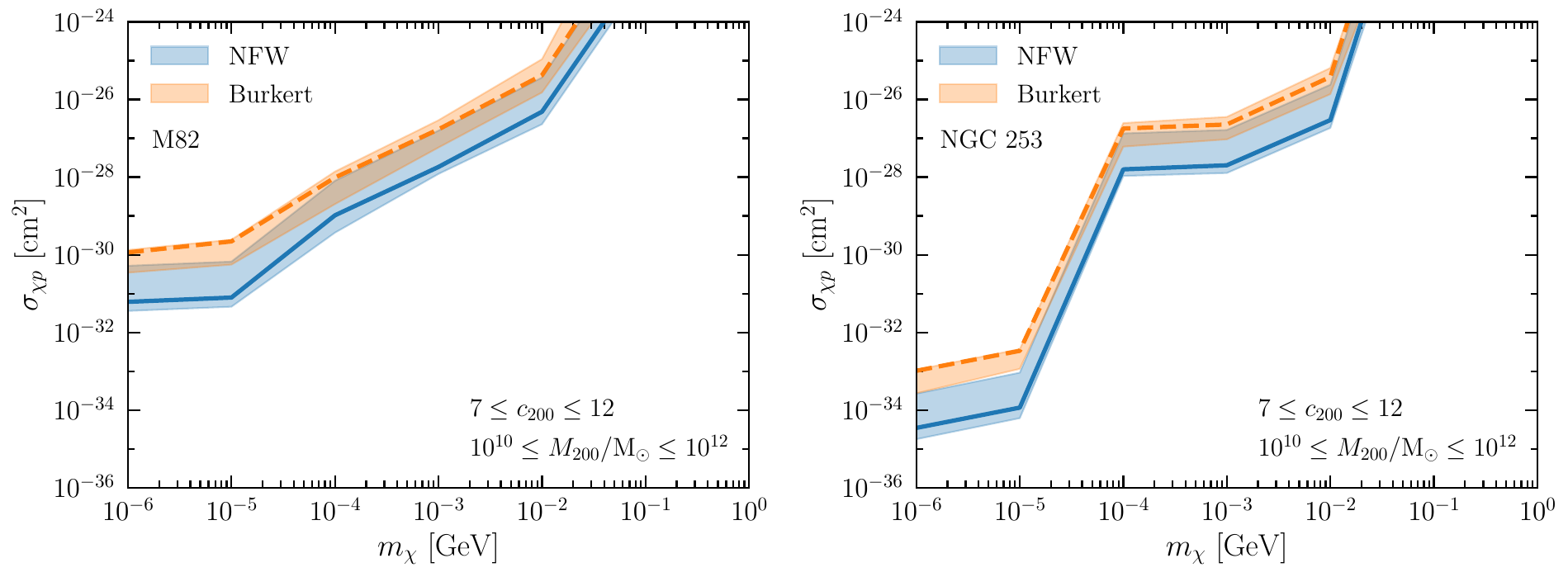}
    \caption{Dependence of the current DM constraints on the halo parameters $c_{200}$ and $M_{200}$ for the NFW (blue band) and Burkert (orange band) profiles, in case of M82 (left panel) and NGC 253 (right panel).}
    \label{fig:bukertconstraint}
\end{figure*}
Fig. \ref{fig:bukertconstraint} shows the constraints bands for M82 (on the left) and NGC 253 (on the right) in case of the Burkert profile (orange band) and the NFW one (blue band). The bands represent the uncertainty due to the different values for $c_{200}$ and  $M_{200}$ within the ranges $[7,12]$ and $[10^{10},10^{12}]~{\rm M}_\odot$, respectively. Under the same values for $(c_{200}, M_{200})$, the bounds imposed using a Burkert profile are usually a factor of 20 weaker than the ones with the NFW profile. The dependence of the constraints on the DM halo parameters is also of one order of magnitude. Hence, we can estimate the variability of the bounds combining the uncertainties from the profile and its parameters to be about two orders of magnitude.

\section{CTA Mock data Generation and Forecast}\label{app:cta}

We generate mock spectral energy distributions (SEDs) for both M82 and NGC 253 under the hypothesis of no DM-CR interactions. Following the approach detailed in Ref.~\cite{Ambrosone:2022fip}, we make only use of the public expected instrument response functions of the detector in ideal conditions~\cite{CTAConsortium:2018tzg} to evaluate the expected number of signal $n_{s}$ and background events $n_{b}$, in an observation time $T_{\rm{obs}}$ of $50$ hours. Starting from this information, we randomly generate a number of events following a Poissonian distribution with a mean value of $n_{\rm{tot}} = n_{s} + n_{\rm{b}}$ and then we subtract the expected number of background to evaluate the empirical number of signal events $\tilde{n}_{s}$. From this quantity, the empirical SED can be evaluated by assuming a generic $E^{-2}$ spectrum as%
\begin{equation}
    {\rm SED}_{i} = \frac{\tilde{n}_{s, i}}{T_{\rm obs} \int_{\Delta E} A_{\rm eff}(E) (E/1\,{\rm GeV})^{-2} \,{\rm d}E}\,,
\end{equation}
where $A_{\rm{eff}}$ is the CTA effective area.  Due to the time-consuming analysis, we only simulate 50 mock data sets as representative for the data variability. For each data sample we perform the same statistical analysis detailed in the previous sections and obtain projected bounds.

\begin{figure*}[t!]
    \centering
    \includegraphics[width=0.95\textwidth]{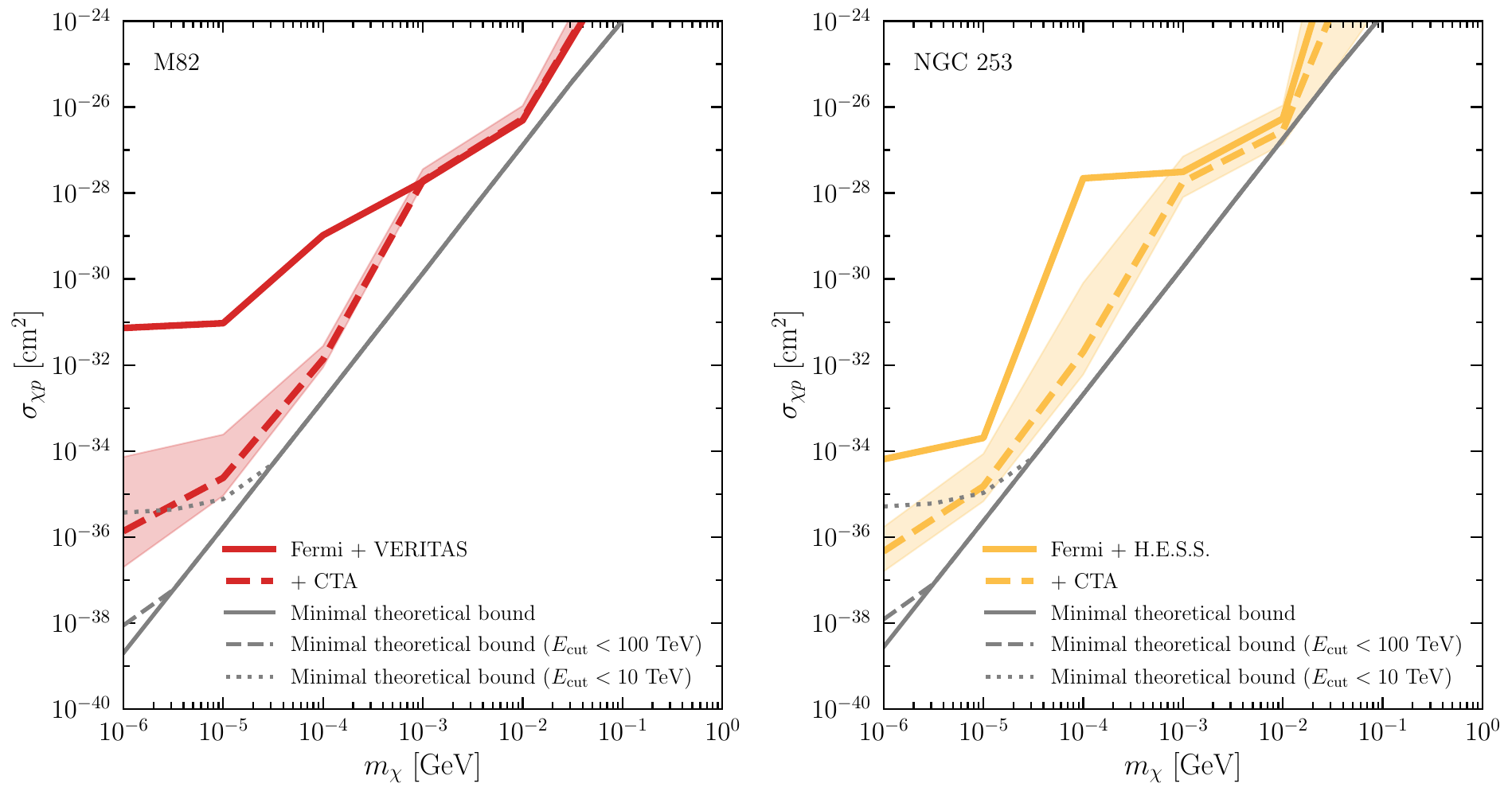}
    \caption{Comparison between the current constraints (solid lines) and the future ones (shaded bands) for the CTA telescope~\cite{CTAConsortium:2018tzg} in case of M82 (left panel) and NGC 253 (right panel). The dashed lines represent the average bounds imposed through current and CTA mock data. The black lines represent the minimal theoretical bounds reachable through SBN according to Eq.~\eqref{eq:min_bound}.}
    \label{fig:forecast}
\end{figure*}

Fig.~\ref{fig:forecast} shows the band of the constraints obtained for CTA from all of the data samples; the width of the bands quantifies the variability of the obtained bounds due to the Poisson fluctuations in the detected photons. In the low mass region, this variability can reach up to more than an order of magnitude for M82, but stays within less than an order of magnitude for NGC 253, due to the larger expected number of events. We also show as black lines  reference minimum theoretical bounds, obtained requiring that the expected number of DM-CR scatterings in the range below a cut energy $E<E_\mathrm{cut}$ is larger than 1, namely
\begin{equation}\label{eq:min_bound}
    \mathrm{min}_{E<E_{\text{cut}}}\left[\tau^\mathrm{el, eff}_{\chi p} \left(\frac{1}{\tau_{\rm esc}} + \frac{1}{\tau^{\rm eff}_{\rm loss}}\right)\right]=1\,.
\end{equation}
This means that the flux below the cut energy suffers by distortions smaller than $50\%$, and therefore cannot be constrained. The cut energy is meant to simulate the energy range accessible by gamma-ray experiments; for example, present gamma-ray data for M82 and NGC 253 are known up to a gamma-ray energy of about 1~TeV, which are hadronically produced by protons with energy of about 10~TeV. We show these theoretical bounds for the three choices $E_\mathrm{cut}=10$~TeV, $100$~TeV, and arbitrarily large $E_\mathrm{cut}$. CTA closely approaches the minimum theoretical bound, but this argument shows that there is still space for improvement in the bounds with an increased precision of the experiment in the gamma-ray energy range below $10$~TeV. All in all, this clearly demonstrate the potentiality for SBN to be complementary tools to collider for constraining DM particle properties.

\section{Dependence of bounds on CR composition}\label{sec:heavier_CR}

The bounds on DM-CR interactions might be affected by a possible contamination from heavier nuclei in the CR spectrum. The current gamma-ray data are not sensitive enough to discriminate the presence of heavier nuclei in SBN and, in general, more complicated spectra. However, similarly to the Milky Way~\cite{Gaisser:2013bla}, at the energies of our interest we expect that the species which could provide a greater contamination are helium nuclei. Hence, for sake of concreteness, we estimate the impact of helium contamination on DM limits by assuming that the interstellar medium is predominantly made of protons and that the helium injection $Q_\mathrm{He}$ follows the same power-law of protons (see also \cite{Ambrosone:2020evo} for details)
\begin{equation}
   Q_{\rm He} (p,R_{SN}) = \frac{\mathcal{N}_{\rm He}\,R_{\rm SN}}{V_{\rm SBN} } p^{-(\Gamma +2)} e^{-p/p^{\rm max}}
\end{equation}
where $R_{\rm SN}$ is the supernovae rate proportional to the star-formation rate $\dot{M}_{*}$, $B_{\rm SBN}$ is the SBN volume, and $p^{\rm max}$ defines the cuf-off in the momentum space. We set $p^{\rm max} = 5\, \rm{PeV}$ which is half the value we assume for protons, since the maximum energy that can be reached vary as a function of the atomic number of the CR species~\cite{Blasi_2012}. However, this value does not impact our analysis, since we focus on gamma-rays up to $\sim 1-10\ \rm{TeV}$. Moreover, we fix the normalization $\mathcal{N}_{\rm He}$ by requiring that helium CRs take $\xi = 10\%$ of the $E_{\rm{SN}}=10^{51}~{\rm erg}$ released by supernovae.

Regarding the dynamical timescales for the helium nuclei, we consider ionization, advection and interactions with the protons in the interstellar medium. We instead neglect the diffusion timescale, since it has a very marginal impact on the CR transport inside SBN and has no impact on our results. For ionization, we consider the same expression for the protons (see~\cite{Peretti:2019vsj}), scaled up by a factor 2 to account for the variation of the energy loss rate by $Z/A$ where $Z$ is the atomic number and $A$ is the mass number. We neglect the Coulomb timescale which is totally negligible above $E>1\, \rm{GeV}/\rm{nucleon}$ (along with the ionization timescale) \cite{Evoli:2008dv}. For advection, we use the same expression as for the protons $\tau_{\rm{adv}} = R_{\rm SBN}/v_{\rm{wind}}$. Regarding the interactions with the interstellar medium, we consider
\begin{equation}\label{eq:tau_phe}
    \tau_{p\rm He} = \left(n_{\rm{ISM}} \,\sigma_{p\rm He}(p) \frac{\kappa}{A} \frac{A+1}{2}\right)^{-1}
\end{equation} 
where the factor $(A+1)/2$ accounts for the helium increase of the multiplicity of pions which carry on average $k_\pi/A$ of the kinetic energy of the parent CR \cite{Joshi:2013aua} with $k_\pi=\kappa/3=17\%$~\cite{Peretti:2019vsj}. Following Ref.~\cite{Luque:2022buq}, we define the total cross section $ \sigma_{p\rm He}(p)$ by rescaling the $pp$ cross section of a factor $A^{2/3}$ and taking into account the different energy threshold with respect to the case of $pp$ interactions~\cite{Kafexhiu:2014cua}.
\begin{figure}[t!]
    \centering
    \includegraphics[width=0.5\textwidth]{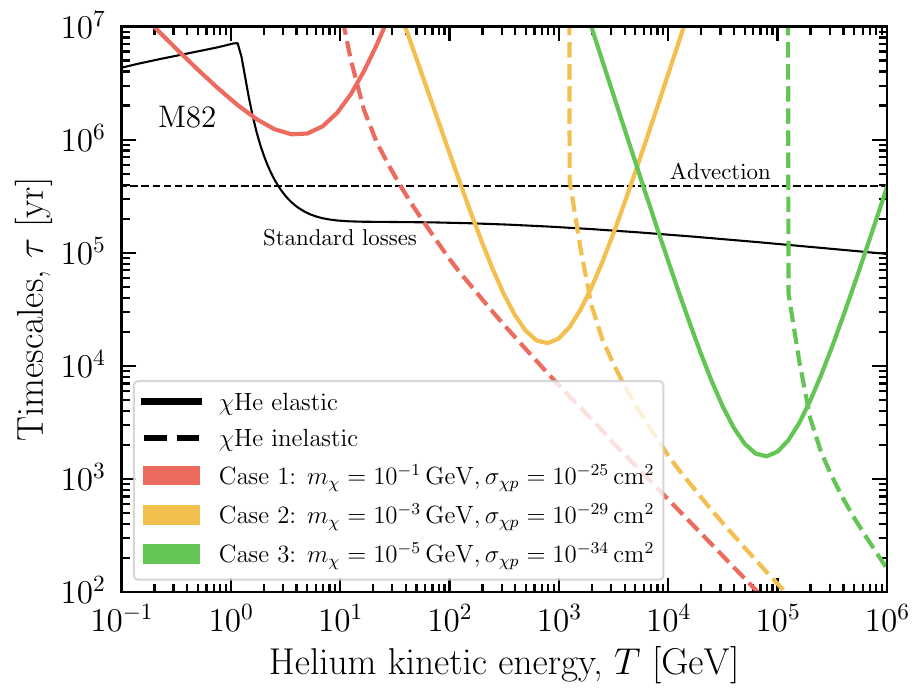}
    \caption{The M82 timescales of the processes involving Helium nuclei as a function fo the Helium kinetic energy $T$. The continuous and dashed black lines represent the standard losses and advection timescales, respectively. The colored continuous (dashed) lines correspond to the elastic (inelastic) DM interactions for three different cases.}
    \label{fig:He_timescale}
\end{figure}

DM particles interact with the nucleons in the Helium nuclei. Hence, the differential DM-He elastic cross section features an $A^2$ enhancement and, in particular, takes the following expression~\cite{Ema:2020ulo}
\begin{equation}\label{eq:differentialcrosssection}
    \frac{{\rm d}\sigma_{\rm{\chi He}}^\mathrm{el}}{{\rm d}T_\chi} = \frac{A^2\,\sigma_{\chi p}}{ T^{\rm max}_\chi} \frac{F_{\rm He}^{2} (q^2)}{16 \,\mu_{\chi \rm He}^2\,s}  (q^2 + 4m^2_{\rm He})(q^2 + 4m^2_\chi)\,,
\end{equation}
where $\sigma_{\chi p}$ is the DM-proton cross section at zero center-of-mass momentum (assuming an isospin-independent interaction), $s = m_{\chi}^{2} + m_{\rm{He}}^{2} + 2Em_{\chi}$ is center-of-mass energy, $F_{\rm He}$ is the helium form factor given by
\begin{equation}\label{eq:formfactor_helium}
    F_{\rm He}(q^2) = \left(\frac{1}{1+ q^2/\Lambda^2}\right)^2 \quad {\rm with}\quad \Lambda = 0.410 \, \rm{GeV}\,,
\end{equation}
and $T^{\rm max}_\chi$ is the maximum recoil DM kinetic energy given by
\begin{equation}\label{eq:energygain}
    T^{\rm max}_\chi = \frac{2T^2 + 4m_{\rm He} T}{m_\chi}\left[\left(1+\frac{m_{\rm He}}{m_\chi}\right)^2 + \frac{2 T}{m_\chi}\right]^{-1} \,,
\end{equation}
with $T = E -m_{\rm He}$. Finally, following the proton case, the timescale for the energy loss from inelastic DM-He collisions is defined as
\begin{equation}\label{eq:inelastic_DM_He}
     \tau^\mathrm{inel}_{\chi \rm He} = \left(\frac{\kappa}{A}\,
     \frac{A+1}{2}\,A^{2/3}\sigma_{\rm inel}\, \frac{\rho_{\chi}}{m_{\chi}} \right)^{-1}\,,
\end{equation}
where we consider the same factors (pion multiplicity and average inelasticity) as for the $p{\rm He}$ interactions and scale the $\chi p$ inelastic cross section by a factor $A^{2/3}$.

In Fig.~\ref{fig:He_timescale} we show the timescales for the helium nuclei inside the M82 starburst galaxy. The black lines correspond to the standard processes (losses and advection), while the colored ones are the timescales for the elastic (solid) and inelastic (dashed) interactions with DM particles. The timescales have similar behaviour as for protons: the only difference is that the elastic $\chi$He have their minimum at higher energies due to the higher CR mass. Similarly to the case analysed in the main text, the inelastic timescale takes over for energies above the dip. As for the protons, we compute the pion production from the $p$He and the $\chi$He inelastic collisions which take the following expressions
\begin{equation}\label{eq:p_He_pion_production_rate}
    q^{p\rm He}_{\pi} (E_{\pi},r) = \frac{A+1}{2}\frac{n_{\rm{ISM}}\,A}{k_{\pi}} \sigma_{p{\rm He}}\,\left(m_{\rm He} +\frac{A\,E_{\pi}}{k_{\pi}}\right)\,n_{\rm He}\left(m_{\rm He}+\frac{A\,E_{\pi}}{k_{\pi}},r\right)\,,
\end{equation}
and 
\begin{equation}\label{eq:pion_production_inelastic_DM_He}
    q^{\chi \rm He}_{\pi} (E_{\pi},r) =\frac{A+1}{2} \frac{\rho_{\chi}(r)}{m_{\chi}}\frac{A}{k_{\pi}}\, A^{2/3}\sigma_{\text{inel}}\left(m_{\rm He} +\frac{A\,E_{\pi}}{k_{\pi}}\right)\,n_{\rm He}\left(m_{\rm He}+\frac{A\,E_{\pi}}{k_{\pi}},r\right)\,.
\end{equation}
Then, we compute the gamma-ray production by inserting Eqs.~\eqref{eq:p_He_pion_production_rate} and~\eqref{eq:pion_production_inelastic_DM_He} in Eq.~\eqref{eq:pi} as done for the protons.
\begin{figure}[t!]
    \centering
    \includegraphics[width=0.95\textwidth]{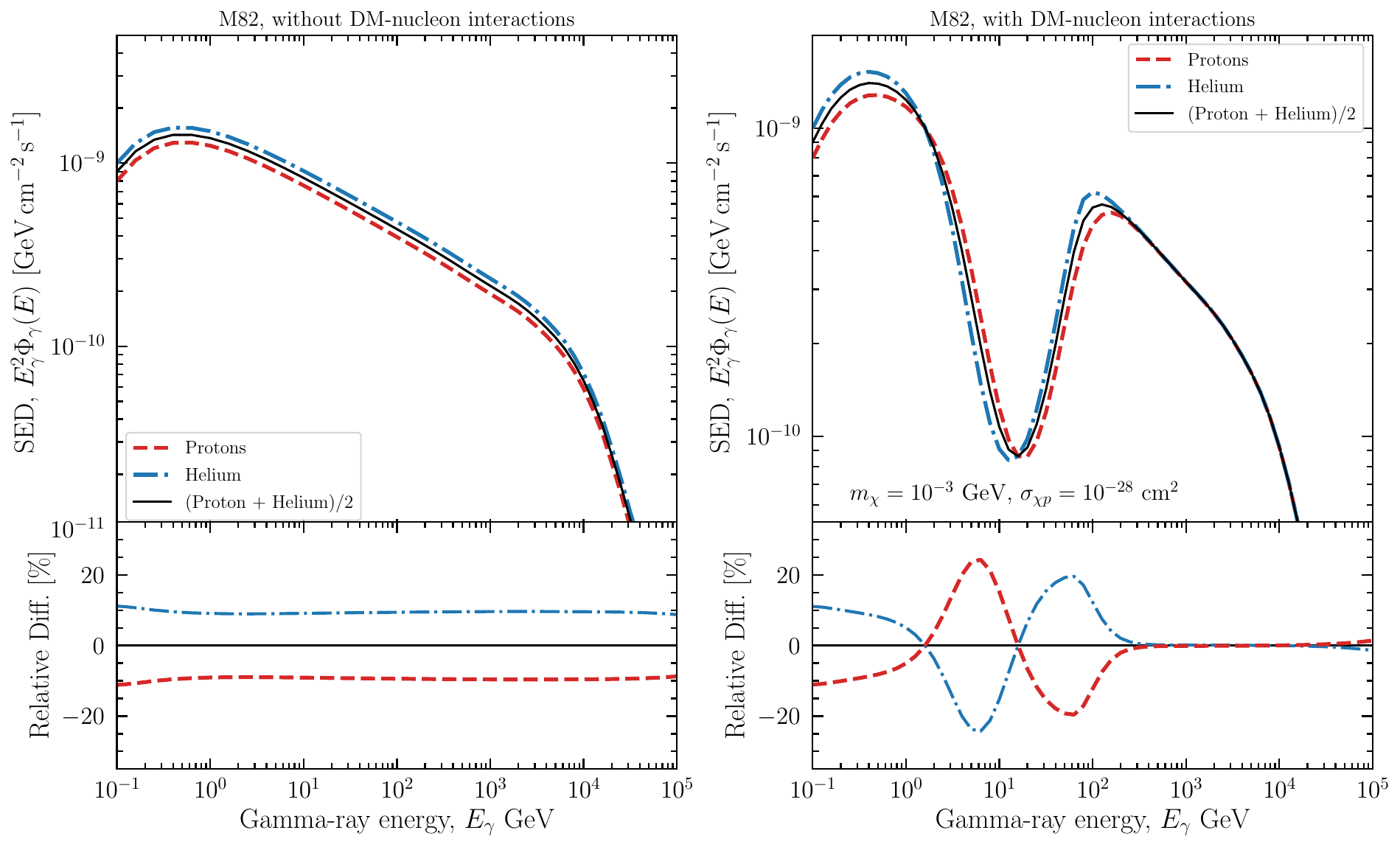}
     \caption{M82 gamma-ray fluxes from protons (blue dot-dashed lines), Helium nuclei (dashed red lines), and their weighted sum (solid black lines) without (left panel) and with (right panel) DM-nucleon interaction. All the fluxes include the marginal leptonic component at low energies. The lower panels display the relative percentage difference of the proton and Helium gamma-ray spectra with respect their weighted sum.}
     \label{fig:gamma_rays_helium}
 \end{figure}

In Fig.~\ref{fig:gamma_rays_helium} we show the gamma-ray spectra obtained without (left panel) and with (right panel) the interaction with DM particles in case of the M82 starburst galaxies. In the latter case, we take $m_{\chi}= 10^{-3}\, \rm{GeV}$ and $\sigma_{\chi p} = 10^{-28}\, \rm{cm}^{2}$. The red and blue lines correspond to the gamma-ray spectra produced by protons and helium nuclei, respectively, while the black lines refer to the total gamma-ray flux divided by a factor of 2 in order to make the comparison easier. Remarkably, we find that the presence of heavier nuclei does not erase the dip in the spectrum, but only shifted in energy, while maintaining more or less the same depth compared to the standard spectrum. We note that the gamma-ray dip for pure helium composition lies at energies slightly lower than the one for protons, even though the dip in the He spectrum is actually at higher energies. This is due to the lower energy fraction carried by pions in the case of heavier nuclei, which moves the dip in the gamma-ray spectrum to lower energies. Moreover, we find that, even in the standard case, the helium nuclei provide a slightly higher gamma-ray spectrum for the same assumption of the star-formation rate, as expected according to a higher pion multiplicity combined with only partial calorimetric condition inside the SBN. On the other hand, at large enough energies, in the regime dominated by DM-CR inelastic scatterings, the gamma-ray spectrum is entirely independent of the chemical CR composition due to the fully calorimetric nature of the DM-CR interactions. 

Hence, the presence of heavier nuclei (e.g. helium nuclei) in the small region of the SBN do not wash out the spectral features induced by the DM-CR elastic and inelastic interactions. By performing the likelihood analysis delineated in the main text with different fraction of the Helium contamination (from 0\% to 50\% with respect to the total CR injection), we have actually found that the constraints on the DM-nucleon cross section $\chi p$ improve by less than an order of magnitude. For these reasons, we conclude that our results very robust against the uncertainty affecting the CR contamination.

\end{document}